\documentclass[12pt]{article}
\usepackage[letterpaper, left=1in, top=1in, right=1in, bottom=1in,nohead,includefoot]{geometry}
\usepackage{color,charter,graphicx,natbib,here,setspace,multirow,amsmath,amssymb,amsfonts,enumitem,fancyvrb}
\usepackage{graphicx}
\usepackage{rotating}
\usepackage{multirow}
\usepackage{booktabs}
\usepackage{color}
\usepackage{setspace}
\usepackage{algorithm2e}
\usepackage{enumitem}
\usepackage[bottom]{footmisc}
\usepackage[colorlinks=true,urlcolor=blue,linkcolor=blue,citecolor=blue]{hyperref}
\usepackage{bigstrut}
\usepackage{threeparttable}
\usepackage{dcolumn}
\usepackage{amsmath}
\usepackage{longtable}
\usepackage{amssymb}
\pdfoutput=1
\VerbatimFootnotes 
\usepackage[title,titletoc]{appendix}
\makeatletter

\renewenvironment{appendices}{%
    \begin{oldappendices}%
    \renewcommand{\thefigure}{\ifnum \c@section>\z@ \thesection.\fi\@arabic\c@figure}%
    \@addtoreset{figure}{section}%
    \renewcommand{\thetable}{\ifnum \c@section>\z@ \thesection.\fi\@arabic\c@table}%
    \@addtoreset{table}{section}%
}{%
    \end{oldappendices}%
}\makeatother

\usepackage{natbib}
\let\natbibcitet\citet
\renewcommand\citet{\bibpunct{(}{)}{,}{a}{,}{,}\natbibcitet}
\let\natbibcitep\citep
\renewcommand\citep{\bibpunct{(}{)}{;}{a}{,}{;}\natbibcitep}
\newcommand{\bi}{\begin{itemize}}
\newcommand{\ei}{\end{itemize}}
\newcommand{\be}{\begin{equation}}
\newcommand{\ee}{\end{equation}}
\defcitealias{ieo14}{IEO, 2014}

\newcommand{\vzero}{\mathbf{0}}
\newcommand{\vone}{\mathbf{1}}

\newcommand{\ve}{\mathbf{e}}
\newcommand{\vh}{\mathbf{h}}
\newcommand{\vu}{\mathbf{u}}
\newcommand{\vv}{\mathbf{v}}
\newcommand{\vx}{\mathbf{x}}
\newcommand{\vy}{\mathbf{y}}

\newcommand{\mB}{\mathbf{B}}
\newcommand{\mE}{\mathbf{E}}
\newcommand{\mI}{\mathbf{I}}
\newcommand{\mW}{\mathbf{W}}
\newcommand{\mX}{\mathbf{X}}
\newcommand{\mY}{\mathbf{Y}}
\newcommand{\mZ}{\mathbf{Z}}

\newcommand{\valpha}{\text{\boldmath{$\alpha$}}}
\newcommand{\vbeta}{\text{\boldmath{$\beta$}}}
\newcommand{\veps}{\text{\boldmath{$\epsilon$}}}
\newcommand{\veta}{\text{\boldmath{$\eta$}}}
\newcommand{\vmu}{\text{\boldmath{$\mu$}}}

\newcommand{\mOmega}{\mathbf{\Omega}}
\newcommand{\mSigma}{\mathbf{\Sigma}}

\newcommand{\IW}{\mathcal{IW}}
\newcommand{\LN}{\mathcal{LN}}
\newcommand{\N}{\mathcal{N}}

\long\def\symbolfootnote[#1]#2{\begingroup%
\def\thefootnote{\fnsymbol{footnote}}\footnote[#1]{#2}\endgroup}

\usepackage{sectsty}
\allsectionsfont{\normalsize}

\makeatletter

\@addtoreset{equation}{section}
\makeatother

\usepackage{caption}
\usepackage[labelformat=simple]{subcaption}
\makeatletter\let\p@subfigure\thefigure\makeatother

\usepackage{cleveref}
\crefname{chapter}{Chapter}{Chapters}
\crefname{section}{Section}{Sections}
\crefname{subsection}{Section}{Sections}
\crefname{subsubsection}{Section}{Sections}
\crefname{figure}{Figure}{Figures}
\crefname{table}{Table}{Tables}
\crefname{equation}{Equation}{Equations}
\crefname{appendix}{Appendix}{Appendices}

\newcolumntype{d}[1]{D{.}{.}{#1}}

\title{
{{How does monetary policy affect income inequality in Japan? Evidence from grouped data}
}} %A Markov switching factor-augmented VAR model for analyzing US business cycles and unconventional monetary policy
\date{}
\usepackage{authblk}

\author[1,2]{Martin Feldkircher}
\author[3]{Kazuhiko Kakamu\thanks{Martin Feldkircher, Vienna School of International Studies (DA), Favoritenstra\ss e 15a, 1040 Vienna, Austria. Email: \texttt{martin.feldkircher@da-vienna.ac.at}.  Kazuhiko Kakamu, Graduate School of Economics, Nagoya City University, Yamanohata 1, Mizuho-cho, Mizuho-ku, Nagoya 467-8501 Japan.
Email: \texttt{kakamu@econ.nagoya-cu.ac.jp}. Any views expressed in this paper represent those of the authors only and not necessarily  of the Oesterreichische Nationalbank. We would like to thank participants of the Seminar in Statistics at Collegio Carlo Alberto, Italy, participants at the  11th International Conference on Computational and Financial Econometrics (CFE 2017), London and Jouchi Nakajima for helpful comments.}}
\affil[1]{Vienna School of International Studies (DA)}
\affil[2]{Oesterreichische Nationalbank (OeNB)}
\affil[3]{Nagoya City University}


\def\equationautorefname~#1\null{%
  Eq.~(#1)\null
}
\def\equationautorefname~#1\null{
Eq.~(#1)\null
}
\begin{document}
\graphicspath{{Figs/}}
\maketitle

\begin{abstract}
We examine the effects of monetary policy on income inequality in Japan. For that purpose, we use a novel Bayesian approach that jointly estimates the Gini coefficient from grouped income data and the dynamics of macroeconomic quantities. Our results indicate different effects on income inequality for different types of households: A monetary tightening decreases inequality when we consider a broad definition of household data that also includes the unemployed and retirees. Higher unemployment and tighter borrowing conditions make the richer (i.e., employed) comparably worse off. The result reverses, if we focus on the sub-sample of households whose head is employed. Through the same channels, inequality increases if monetary policy is tightened. A counterfactual analysis reveals that indeed the financial channel and the job destruction channel are the most important transmission mechanisms.
\end{abstract}

\bigskip
\begin{tabular}{p{0.2\hsize}p{0.65\hsize}} %0.15
\textbf{Keywords:}  &Income inequality; Monetary policy; Grouped data; Bayesian analysis.
\end{tabular}

\smallskip

\begin{tabular}{p{0.2\hsize}p{0.4\hsize}}
\textbf{JEL Codes:} & C30, E52, F41, E32. \\
\end{tabular}
\vspace{0.6cm}
%\begin{center}
%\date{\small{\today }}
%\end{center}
 
\bigskip
%This draft: \today

\newpage
\section{Introduction}

While there is a long-standing literature on the nexus between income inequality and economic growth, income distribution and monetary policy have been traditionally considered separate issues. The outbreak of the global financial crisis and the subsequent recession led  major central banks to loosen their monetary policy stance. 
This spiked new interest in the relationship between monetary policy and income inequality -- also since a lot of these measures comprised asset purchases by the central bank and it is known that price movements in assets should impact inequality. 
Hitherto, empirical work did not provide clear evidence, also since transmission channels seem complex and depend on the distribution of financial assets and liabilities in the population - at best, a modest effect of monetary policy on inequality can be found in the data \citep{Amaral2017}. In recent contributions,  \citet{Coibion2017} for the USA, \citet{Mumtaz2017} for the UK and \citet{Furceri2017} for a broader set of countries, a systematic impact of monetary policy on inequality has been established: If monetary policy is tightened, inequality increases.

One reason why the literature on the monetary policy and inequality nexus is scarce might be that effects of monetary policy in a macroeconomic context are frequently addressed within a time series framework such as a vector autoregression (VAR). Income data, however, is often collected in surveys that are available only over a short time period, or isolated years. For example, the Japanese National Survey of Family Income and Expenditure offers very detailed information on household income and expenditures but is carried out every five years only. To circumvent the paucity of continuous time series on income data, researchers have used  grouped data that are constructed by aggregating individual observations of a variable into income groups and are available over a longer time horizon. Using these data directly, however,  would imply the  stark assumption that all households within a group have the same income \citep{Chotikapanich2007}. A better alternative is to estimate the underlying distribution using suitable econometric tools \citep[see, e.g.,][]{Chotikapanich2007a}.

In this paper, we investigate the effects of monetary policy on income inequality in Japan. The Bank of Japan (BoJ) can be considered a front runner in deploying different forms of monetary policy besides interest rate changes \citep[e.g., see][]{shirakawa2010roles, bernanke2017some}. From the perspective of other central banks, Japanese data can thus provide useful insights on how these measures affect income inequality.  We propose a novel econometric approach that jointly models the Gini coefficient based on grouped income data of households and the dynamics of macroeconomic quantities. In that sense we combine the predominant framework used in monetary economics, VAR models, and the methods used to estimated income distributions. The proposed model is an extension of the work of \citet{Nishino2012} to the VAR case. More precisely, \citet{Nishino2012} estimate income inequality using a state-space representation with the state variable reflecting income inequality. In this paper, we take this approach one step further and include the state variable as an endogenous variable into a VAR system. In other words,  our model allows for a coherent and joint estimation of macroeconomic key variables and income inequality. This is a significant improvement over existing approaches that calculate measures of inequality in a first step and treat them as observed data in the subsequent estimations. As pointed out in \citet{Carriero2018} this might lead to a severe  bias and flawed inference.  

Our main results are as follows. First, we find that the effect of monetary policy on income inequality depends on the household under consideration. A monetary tightening leads to a decrease in inequality if we consider a broader measure of income that also covers households that are unemployed or retired. The Gini decreases since the monetary tightening leads to more unemployment and worse borrowing conditions. Both affect rather the employed households which -- compared to the unemployed / retired -- makes income more equal. The same channels are at play if we consider the sub-sample of households whose head is employed. However, here, more unemployment and tighter financing conditions lead to more inequality since it affects poorer households -- in that sample -- more strongly. A counterfactual analysis corroborates that monetary policy transmits most strongly through the unemployment rate and long-term interest rates, while other potential transmission channels such as through inflation or the exchange rate are less important. Our results are robust to different ways of identifying the monetary policy shock and more generally measures of monetary policy.

The paper is structured as follows: The next section introduces the model, while section \ref{sec:data} summarizes the data. Section \ref{sec:emp} presents the main findings from the estimation, while section \ref{sec:concl} concludes.

\section{The model}\label{sec:model}
In this section we describe the econometric approach. We start with outlining the basic idea of how to estimate an income distribution from grouped data. We follow \citet{Nishino2011} and assume that income $x$ is log-normally distributed\footnote{Alternatively one could estimate a Pareto distribution or more generally a generalized beta distribution, for the latter, however, it is not straightforward to calculate the Gini coefficient; see for example \cite{McDonald2008}. More recently,  \citet{Kobayashi21} proposed a method to estimate Lorenz curves assuming more flexible distributions than the log-normal distribution. However, their method is not directly applicable to our modeling framework. That said, it should be noted that there is empirical evidence that the log-normal distribution fits Japanese income data well \citep{Nishino2011}.}, $x \sim \LN(\mu,\sigma^{2})$. The associated probability density
 function (PDF) is then given by
\begin{eqnarray*}
	f(x | \mu, \sigma^{2}) = \frac{1}{x \sqrt{2 \pi \sigma^{2}}} \exp\left\{ -\frac{(\ln x - \mu)^{2}}{2 \sigma^{2}} \right\}.
\end{eqnarray*}
The cumulative density function (CDF) is 
\begin{eqnarray*}
	F(x) = \Phi\left( \frac{\ln x - \mu}{\sigma} \right),
\end{eqnarray*}
where $\Phi(\cdot)$ denotes the CDF of the standard normal distribution.
Then, the Gini coefficient based on the lognormal distribution is defined as
\begin{eqnarray*}
	G = 2 \Phi\left( \frac{\sigma}{\sqrt{2}} \right) - 1,
\end{eqnarray*}
and we notice that the Gini coefficient is a monotone function of $\sigma$. 

Assume that the income distribution is $x \sim \LN(\mu, \sigma^{2})$ with $p_{i}$th quantiles ($i = 1,\ldots, k$), where $\displaystyle p_{i} = \frac{n_{i}}{n}$ denote the relative frequencies and $1 \le n_{1} < n_{2} < \ldots < n_{k} \le n$, are observed as $(x_{1}, x_{2},\ldots, x_{k})$ from $n$ observations. In the empirical application, $x_k$ is going to denote the endpoint of each income category (as opposed to the mean value).
Let $\vx = (x_{1}, x_{2},\ldots, x_{k})^{\prime}$, then by the asymptotic theorem of the selected order statistics,
\begin{eqnarray*}
	\sqrt{n}( \ln \vx - \mu - \sigma \vu ) \sim \N\left( \vzero_{k}, \sigma^{2} \mW \right)\quad \text{as $n \to \infty$},
\end{eqnarray*}
where $\vu = (u_{1}, u_{2}, \ldots,u_{k} )^{\prime}$, $u_{i} = \Phi^{-1}(p_{i})$, $\Phi^{-1}(\cdot)$ is the inverse distribution function, $\vzero_{k} = (\underbrace{0, 0, \ldots, 0}_{k})^{\prime}$ and the $ij$th element of $\mW$ is defined as
\begin{eqnarray*}
	w_{ij} = w_{ji} = \frac{p_{i}(1-p_{j})}{\phi\left( \Phi^{-1}(p_{i}) \right) \phi\left( \Phi^{-1}(p_{j}) \right)},\quad (i \le j),
\end{eqnarray*}
where $\phi(\cdot)$ is the PDF of the standard normal distribution. The proof of these asymptotic properties is provided in \citet{Nishino2011}. The assumption of a constant variance of the income distribution is relaxed in \citet{Nishino2012}. \citet{Nishino2015} show that the income inequality in Japan is indeed persistent and can be modeled using a stochastic volatility model. Here, we follow \citet{Nishino2015} and allow the variances of the income distribution to drift over time.

Let $\vx_{t} = (x_{1,t}, x_{2, t}, \ldots, x_{k, t})^{\prime}$ and $\vy_{t} = (y_{1, t}, y_{2, t}, \ldots, y_{m, t})^{\prime}$ be the income data and macroeconomic variables of the $t$th period ($t = 1, 2, \ldots, T$), respectively.
Then, we propose the  following joint model:
\begin{eqnarray}
	\ln \vx_{t} &=& \mu_{t} + \exp( h_{t} / 2 ) \vu_{t} + \exp(h_{t} / 2) \veps_{t},\quad \veps_{t} \sim \N\left( \vzero_{k}, \mW_{t}^{*} \right),
	\label{eq:model1}\\
	\vy_{t} &=& \valpha + \mB \vy_{t-1} + \veta_{t},\quad \veta_{t} \sim \N\left( \vzero_{m}, \mSigma \right),
	\label{eq:model2}
\end{eqnarray}
where $\vu_{t} = (u_{1,t}, u_{2,t}, \ldots, u_{k,t})^{\prime}$, $\displaystyle \mW_{t}^{*} = n_{t}^{-1}\mW_{t}$, $y_{1,t} = h_{t}$, $\valpha$ is an $m \times 1$ vector of parameters, and $\mB$ and $\mSigma$ are $m \times m$ matrices of parameters, and we assume $\vy_{0} = \vzero_{m}$ for simplicity.\footnote{It should be mentioned that this model is a straightforward extension of the simple stochastic volatility model. The state variable $h_{t}$ means the income inequality and this variable is used as a variable in the VAR system. Therefore, it can be seen as the joint estimation of a stochastic volatility model and VAR model. In the recent financial time series analysis, the leverage effect in the stochastic volatility model, which is a drop in the return followed by an increase in the volatility, plays an important role. However, in the context of income data, stylized facts do not report such a leverage effect. Thus, we assume independence between the variance and the distribution over time. Moreover, the estimation methods follow the previous standard methods in the estimation of the stochastic volatility model and VAR model.} In the empirical section, we use 1 lag.

The likelihood function of this model is written as
\begin{eqnarray}
	\lefteqn{L(\mX | \mY, \vmu, \valpha, \mB, \mSigma)}\nonumber\\ 
	&\propto& \int \cdots \int \prod_{t=1}^{T} \exp\left(-\frac{h_{t}}{2}\right) \exp\left( -\frac{\vv_{t}^{\prime} \mW_{t}^{*-1} \vv_{t}}{2 \exp(h_{t})} \right)|\mSigma|^{-\frac{1}{2}}\exp\left( -\frac{\ve_{t}^{\prime} \mSigma^{-1} \ve_{t}}{2} \right) d h_{1} \ldots d h_{T},\nonumber\\
    &&
	\label{likelihood}
\end{eqnarray}
where $\mX = (\vx_{1}, \vx_{2},\ldots, \vx_{T})$, $\vmu = (\mu_{1}, \mu_{2}, \ldots, \mu_{T})^{\prime}$, $\vv_{t} = \ln \vx_{t} - \mu_{t} - \exp(h_{t}/2)\vu_{t}$, $\ve_{t} = \vy_{t} - \valpha - \mB \vy_{t-1}$, $\vh = (h_{1}, h_{2}, \ldots, h_{T})^{\prime}$, and $\mY = (\vy_{1}, \vy_{2}, \ldots, \vy_{T} ) = (\vh, \mY^{*\prime})^{\prime}$.

It is noteworthy to stress that our proposed model reflects a \textit{joint} estimation strategy of the measure of income inequality and the macroeconomic vector autoregressive model. An alternative often pursued in the literature would be to treat estimated Gini coefficients as observed data thereby neglecting inherent estimation uncertainty. As pointed out in \citet{Carriero2018} these ``two-step'' procedures potentially lead to flawed inference. More precisely and in the context of income inequality, in case the Gini is calculated by descriptive statistics using grouped data, inequality tends to be underestimated. This point is illustrated in \autoref{fig:lorenz} below:

\begin{center}[INCLUDE \autoref{fig:lorenz} HERE]\end{center}

\begin{figure}[p]
	\caption{Lorenz curve and its Gini coefficient}
	\label{fig:lorenz}
	\centering
	\includegraphics[scale = 0.4]{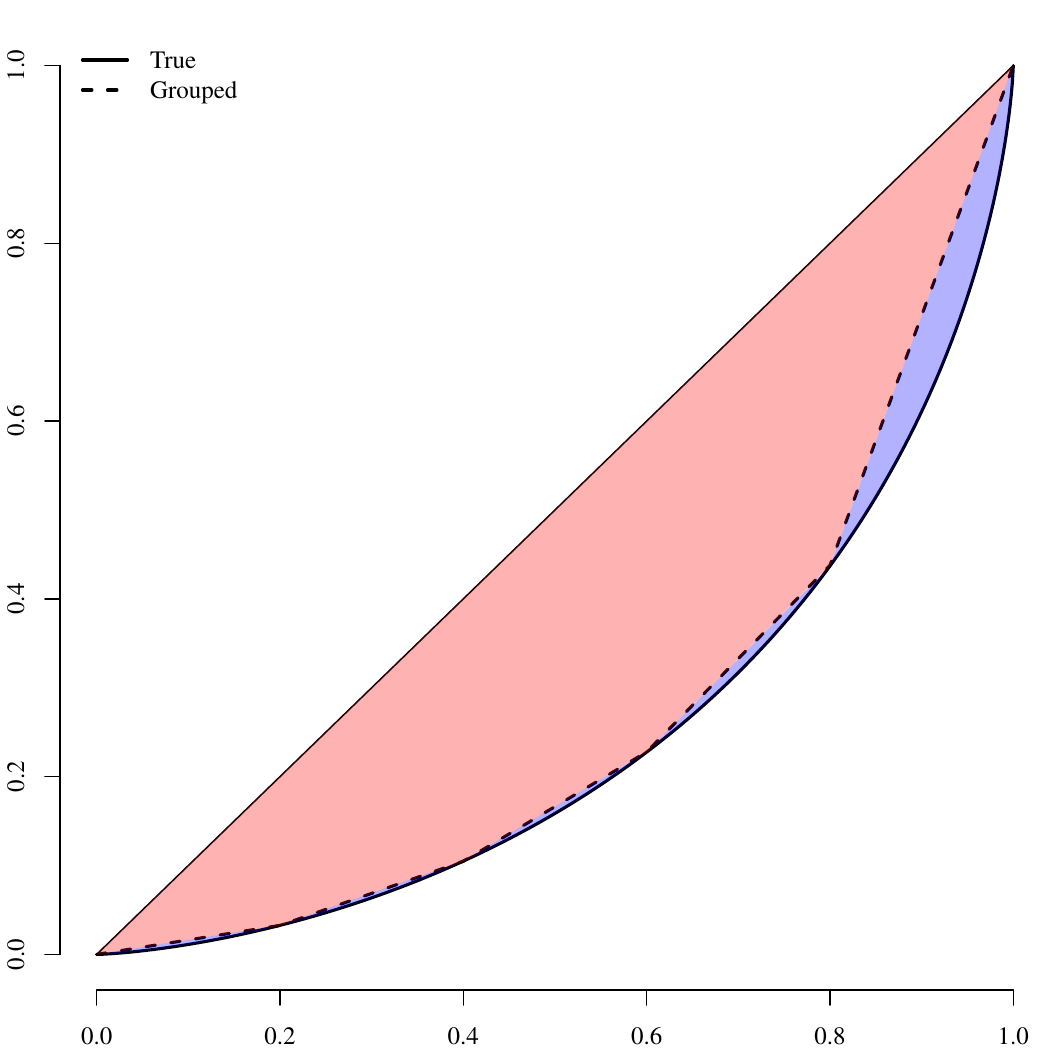}
	\begin{minipage}{16cm}
\footnotesize \textit{Notes}: The plot shows the Gini area based on simulated data for two cases: In red, the Gini is calculated using grouped data from the simulation, whereas in red and blue the Gini is derived directly from the simulated data. The blue area shows the difference between the two.
\end{minipage}%    
\end{figure}

For the true Lorenz curve (solid line) we assume a log-normal distribution;  the Lorenz curve based on  grouped data  (dotted line) is based on repeated draws from a log-normal distribution.\footnote{More precisely, the simulated data is constructed as follows. First, $100,000$ observations are simulated from the log-normal distribution. The observations are sorted in an ascending order, and are divided into quintile groups. Then, we calculate the class income means. Using the frequencies and class income means, we can finally draw the Lorenz curve from grouped data.} The Gini coefficient derived from grouped data is shown by the red area, while the one for the true distribution covers the red and blue area. The graph illustrates that the Gini coefficient from grouped data (red area) cannot capture the full area (blue area). The reason behind is the underlying assumption of equal income within each group and this assumption leads to an underestimation of inequality, especially in the top income group.

In what follows, we are going to use Bayesian methods to estimate the joint model outlined in equations \eqref{eq:model1} and \eqref{eq:model2} and assume the subsequent prior distributions:
$\mu_{t} \sim \N(\mu_{0},\tau_{0}^{2})$, $t=1,2,\ldots,T$,
$\vbeta \sim \N(\vbeta_{0},\mOmega_{0})$,
$\mSigma \sim \IW(\nu_{0}, \mSigma_{0})$.
In addition, we set hyper-parameters as $\mu_{0} = 0$, $\tau_{0}^{2} = 100$, $\vbeta_{0} = \vzero$, $\mOmega_{0} = 100 \cdot \mI$, $\nu_{0} = m + 1$ and $\mSigma_{0} = 0.01 \cdot \mI$, which make the variance of prior distributions diffuse.
In the empirical analysis, we run the MCMC algorithm, which is described in Appendix \ref{sec:PA}, for $10,000$ iterations after a burn-in phase of $10,000$ iterations.

\section{Data}\label{sec:data}

Following \citet{Lise2014} and \citet{Inui2017} we derive the Gini coefficient using the Japanese household survey called Family Income and Expenditures Survey (FIES), which is compiled by the Statistics Bureau of the Ministry of Internal Affairs and Communications. The survey is carried out on a monthly basis and collects data on earnings, income and consumption expenditures for about 9,000 households and is appropriately adjusted to make the sample size to $n_{t} = n = 10,000$ in order to compensate the difference in sampling ratios for strata. The survey unit are households in the entire area of Japan.\footnote{The survey excludes one-person student households, inpatients in hospital, inmates of reformatory institutions, households which manage restaurants, hotels, boarding houses or dormitories, sharing their dwellings, households which serve meals to the boarders even though not managing boarding houses as an occupation,  households with 4 or more living-in employees, households whose heads are absent for a long time (three months or more) and foreigner households. For more details, see \url{http://www.stat.go.jp/english/}.}  We use the monthly subset of the survey that covers  household incomes and expenditures. Each wave of the survey offers among other data on consumption expenditure, the frequency distribution over 18 groups of yearly pre-tax income: $\vx_{t} = \vx = (200,250,300,350,400,450,500,550,600,650,700,750,800,900,$ \\
$1000,1250,1500)^{\prime}$, which are collected from Table 8-2 ``Amounts of Savings and Liabilities Held per Household by Yearly Income Group''. Income data refers to actual income including tax (i.e., the sum of cash income of all household members), cash on hand carried over from previous months, and so-called ``spurious income''. The latter is defined as an increase in cash accompanied by a decrease in assets and increase in liabilities. As an example, the sale of a house generates cash, but decreases households assets. By the same token, if a household borrows money from a bank, cash increases in parallel with household's liabilities. Both cash increases would be labeled spurious income and are included in overall pre-tax income of households. The 10,000 surveyed households can be roughly divided into two groups: \textit{workers'} households are those whose heads are employed as clerks or wage earners by public or private enterprises, such as government office, private companies, factories, schools, hospitals, and shops.  Workers' households account for roughly half of the 10,000 observations in each survey wave. \textit{All} households include on top of \textit{workers'} households, those whose heads are  individual proprietors (e.g., merchants, artisans, administrators of unincorporated enterprise, farmers, about 10\% of all observations) and those with no occupation (unemployed and retired, about 35\% of all observations).\footnote{The remaining 5\% are accounted for by other professional services and corporative administrators.} Due to changes in the sample design and to get a consistent income series our data starts from 2002Q1. 

%In the recent survey, sample design has changed twice. One is in July 1999 and the other is in January 2002. In the first change, the households engaged in agriculture, forestry and fisheries were included in the coverage of the Survey, and the results of the households engaged in agriculture, forestry and fisheries have been available since January 2000. In the latter change, one person households were incorporated into the coverage of the FIES, which had been independently surveyed by the Income and Expenditure Survey for one-person households from 1995 until 2001, and the Survey of Savings and Liabilities was introduced. The sample of the FIES has been enlarged to about 9000 households as a consequence. Therefore, our dataset starts from 2002Q1. 

Pre-tax income data are complemented by macroeconomic data such as real GDP (rgdp),   the unemployment rate  (unempl), 10-year government bond yields (ltir), the real effective exchange rate (reer) and stock market prices (eq) measured by the NIKKEI225 index. As pointed out in \citet{Amaral2017}, inflation expectations are a crucial determinant of inequality. We measure inflation expectations, as one-year ahead projections of consumer price inflation (p). Inflation forecasts are obtained from the IMF's biannual world economic outlook database and converted to quarterly frequency by reusing the biannual observations over the quarterly frequency domain. Last, we include several measures for the monetary policy stance of the BoJ. These include the the shadow rate (ssr) of \cite{Wu2016}, the 2-year government bond yields (2ygby) as well as the recently proposed target and path factors of \citet{Kubota2020}. Exchange rate data stem from the BIS and an increase denotes an appreciation in real terms; all other macroeconomic data are obtained via the FRED database, \url{https://fred.stlouisfed.org/}.\footnote{Data mnemonics are as follows, rgdp:\verb+JPNRGDPEXP+, ltir:\verb+IRLTLT01JPM156N+, eq:\verb+NIKKEI225+ and m3:\verb+MABMM301JPM189S+.} Our sample for the combined income and macro data spans the period from 2002Q1 to 2018Q4.

\begin{figure}[t]
\caption{Estimated Gini coefficient}\label{fig:GiniW}
	\begin{subfigure}{.55\textwidth}
    \caption{All households}
    \centering
	\includegraphics[width=.75\linewidth]{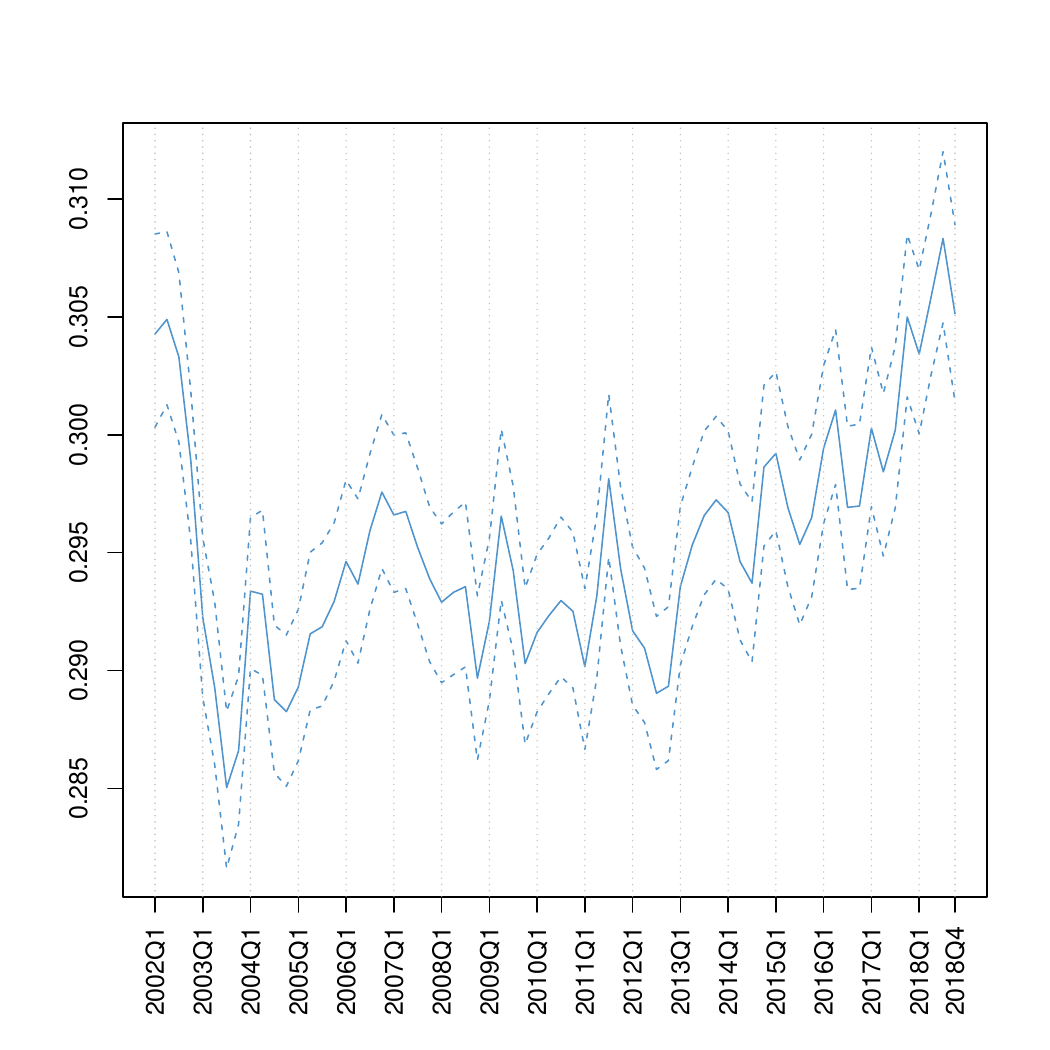}
    \end{subfigure}
    \begin{subfigure}{.55\textwidth}
     \caption{Workers' households}
     \centering
	\includegraphics[width=.75\linewidth]{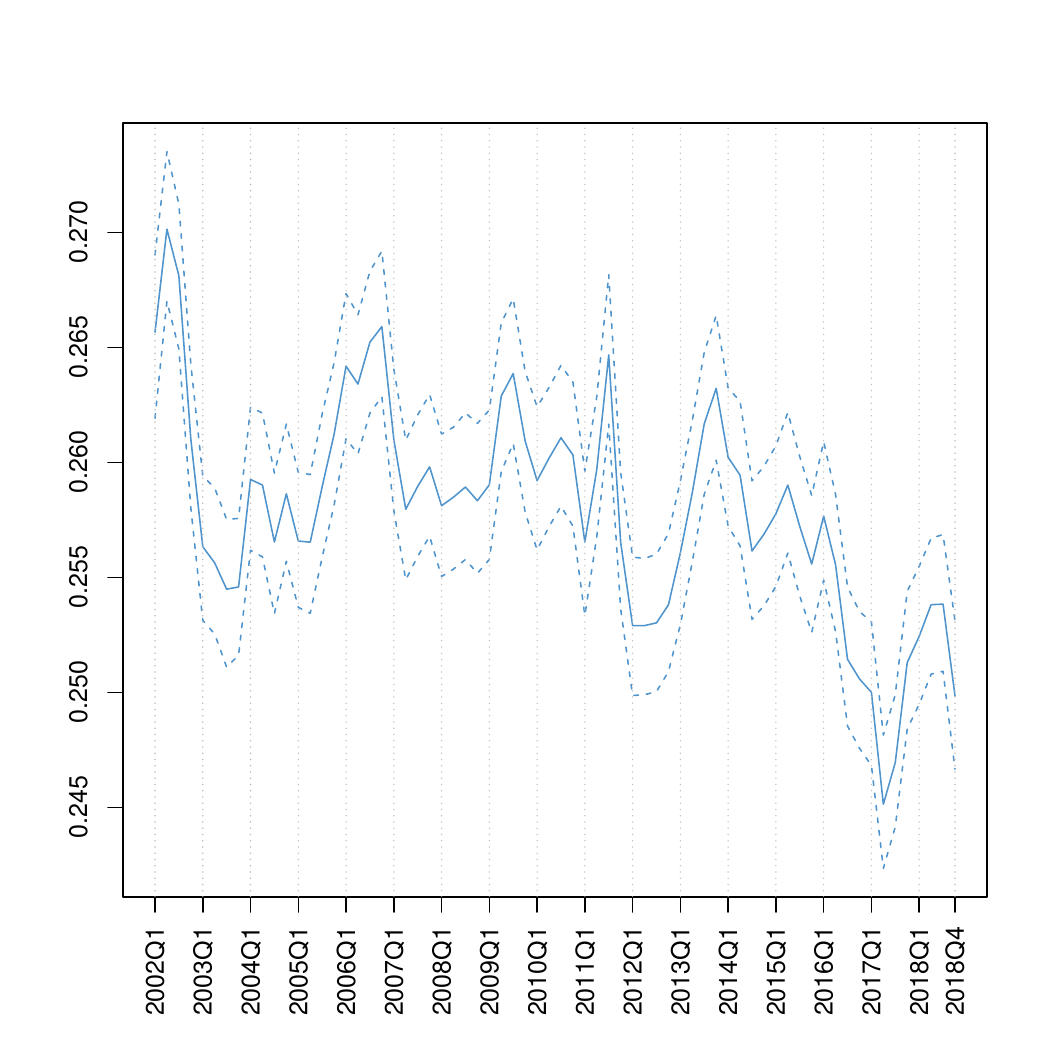}
    \end{subfigure}
\begin{minipage}{16cm}
\footnotesize \textit{Notes}: The plot shows the estimated Gini coefficient for all households (left panel) and workers' households (right panel). Posterior median along with 95\% credible intervals shown. 
\end{minipage}%    
\end{figure}

In a first step and to assess the accuracy of the proposed modeling framework, we plot the estimated Gini coefficient over time.  The results are depicted in \autoref{fig:GiniW}. The figure reveals some interesting differences of the development of income inequality between the two sub-samples we use. Inequality for all households was very high at the beginning of the sample period, more modest during the period  from 2004 to 2013 after which the Gini picked up again. The latest observations from 2018 indicate a significantly higher level of inequality than during the middle of our sample period, surpassing even the peak value of 0.3 at the beginning of the sample. The Gini coefficient based on data for workers' households shows a different picture: While the estimated Gini coefficient was high at the beginning of the sample period (as for all households), inequality has later embarked on a declining trend. More so, the Gini coefficient was significantly smaller in the most recent period compared to the beginning of the sample. Our estimates are well in line with other Gini coefficients derived from descriptive statistics as in \cite{Inui2017} and \citet{Saiki2014}. %Also and based on micro data, the World bank estimates a Gini of about 0.3 for 2008, which is also close to values obtained in this paper.

\section{The effects of monetary policy on inequality}\label{sec:emp}

To assess the effects of monetary policy shocks on income inequality we have estimated the vector autoregression in equ. \ref{eq:model2} that jointly models the dynamics of income inequality (i.e., the Gini coefficient) and the macroeconomic variables.  

There are three prominent channels through which monetary policy might affect income inequality. First, the \textit{inflation tax channel} postulates that a decrease in inflation, after a monetary policy tightening  disproportionately affects purchasing power of households at the lower end of the income distribution since these typically hold more cash \citep{Erosa2002}. This would imply less inequality. Second, the earnings heterogeneity or \textit{job destruction channel} states that income / occupation of poorer households is more strongly shaped by changes in the unemployment rate than that of richer households for which hourly wages are a more prominent factor as a determinant of income \citep{Amaral2017}. This implies that if unemployment rises due to the monetary policy tightening, poorer households are disproportionately more strongly affected and income inequality increases \citep{Inui2017}. Last, the \textit{income composition channel} applies if households in different segments of the income distribution rely on different sources of income \citep[e.g., labor, business and capital, see][for more details.]{Amaral2017}. Depending on the distribution of income sources, the effect of an interest rate increase might be either inequality reducing or enhancing and thus remains to be empirically assessed. Our data allows us to examine these channels also with respect to different occupations. More specifically, we can examine whether the income distribution of unemployed and retirees (contained in all households data) responds differently from the employed population (workers households). That the occupational status of the household can trigger different effects of how monetary policy affects income inequality has been stressed in e.g., in \citet{Gornemann2012}.

Against the backdrop of prolonged periods of near zero interest rates, choosing a suitable policy instrument to model monetary policy in Japan is not straightforward. While in the early studies, monetary policy was thought to mainly work through short-term interest rates \citep{Miyao2000, Miyao2002}, other measures such as shadow rates \citep{Iwasaki2017}, longer-term yields \citep{Saiki2014, Spiegel2018} or external instruments \citep{Honda2006, Kubota2020} have been considered more recently. Our approach to tackling uncertainty about the policy instrument is to provide a baseline instrument along with several robustness estimations.  As a baseline measure, we use the shadow rate of \citet{Wu2016}. This shadow rate is extracted using a term structure model that allows the relaxation of the zero lower bound of nominal short-term rates. During normal times the shadow rate mimics short-term interest rates; in times short-term rates are stuck at the zero lower bound and the central bank provides additional stimulus, the shadow rate can become negative. It thus indicates the monetary policy stance of a central bank even if actual interest rates are zero, which is the case for the sample period we cover. In a robustness exercise we follow \citet{Spiegel2018} who use 2-year Japanese government bond yields as the policy instrument. We also consider two recently proposed external instruments, namely the target and the path factor of \citet{Kubota2020}. Both instruments exploit high frequency variation of euro-yen Tokyo interbank offered rates (TIBOR) at different maturities, around policy announcements of the BoJ. The target factor is more reminiscent of conventional / interest rate monetary policy, whereas the path factor tends to affect the longer end of the yield curve reflecting unconventional monetary policies. 

As outlined above, all three instruments can be considered as broad monetary policy measures and hence also cover aspects of unconventional monetary policy. This is essential given the macroeconomic environment in Japan over the sample period. That said, one could argue that unconventional monetary policy  rather affects wealth than income inequality: bringing down long-term yields mostly impacts financial markets and wealth of households holding these assets. Nevertheless, some effects should be also seen on the distribution of income. Foremost, the spurious income component of our income data is naturally affected by asset / house price movements. Moreover, since (longer-term) interest rates are moved, some of the channels mentioned above could also be operational and income inequality affected.\footnote{In theory, since older households tend to depend more strongly on interest rate income -- which is not covered in the income data available for this study -- quantitative easing should decrease income inequality \citep{Amaral2017}.} Also, \citet{Saiki2014} who examine the effects of the BoJ's asset purchases, report a strong empirical correlation between income and wealth inequality in Japan and \citet{Inui2017} find that distributions of households’ financial assets and liabilities do not play a significant
role in the distributional effects of monetary policy. Hence an assessment of monetary policy on income inequality in Japan seems warranted.

To identify the monetary policy shock, we rely on a simple Cholesky decomposition, which requires categorizing the set of macroeconomic variables into slow and fast moving variables. This ``recursiveness'' assumption has been frequently used in the literature on Japanese monetary policy shocks \citep{Miyao2000, Miyao2002} and its implications are discussed more generally in, among others, \citet{Christiano1999}. We assume the following ordering of variables:

\begin{equation*}
\vy_{t} = \left(\begin{array}{c} 
\text{gini} \\ 
\text{rgdp} \\ 
\text{p} \\
\text{unempl} \\
\text{shadow rate} \\ 
\text{ltir} \\ 
\text{reer} \\ 
\text{eq}
\end{array} \right).
\end{equation*}

 The ordering implies that the Gini coefficient, real GDP, inflation expectations and the unemployment rate do not react within the same quarter to an unexpected increase in the policy rate.  By contrast, long-term interest rates, the exchange rate and equity prices are allowed to react immediately to a monetary policy shock.\footnote{A similar ordering has been applied in the robustness section of \citet{Mumtaz2017}.} The shock is calibrated as a +100bp increase in the policy instrument. Note that in what follows we report estimates of $\sigma$ as the ``Gini coefficient'' rather than its monotone transform $G$. 

\begin{figure}[p]
\caption{Responses to an increase in the shadow rate - all households}\label{fig:ssr_all}
\begin{subfigure}{.329\textwidth}
\caption{Gini}
\includegraphics[width=\textwidth]{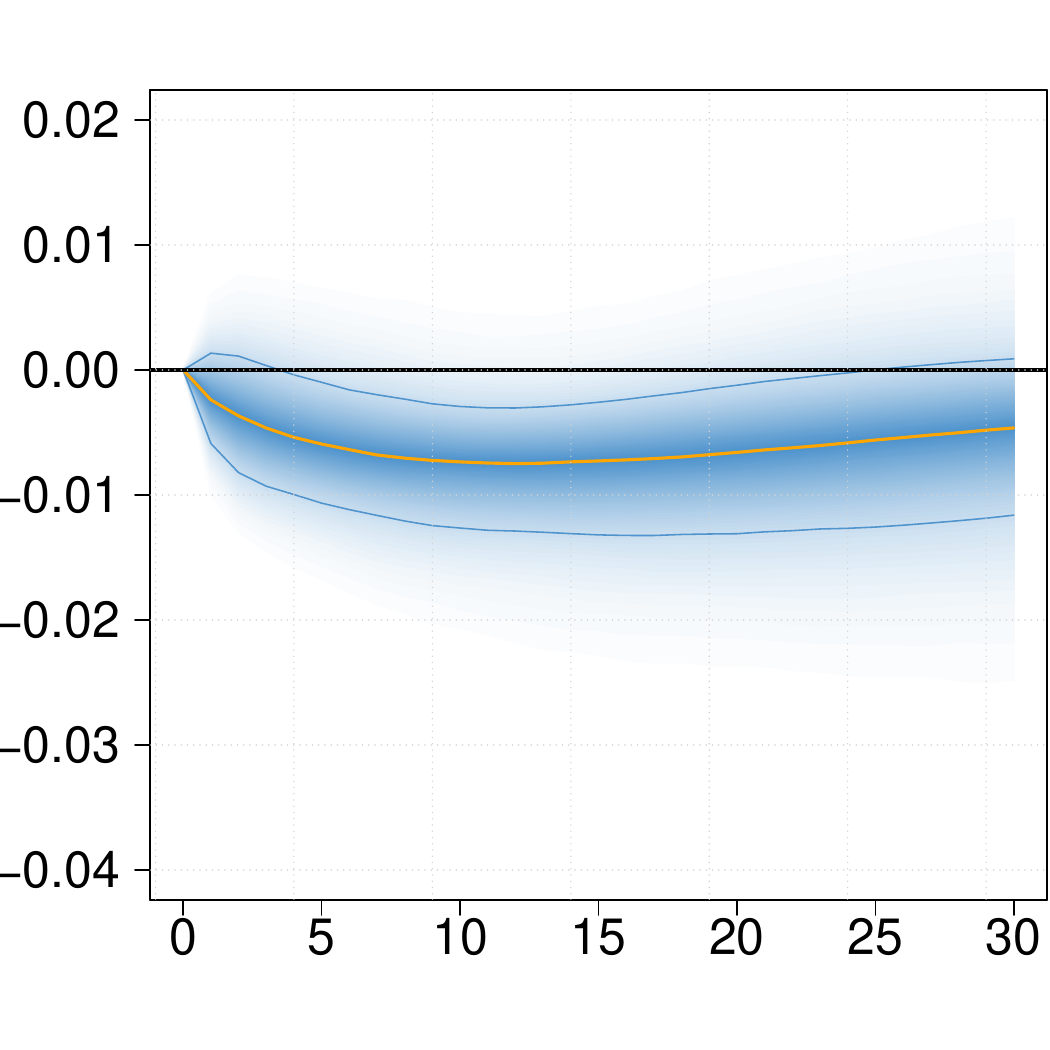}
\end{subfigure}
\begin{subfigure}{.329\textwidth}
\caption{Real GDP}
\includegraphics[width=\textwidth]{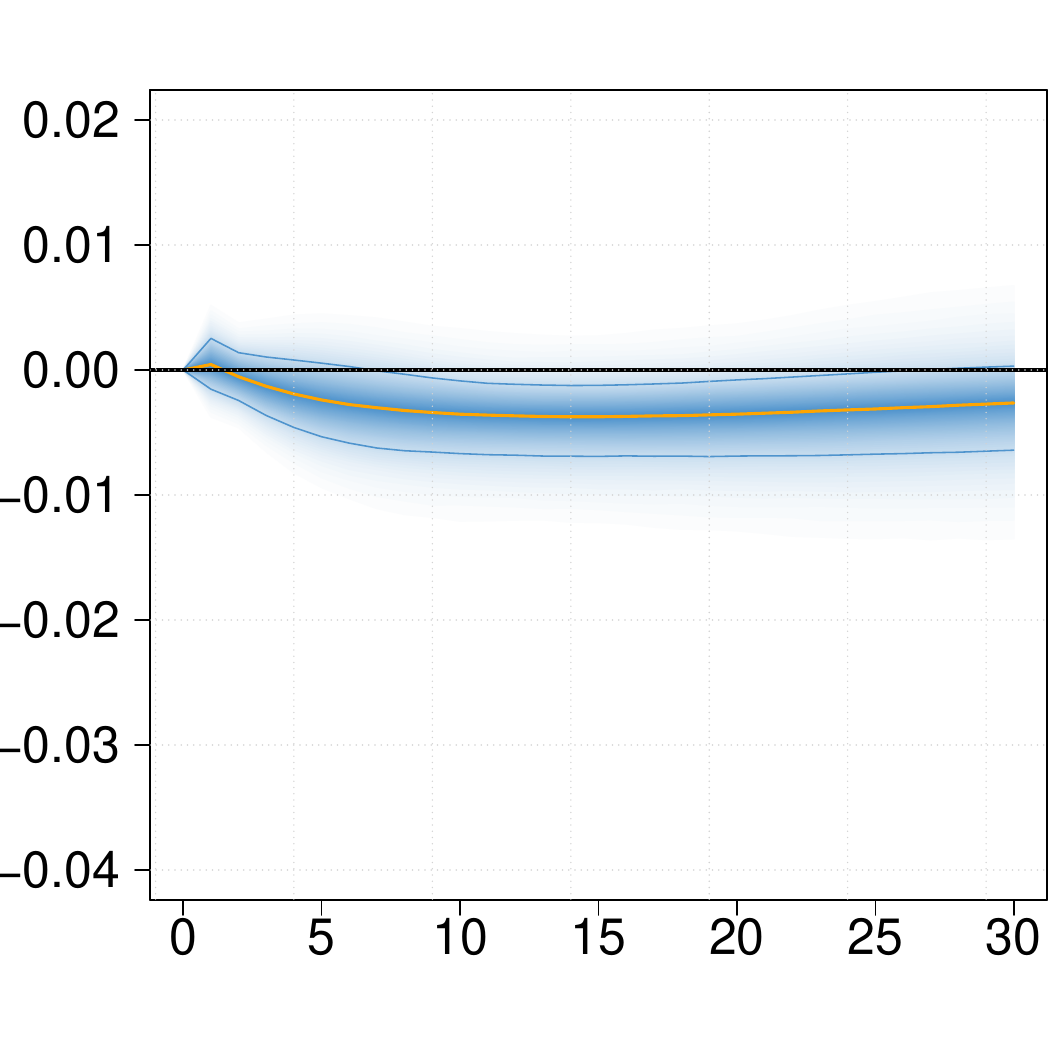}
\end{subfigure}
\begin{subfigure}{.329\textwidth}
\caption{Inflation expectations}
\includegraphics[width=\textwidth]{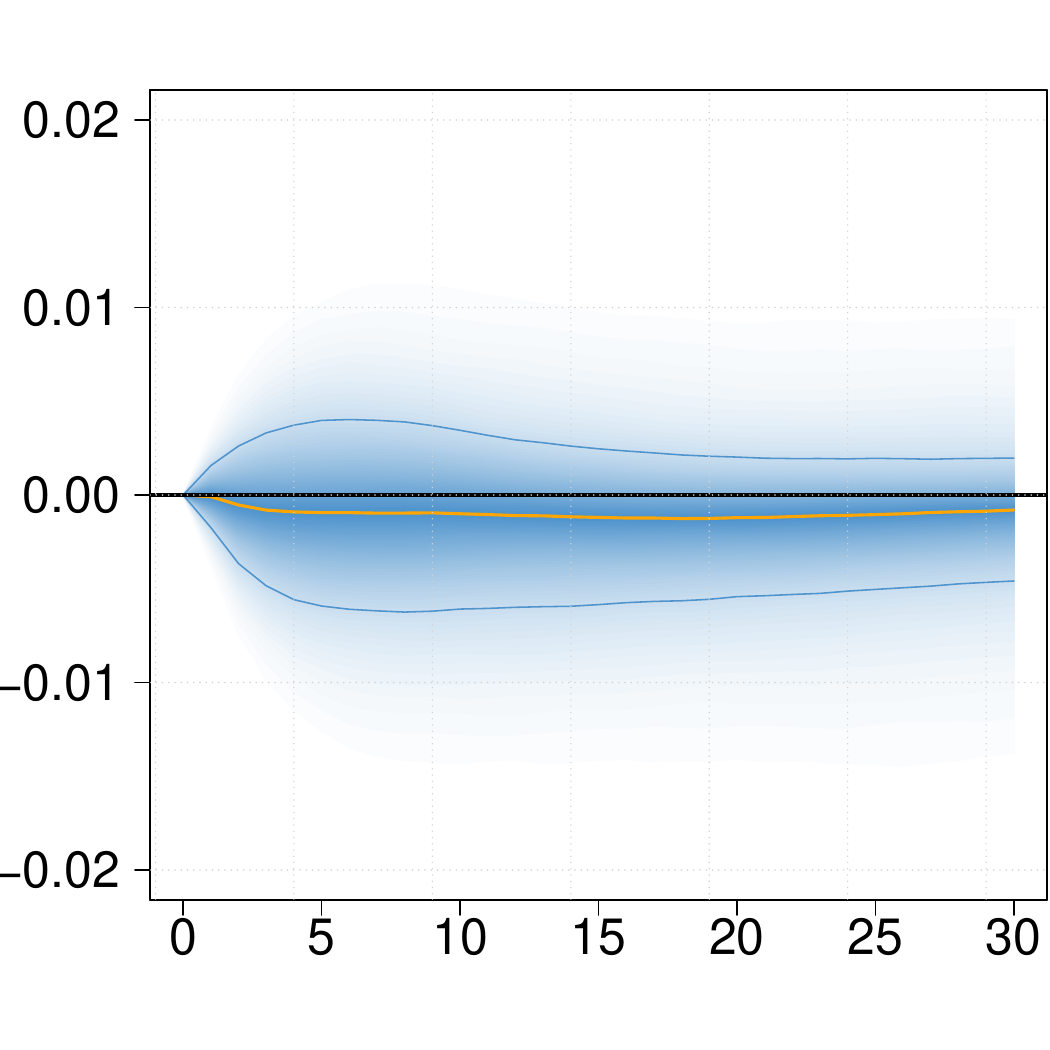}
\end{subfigure}\\
\begin{subfigure}{.329\textwidth}
\caption{Unemployment rate}
\includegraphics[width=\textwidth]{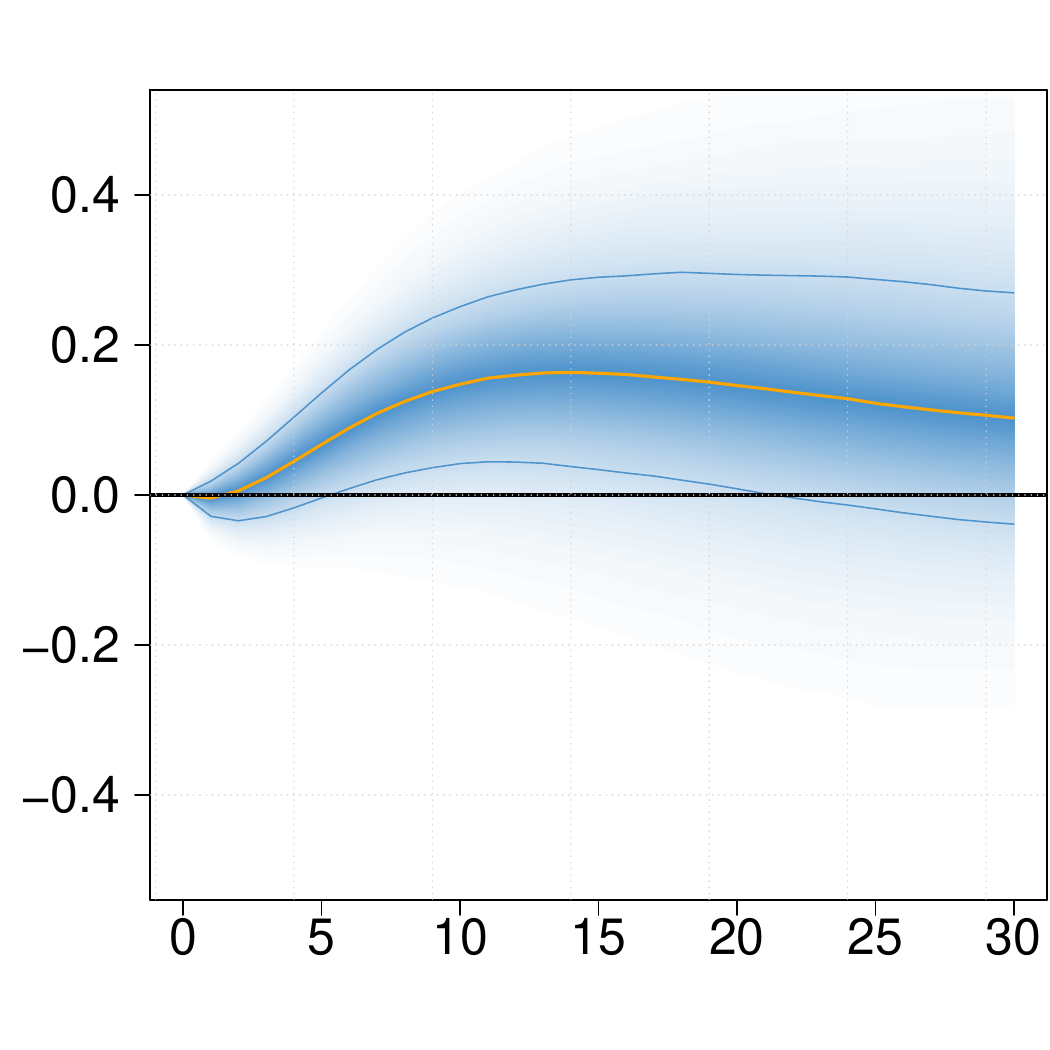}
\end{subfigure}
\begin{subfigure}{.329\textwidth}
\caption{Shadow rate}
\includegraphics[width=\textwidth]{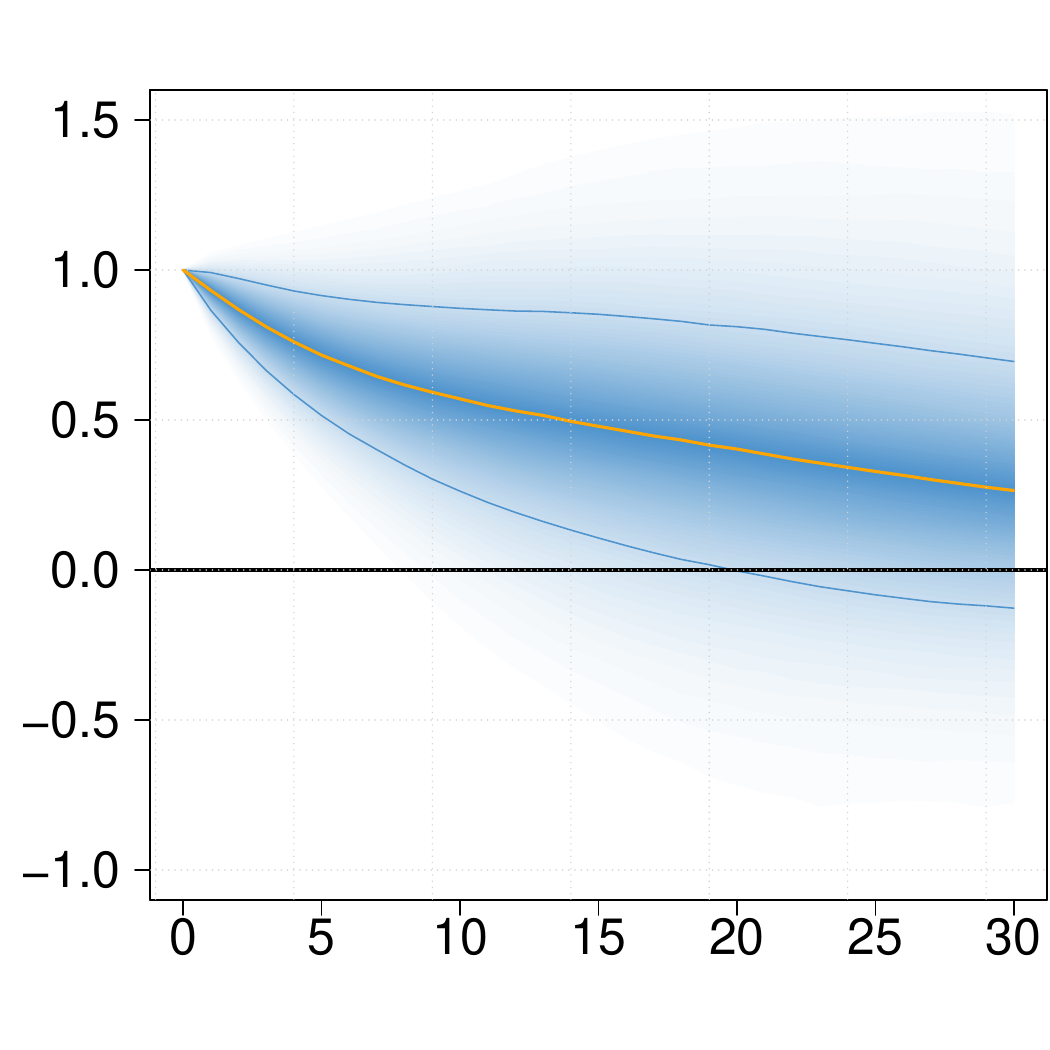}
\end{subfigure}
\begin{subfigure}{.329\textwidth}
\caption{Long rates}
\includegraphics[width=\textwidth]{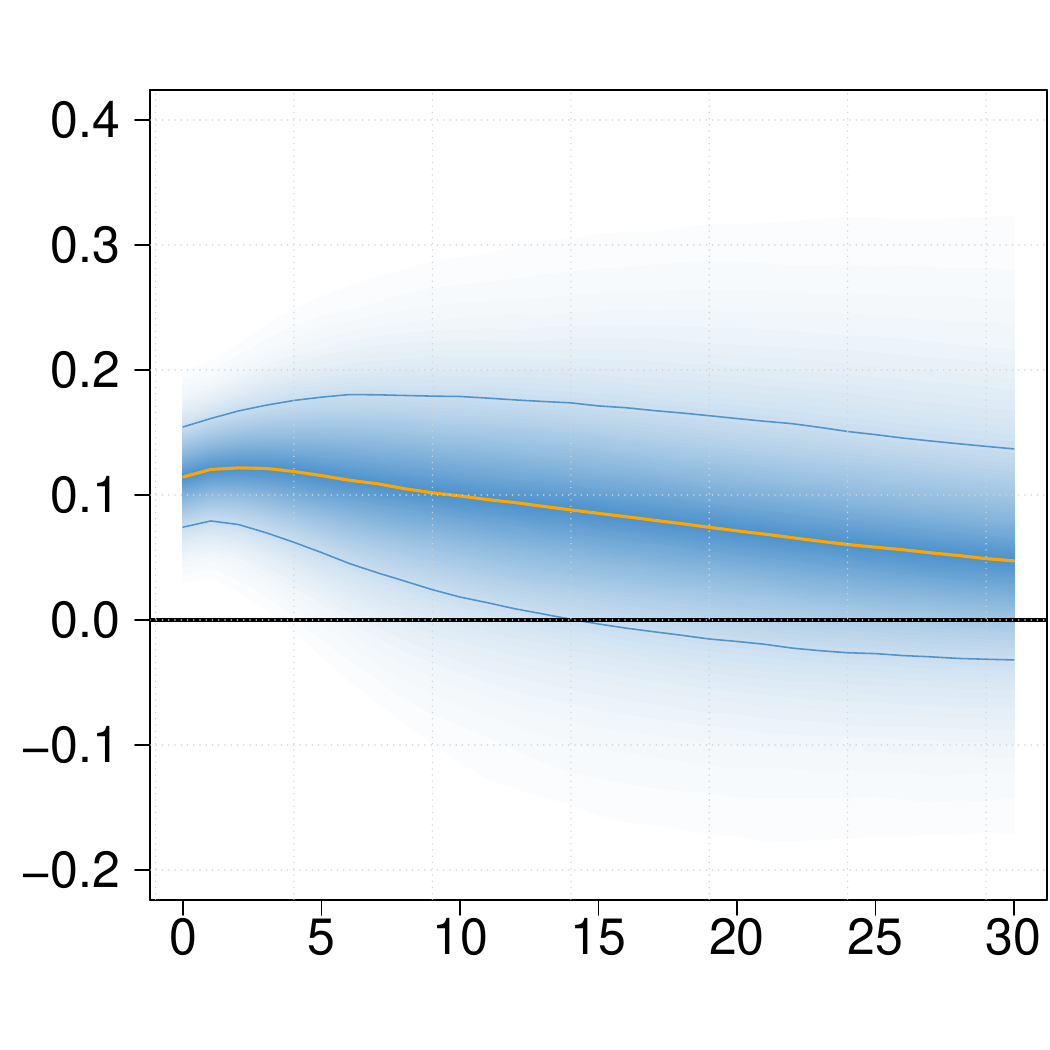}
\end{subfigure}\\
\begin{subfigure}{.329\textwidth}
\caption{Real eff. exchange rate}
\includegraphics[width=\textwidth]{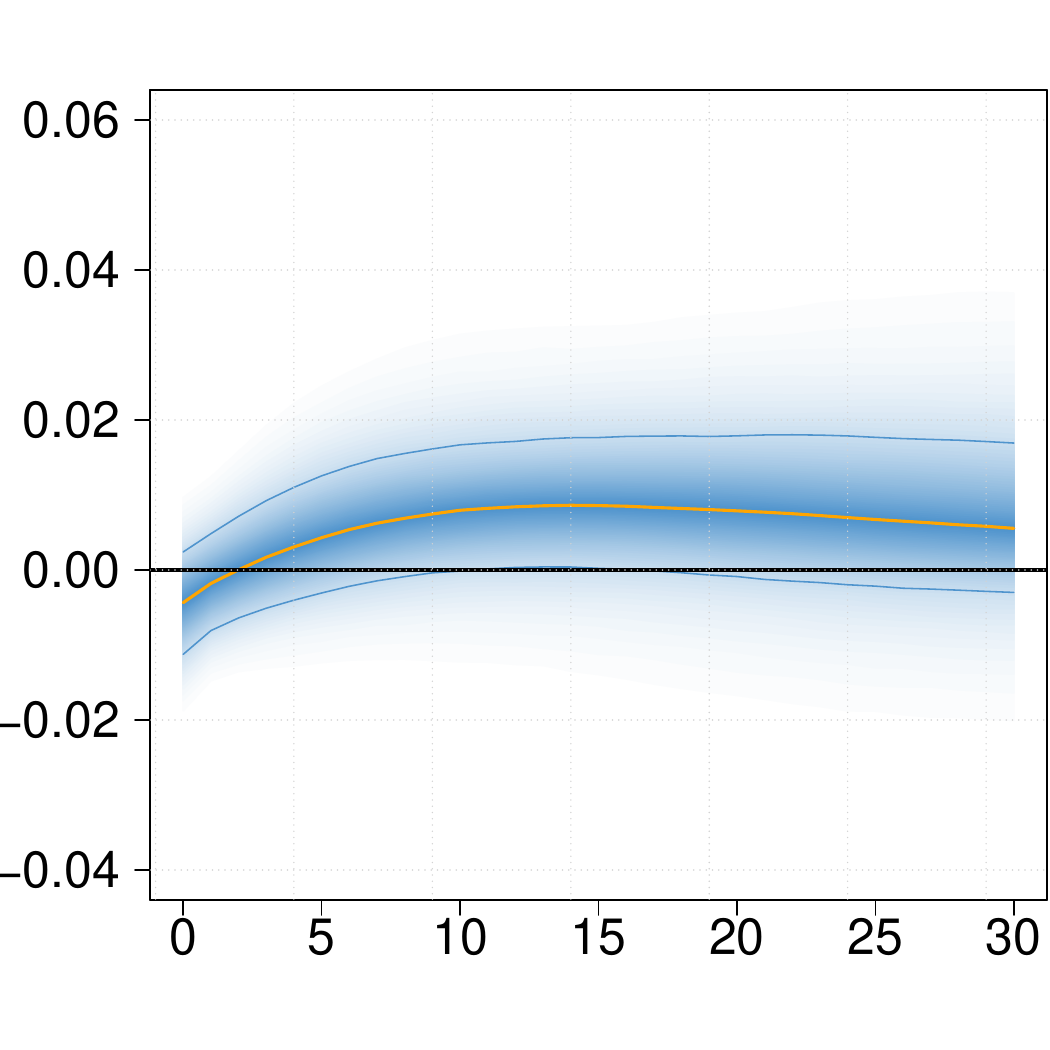}
\end{subfigure}
\begin{subfigure}{.329\textwidth}
\caption{Equity prices}
\includegraphics[width=\textwidth]{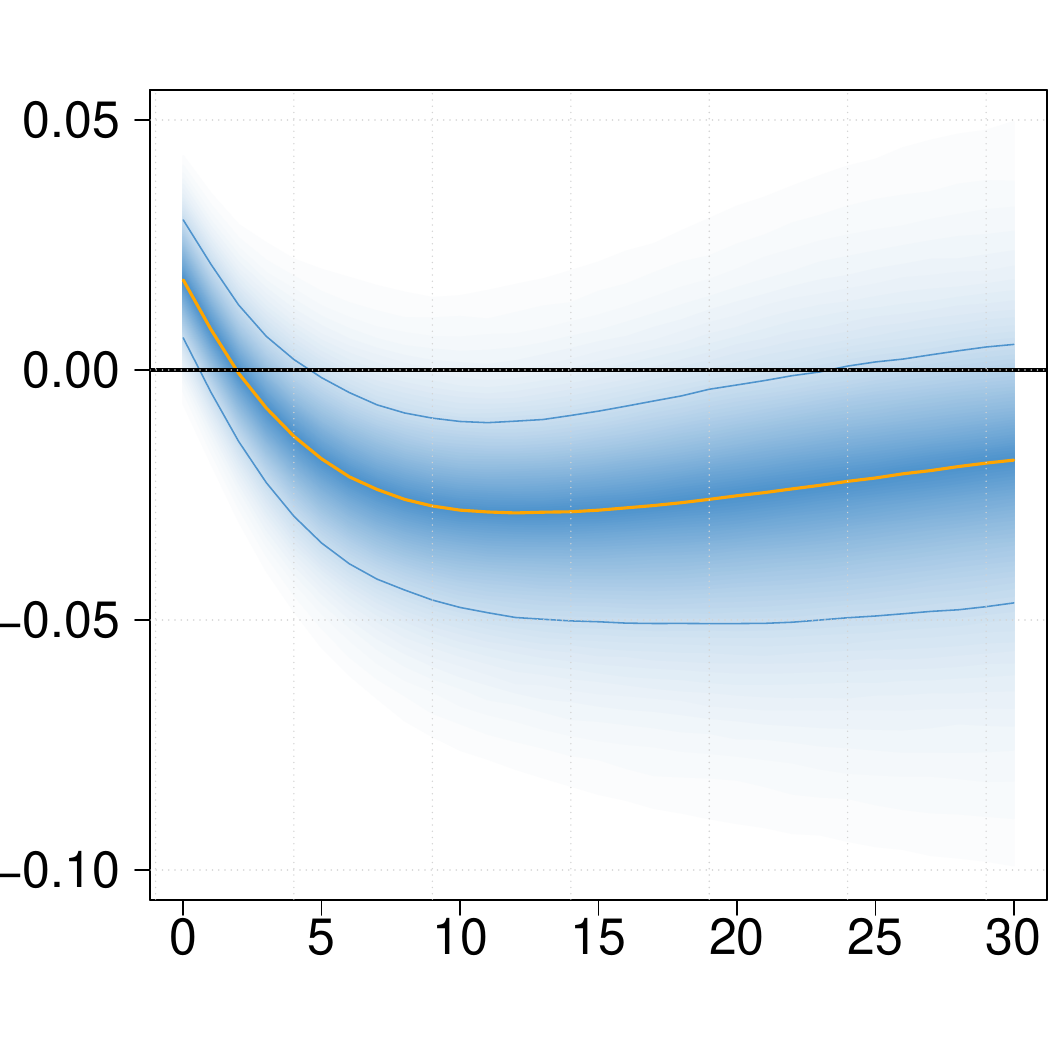}
\end{subfigure}\\
\begin{minipage}{17cm}
\footnotesize \textit{Notes}: The plot shows the full posterior distribution of structural responses to a +100bp increase in the shadow rate using a Cholesky decomposition. Posterior median in orange along  68\% credible sets (in blue). An increase in the real effective exchange rate corresponds to an appreciation. 
\end{minipage}%
\end{figure}

\begin{figure}[p]
\caption{Responses to an increase in the shadow rate - workers' households}\label{fig:ssr_worker}
\begin{subfigure}{.329\textwidth}
\caption{Gini}
\includegraphics[width=\textwidth]{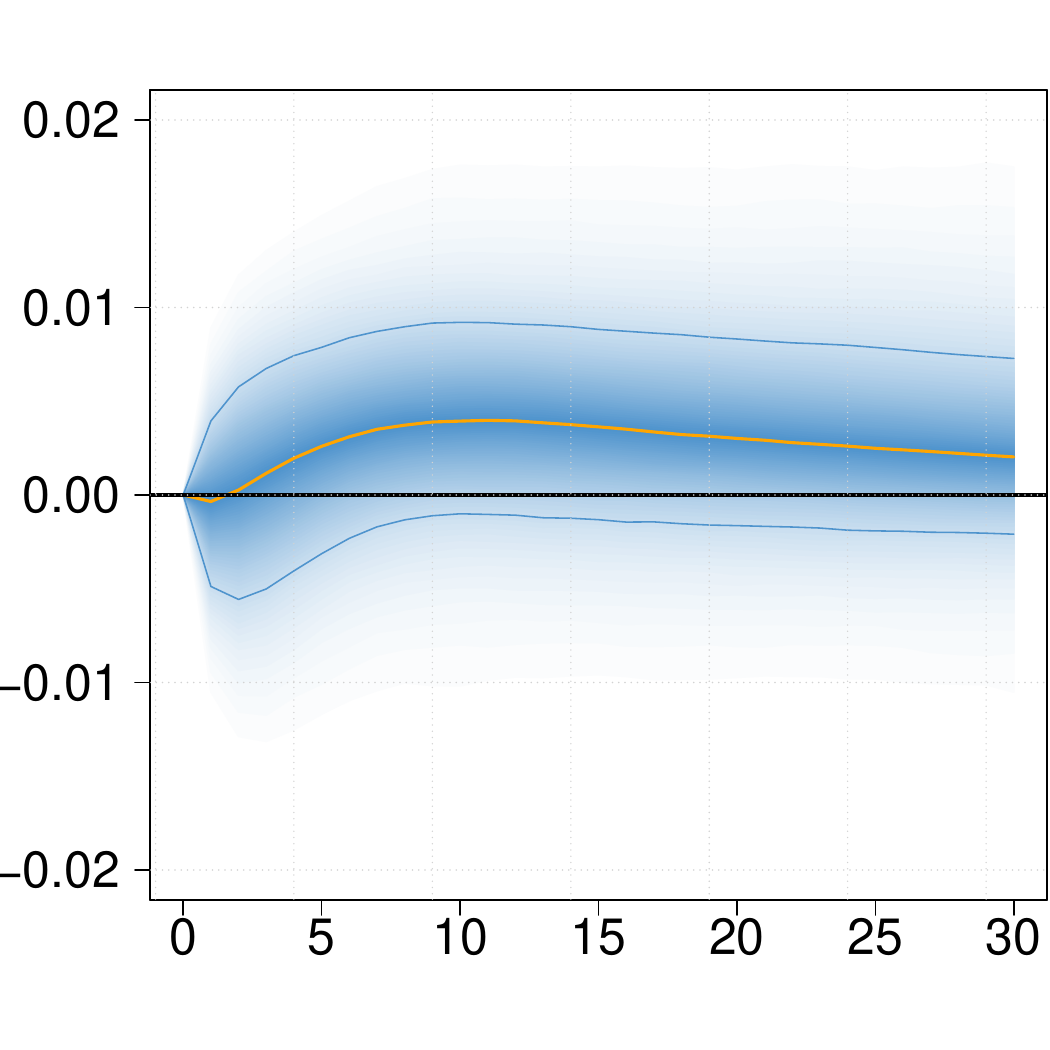}
\end{subfigure}
\begin{subfigure}{.329\textwidth}
\caption{Real GDP}
\includegraphics[width=\textwidth]{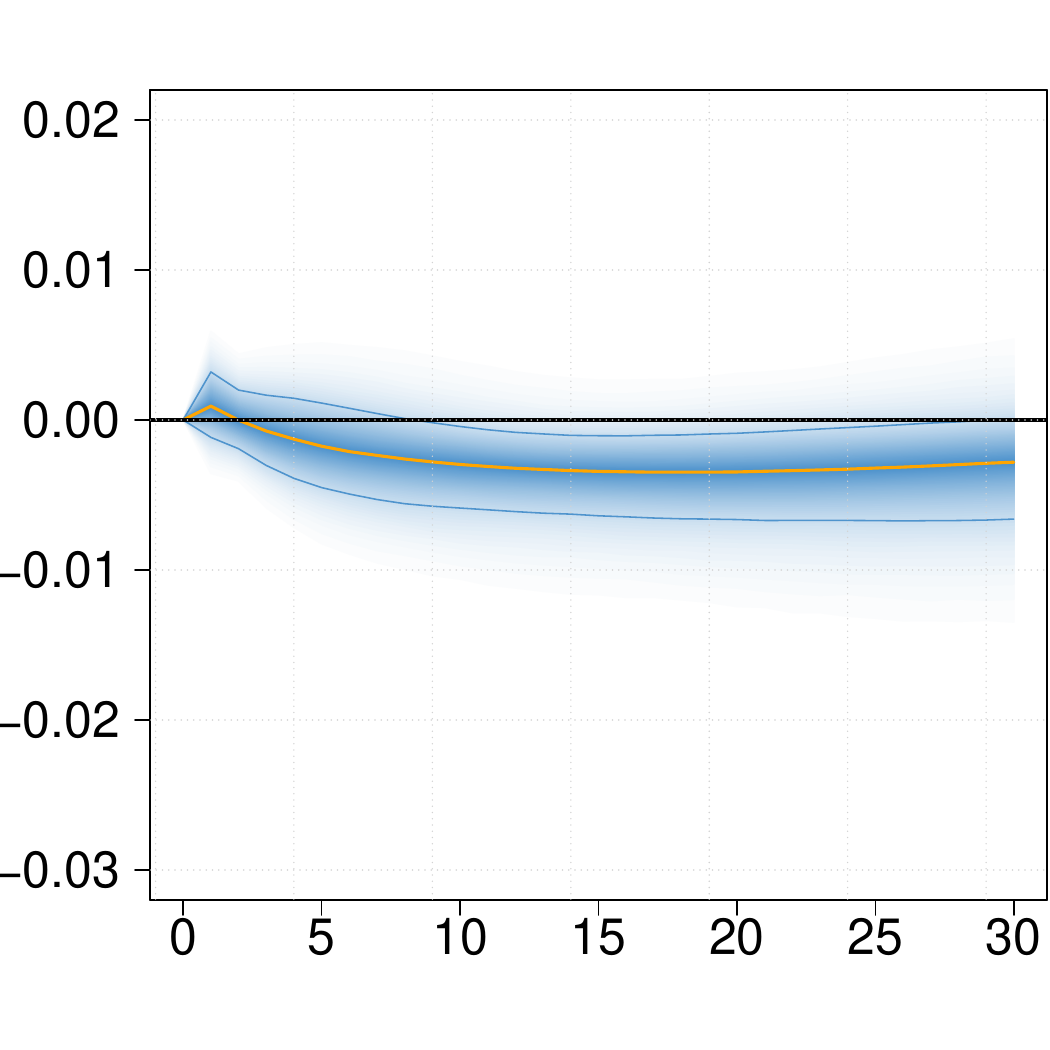}
\end{subfigure}
\begin{subfigure}{.329\textwidth}
\caption{Inflation expectations}
\includegraphics[width=\textwidth]{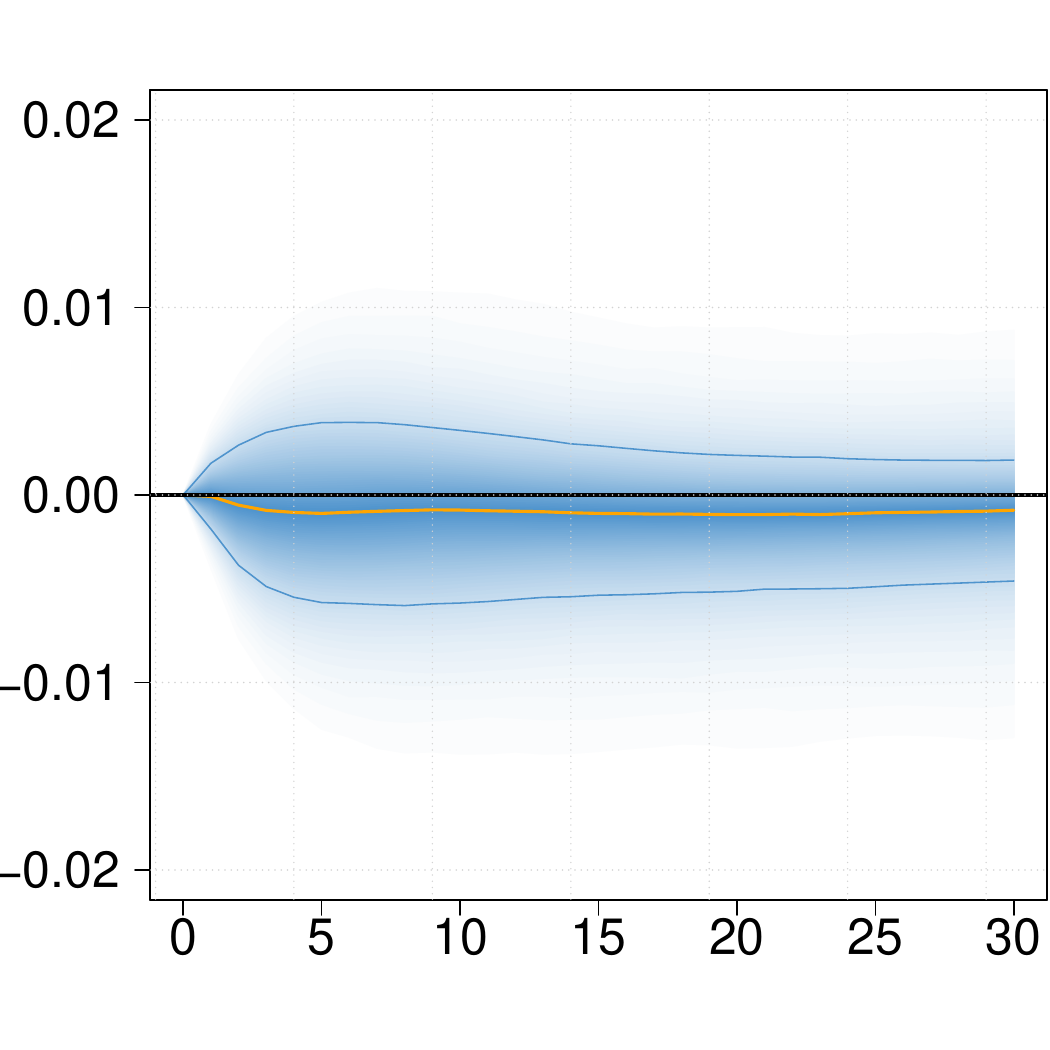}
\end{subfigure}\\
\begin{subfigure}{.329\textwidth}
\caption{Unemployment rate}
\includegraphics[width=\textwidth]{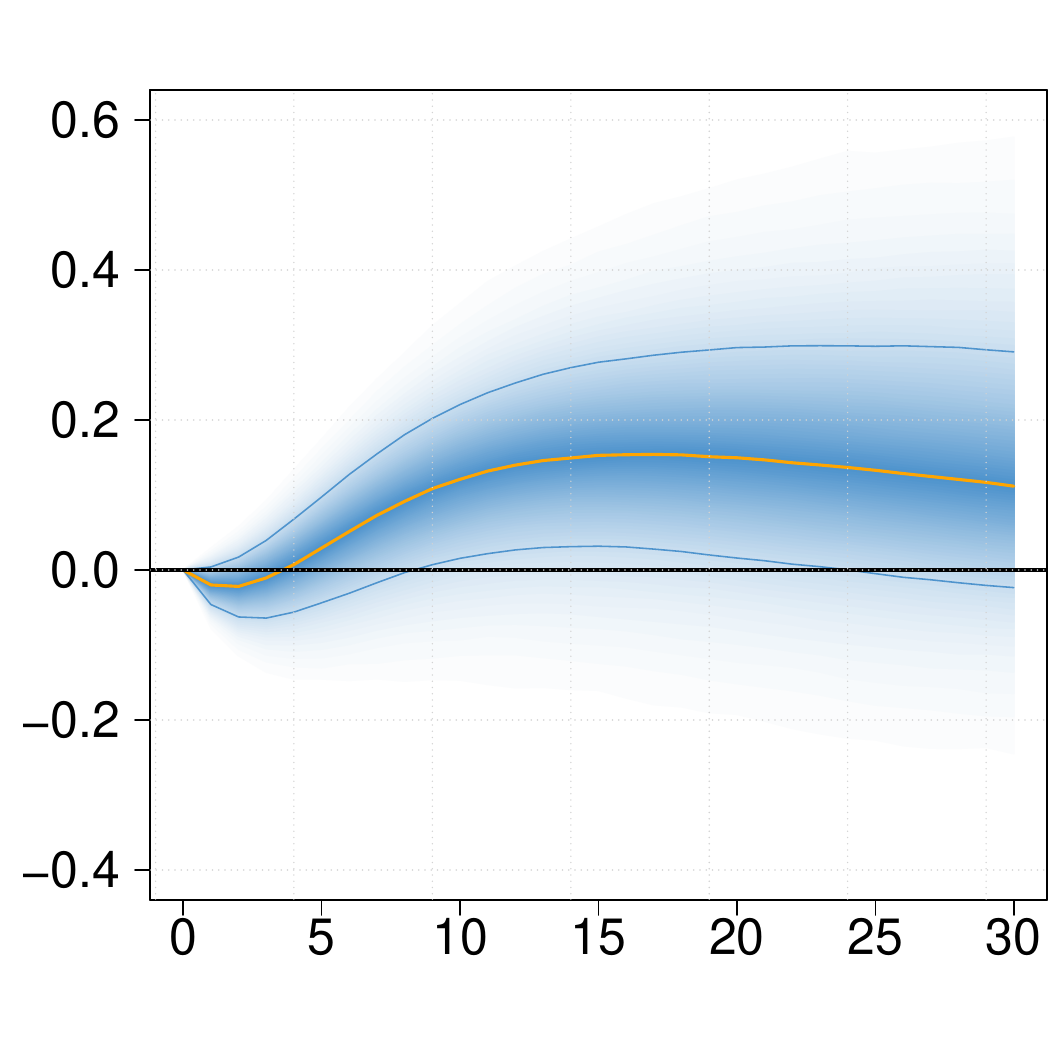}
\end{subfigure}
\begin{subfigure}{.329\textwidth}
\caption{Shadow rate}
\includegraphics[width=\textwidth]{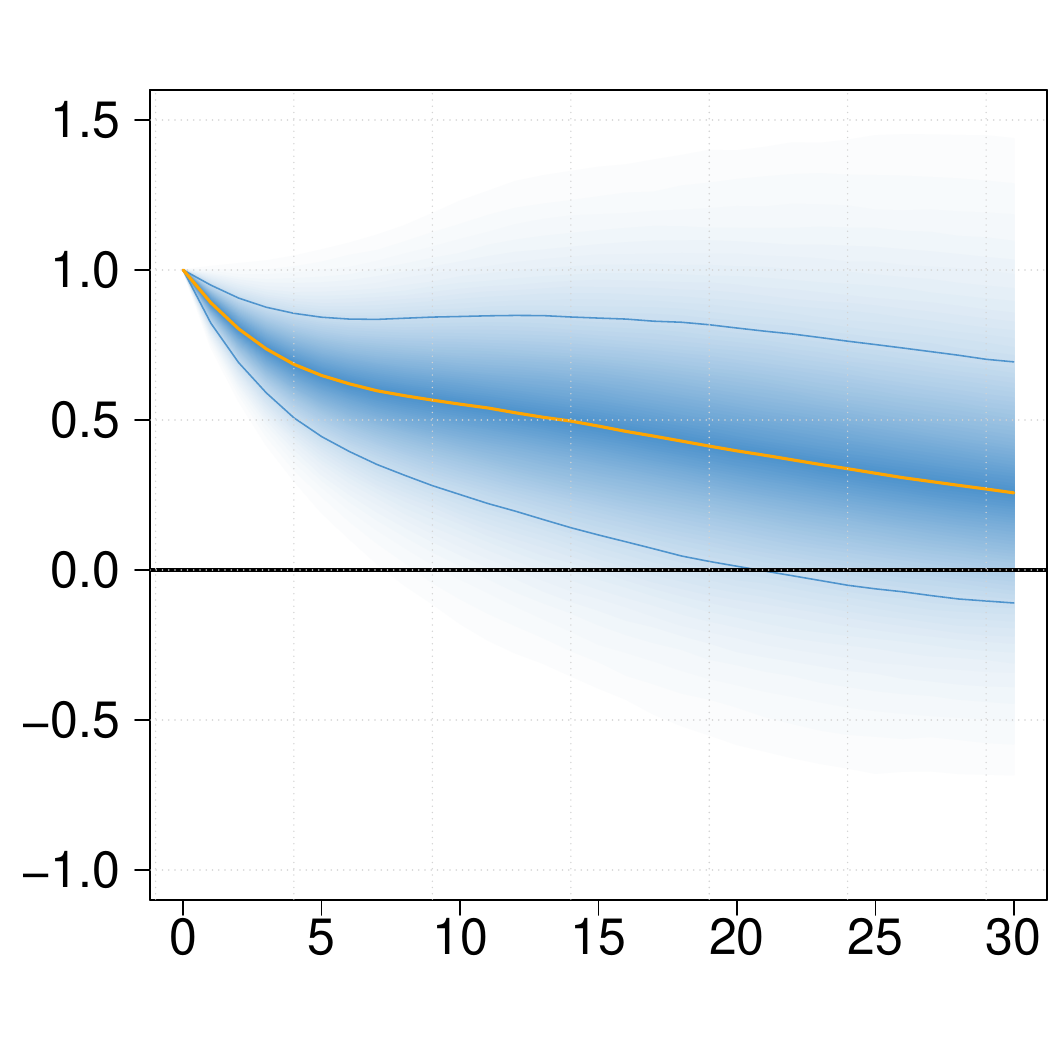}
\end{subfigure}
\begin{subfigure}{.329\textwidth}
\caption{Long rates}
\includegraphics[width=\textwidth]{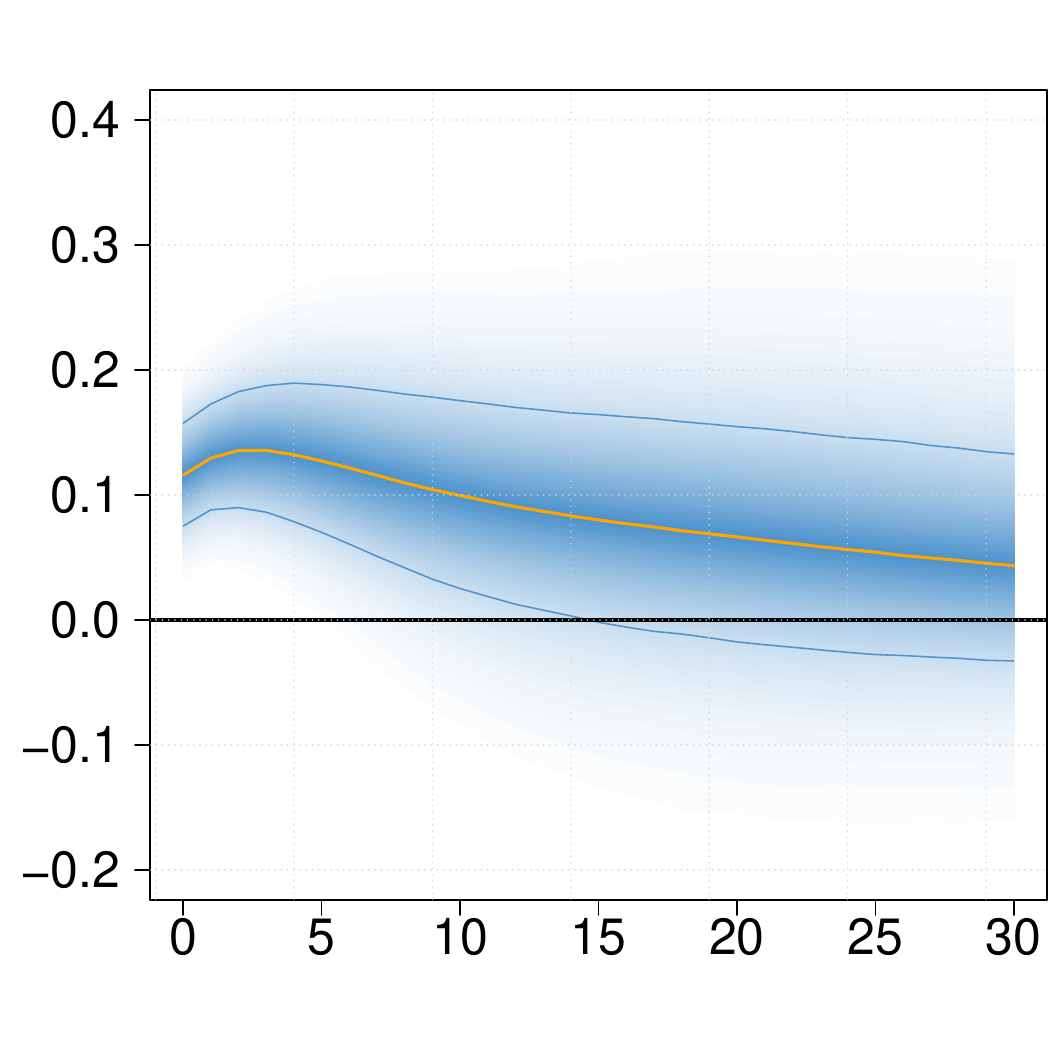}
\end{subfigure}\\
\begin{subfigure}{.329\textwidth}
\caption{Real eff. exchange rate}
\includegraphics[width=\textwidth]{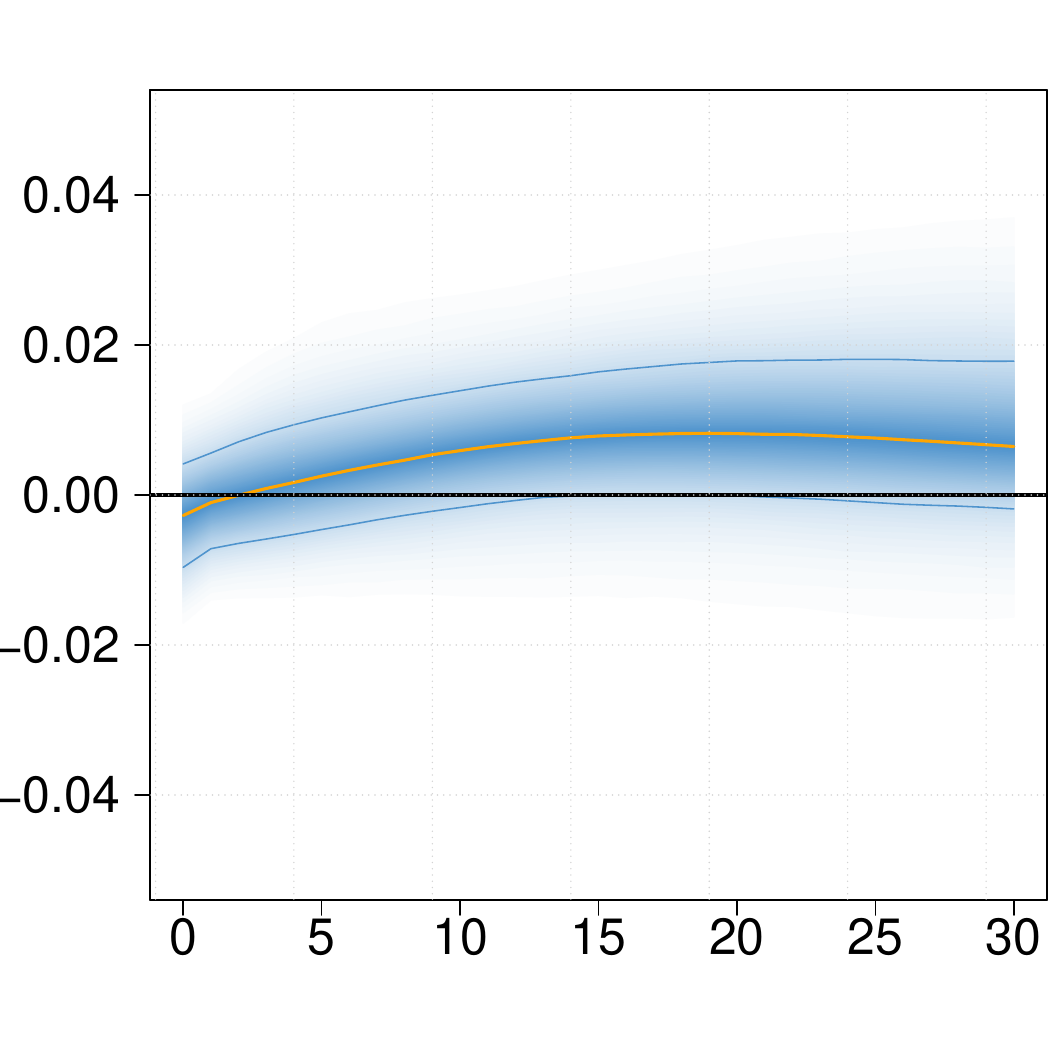}
\end{subfigure}
\begin{subfigure}{.329\textwidth}
\caption{Equity prices}
\includegraphics[width=\textwidth]{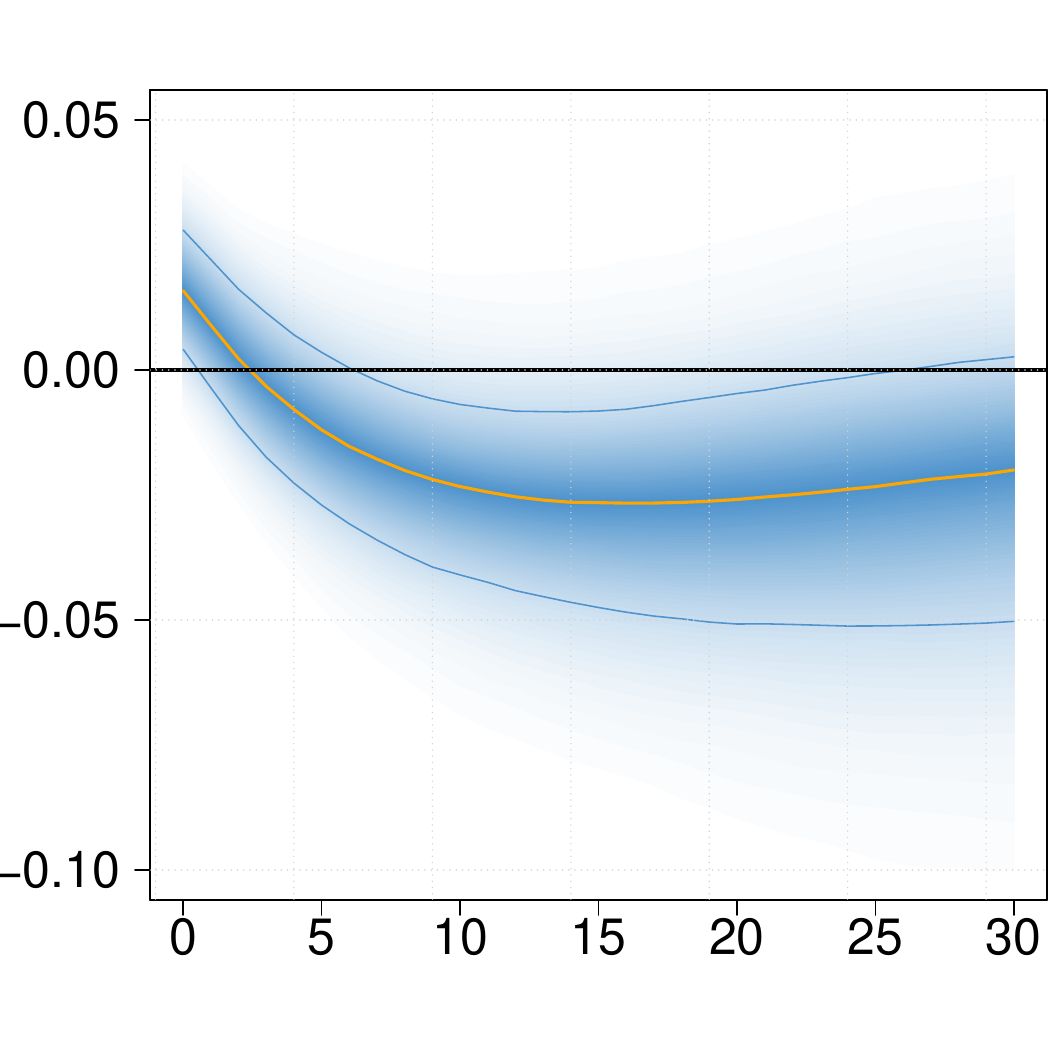}
\end{subfigure}\\
\begin{minipage}{17cm}
\footnotesize \textit{Notes}:  The plot shows the full posterior distribution of structural responses to a +100bp increase in the shadow rate using a Cholesky decomposition. Posterior median in orange along  68\% credible sets (in blue). An increase in the real effective exchange rate corresponds to an appreciation. 
\end{minipage}%
\end{figure}

Our main results are provided in Figures \ref{fig:ssr_all} and \ref{fig:ssr_worker}. The figures show the whole posterior distribution of impulse responses along the median and 68\% credible intervals. 

\begin{center}[INCLUDE \autoref{fig:ssr_all} HERE]\end{center}

\begin{center}[INCLUDE \autoref{fig:ssr_worker} HERE]\end{center}

We first consider our results for the larger sample that includes all households, depicted in \autoref{fig:ssr_all}. Here, we find that in response to the monetary tightening, output declines and unemployment rises over the medium-term. We also find a small but not precisely estimated negative effect on inflation expectations. Furthermore the monetary tightening induces an  appreciation of the real effective exchange rate and an increase in long-term yields.  Stock prices show a very similar picture to real output: while they initially pick up, there is a persistent decline up to about 25 quarters. 

Next, we focus on the sub-sample of workers' households, depicted in \autoref{fig:ssr_worker}. Here, the results are very similar to the ones based on the broader sample: Real GDP declines over the medium term, the unemployment rate rises and inflation expectations fall but not significantly so. We also find evidence for a decrease in stock market prices and a real appreciation of the effective exchange rate. 

The fact that the overall responses are very similar for both sub-samples is not surprising since both estimations rely on the same macroeconomic data. Moreover, the direction of responses discussed above are in line with economic theory and qualitatively similar to those provided in \citet{Nakajima2011}, who use a more complex time-varying parameter VAR with stochastic volatility. In this sense we do not find evidence for counterintuitive output and price responses often reported for Japanese data.\footnote{For example  \citet{Nakashima2017} attribute these short-run anomalies in output and price behavior to revision of expectations of market participants regarding future economic conditions.}

Turning now to the focal variable, a striking difference between the two samples emerge: The Gini coefficient declines when considering the broader definition of income data (all households, \autoref{fig:ssr_all}) implying that a monetary policy tightening decreases inequality and rather persistently so. A theoretical explanation for this finding could be through the tax inflation channel, less inflation benefiting the poorer who tend to use more cash. But the response of inflation expectations is small and not precisely estimated. Another and more plausible explanation might be  through job destruction, as indicated by a rise in unemployment. Since the sample contains about 35\% of unemployed and retirees, more unemployment leads to less inequality. It makes the whole sample poorer. Also, as long-term interest rates increase implying a tightening in financing conditions, spurious income through new loans, declines. This will also disproportionately affect the part of households which have an active income source, leading to less inequality.

The result reverses, if we consider income data  based on households whose head is employed (workers' households, \autoref{fig:ssr_worker}). More specifically, the effect on the Gini is positive, but not precisely estimated. A positive effect on income inequality after a monetary tightening would be in line with estimates of \citet{Coibion2017} for the USA and theoretical predictions based on a New Keynesian model provided in \citet{Gornemann2012}. The rise in unemployment, tighter financing conditions through higher long-term rates and a real appreciation, which causes firms in the tradable sector to either lay of workers or cut wages, all could lead to more inequality among the Japanese workforce. We elaborate more on the relative strength of these transmission channels in the next section. 

%\subsection{Through which channels is income inequality affected?}

Summing up, we find that a tightening of the monetary policy stance, decreases  output and equity prices, while the real effective exchange rate appreciates and unemployment rises. These findings hold true for both groups of households and are in line with common economic reasoning, which ensures overall confidence in our estimation strategy. It turns out that these macroeconomic developments have very distinct effects depending on the employment status of the household's head. In the larger panel, which also contains the unemployed and retirees, income inequality decreases once monetary policy is tightened. This finding is not present, once considering the smaller group of households who are employed. Here, the Gini has a tendency to increase but effects are not precisely estimated.

\subsection{Through which channels does monetary policy affect income inequality?}

In this section, we investigate the relative strength of the various transmission channels. For that purpose, we construct a sequence of counterfactual shocks that neutralize effects through a particular variable. The difference between the counterfactual response and the unconditional response (i.e., with all channels open) indicates the relative importance of a particular channel. 

Recall that the contemporaneous relationships are captured by the lower triangular, structural $m \times m$ matrix $\boldsymbol{Q}$ and $\mSigma = \boldsymbol{Q} \boldsymbol{Q}^{\prime}$ being a Cholesky decomposition of $\mSigma$. We are now interested in the effect of a monetary policy shock on the Gini coefficient shutting down effects through a particular channel / variable $j$. Here, we can distinguish between a \textit{direct effect} and an \textit{indirect} effect of the monetary policy shock on the Gini. The direct effect can be obtained by  $q_{1,5}$, denoting the element in the $k=1$st row and fifth column of $\boldsymbol{Q}$. It is the fifth column since we have ordered the shadow rate as the fifth variable in the VAR, and the first row since the Gini is ordered first. Similarly, the effect of monetary policy on any other variable $j$ is denoted by $q_{j,5}$. The \textit{indirect effect} of the monetary policy shock on the Gini \textit{through} variable $j$ is consequently measured by $q_{1,5} \times q_{j,5}$ and  additional higher order effects that can arise through the lag structure in the system \citep{Bachmann2012}.  
Following \citet{Bachmann2012} and \citet{Wong2015}, we proceed by constructing a hypothetical shock sequence for equation $j$ that equalizes the response of $j$ to the monetary policy shock. The resulting impulse responses correspond to a hypothetical scenario under which the constructed shock completely offsets the indirect effects through a particular channel. Manipulating the error terms in that way has the advantage of not having to estimate a new structural model.\footnote{As an alternative, consider the approaches of  \citet{Baumeister2013} and \citet{Ludvigson2002} who manipulate the VAR coefficients and corresponding elements in the variance covariance matrix directly to offset the effects of a particular variable.} 

\begin{sidewaysfigure}[p]
\caption{Responses of the Gini coefficient to an increase in the shadow rate -- different channels switched off}\label{fig:shut}
\begin{minipage}{1\linewidth}~\\
\centering \textbf{All households}
\end{minipage}\\
\begin{minipage}[b]{00.162\linewidth}
No real GDP
\centering \includegraphics[scale=0.2]{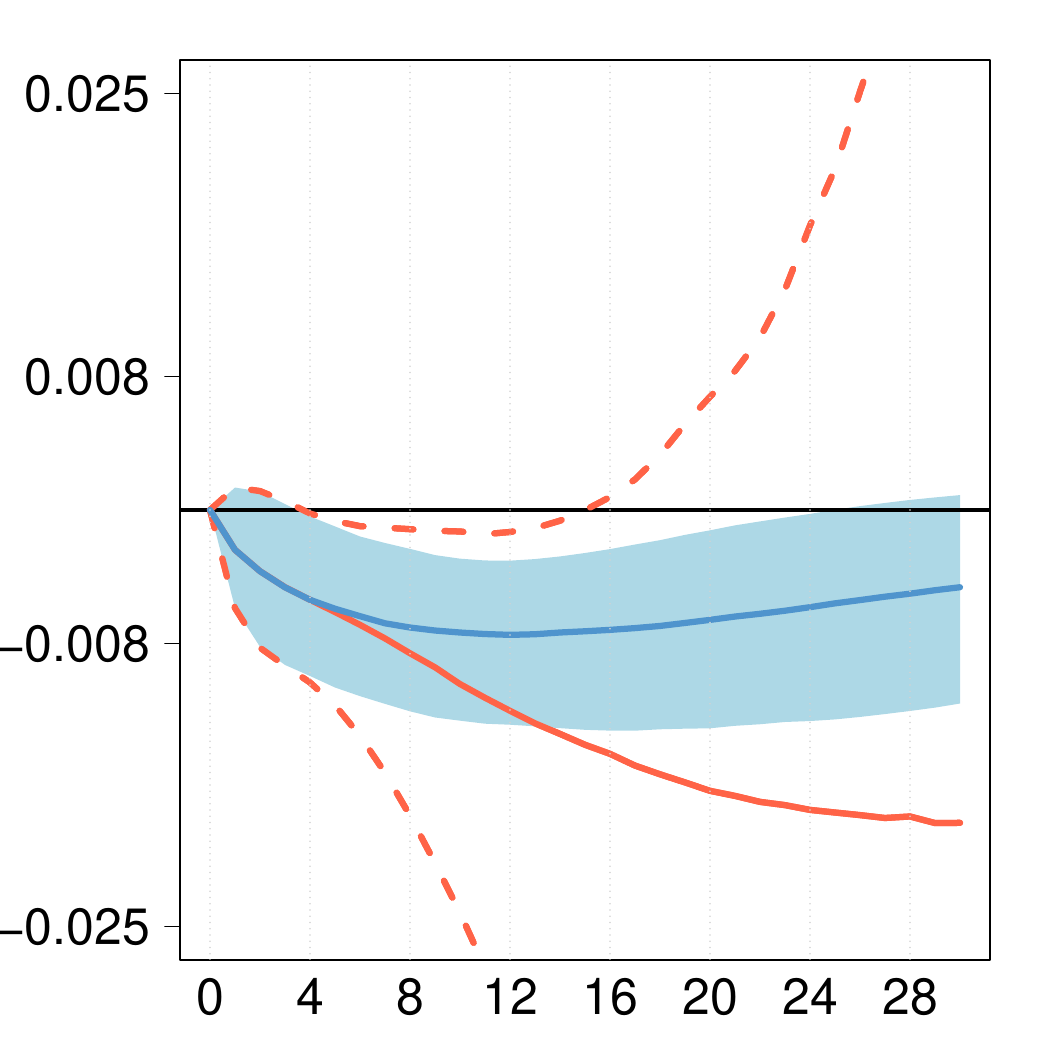}
\end{minipage}%
\begin{minipage}[b]{00.162\linewidth}
No inflation expectations
\centering \includegraphics[scale=0.2]{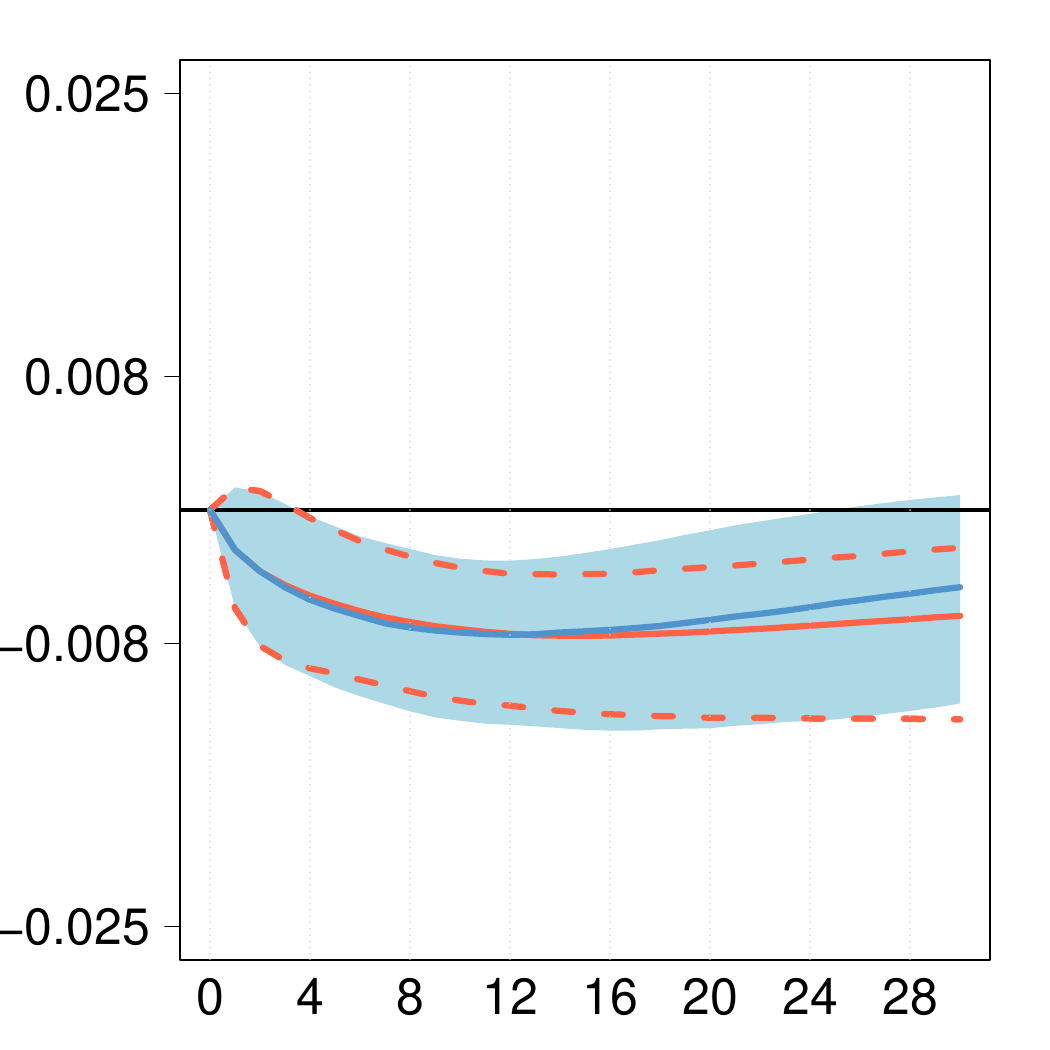}
\end{minipage}
\begin{minipage}[b]{00.162\linewidth}
No unemployment
\centering \includegraphics[scale=0.2]{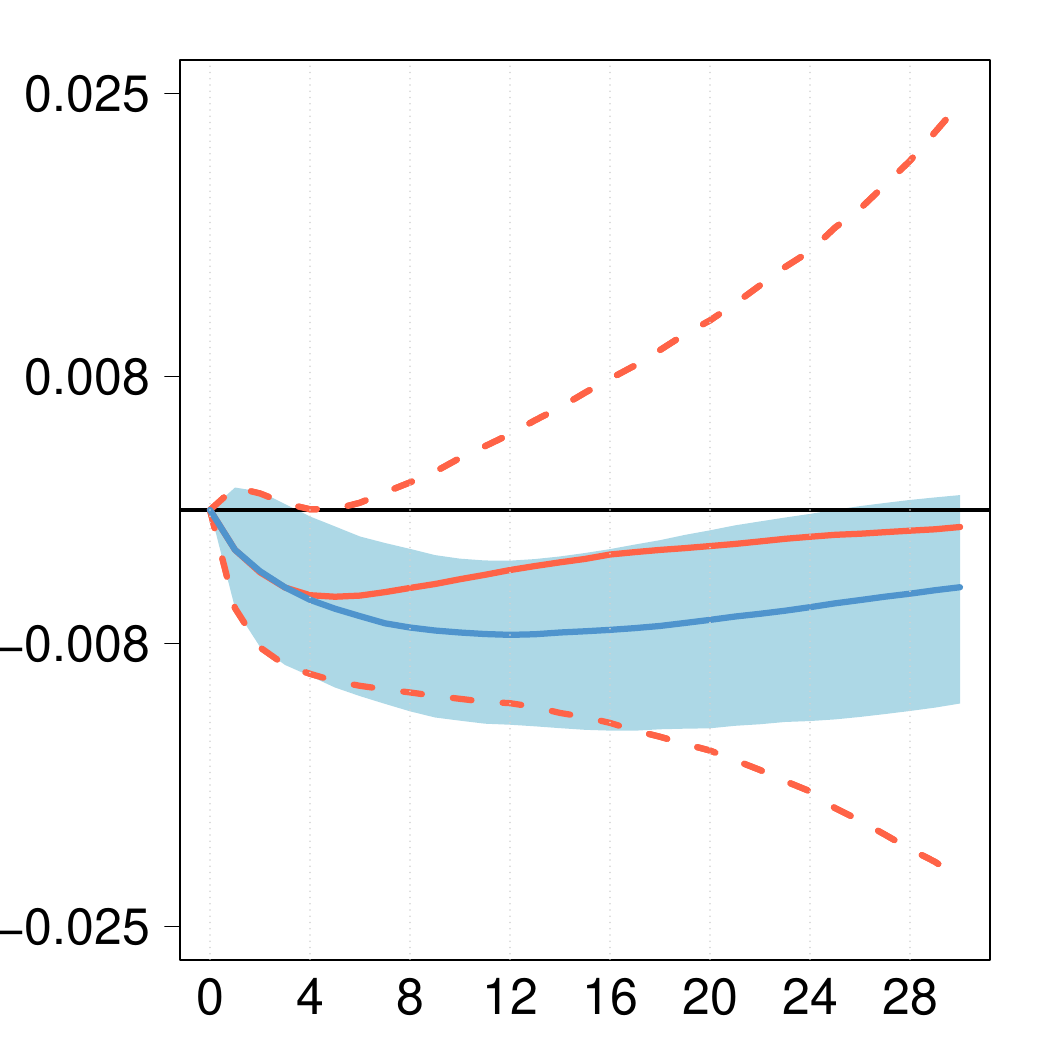}
\end{minipage}
\begin{minipage}[b]{00.162\linewidth}
No long rates
\centering \includegraphics[scale=0.2]{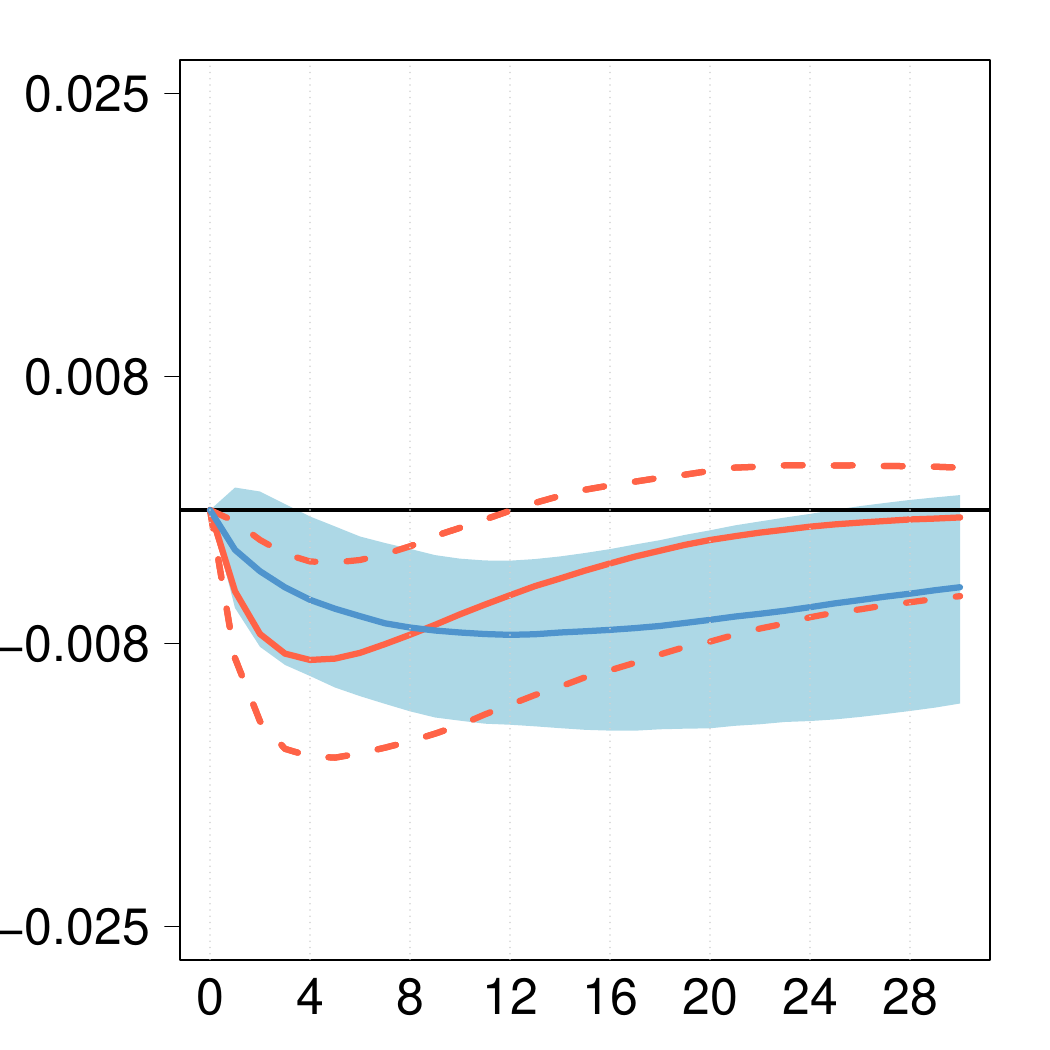}
\end{minipage}%
\begin{minipage}[b]{00.162\linewidth}
No real eff. exchange rate
\centering \includegraphics[scale=0.2]{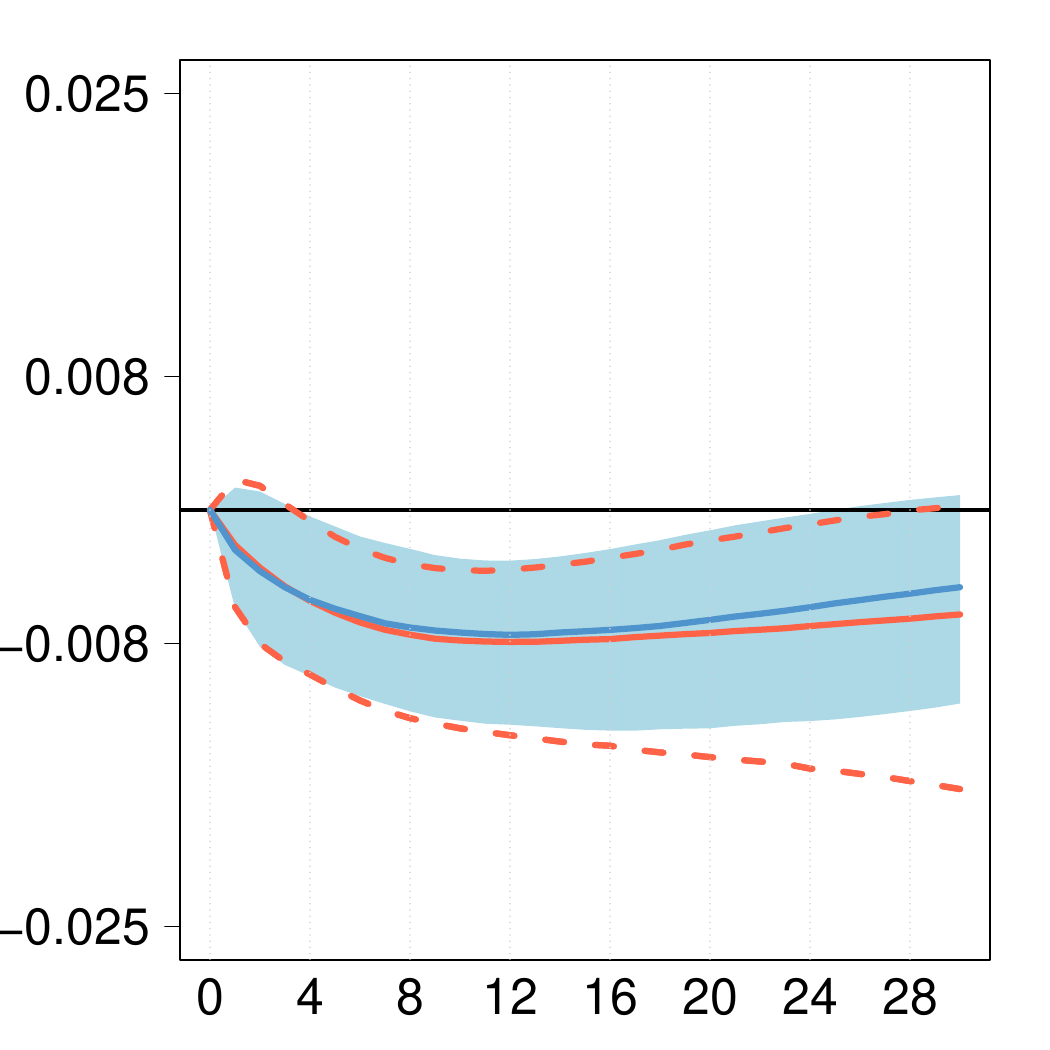}
\end{minipage}
\begin{minipage}[b]{00.162\linewidth}
No equity prices
\centering \includegraphics[scale=0.2]{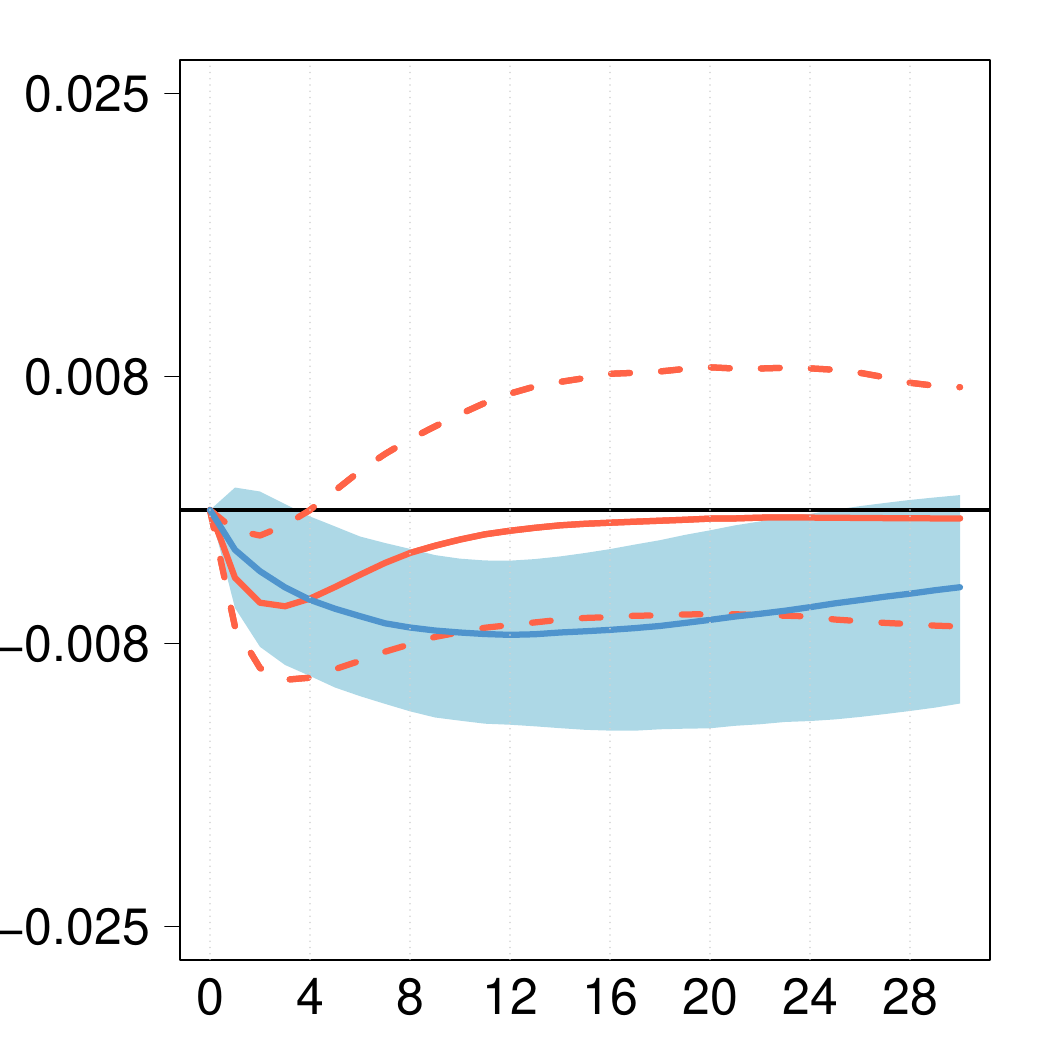}\\
\end{minipage}\\
\begin{minipage}{1\linewidth}~\\
\centering \textbf{Workers' households}
\end{minipage}\\
\begin{minipage}[b]{00.162\linewidth}
No real GDP
\centering \includegraphics[scale=0.2]{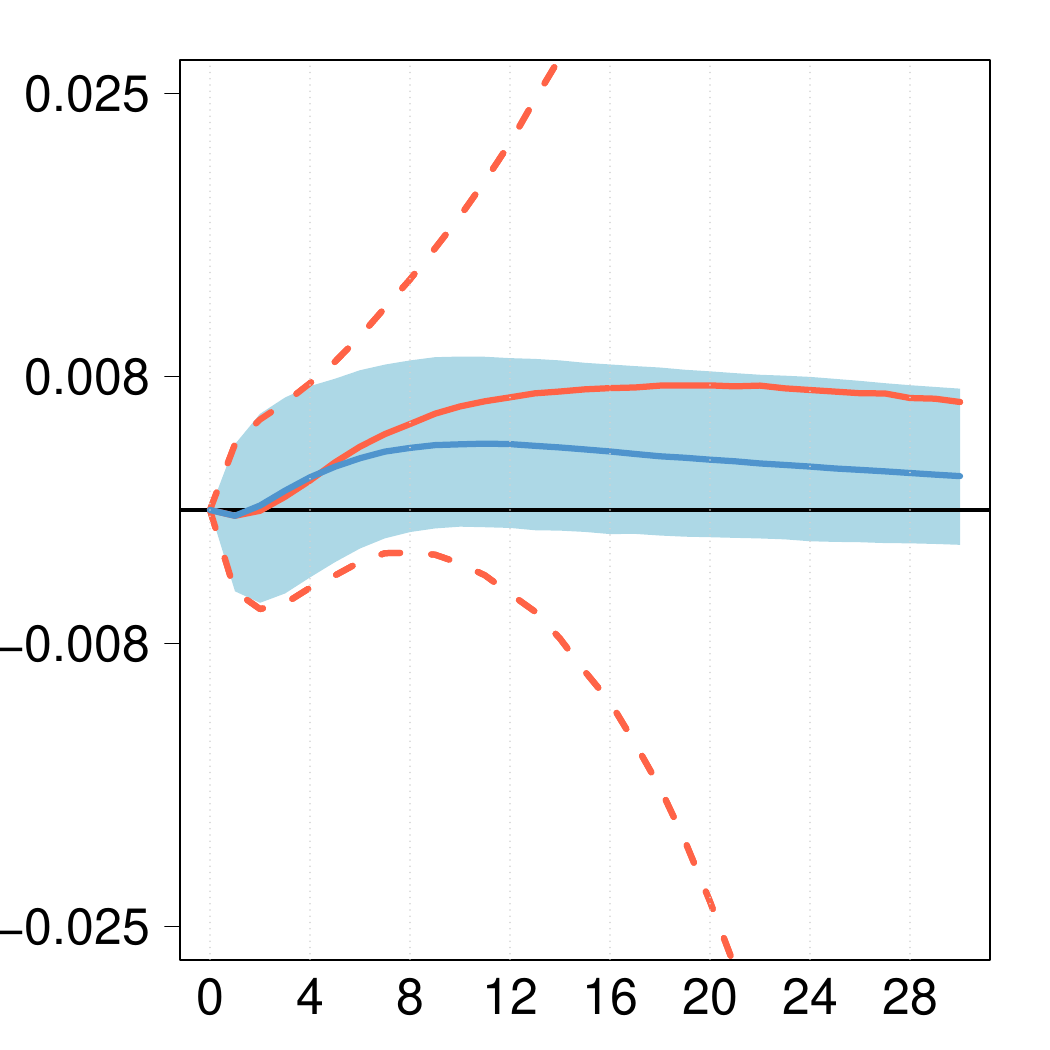}
\end{minipage}%
\begin{minipage}[b]{00.162\linewidth}
No inflation expectations
\centering \includegraphics[scale=0.2]{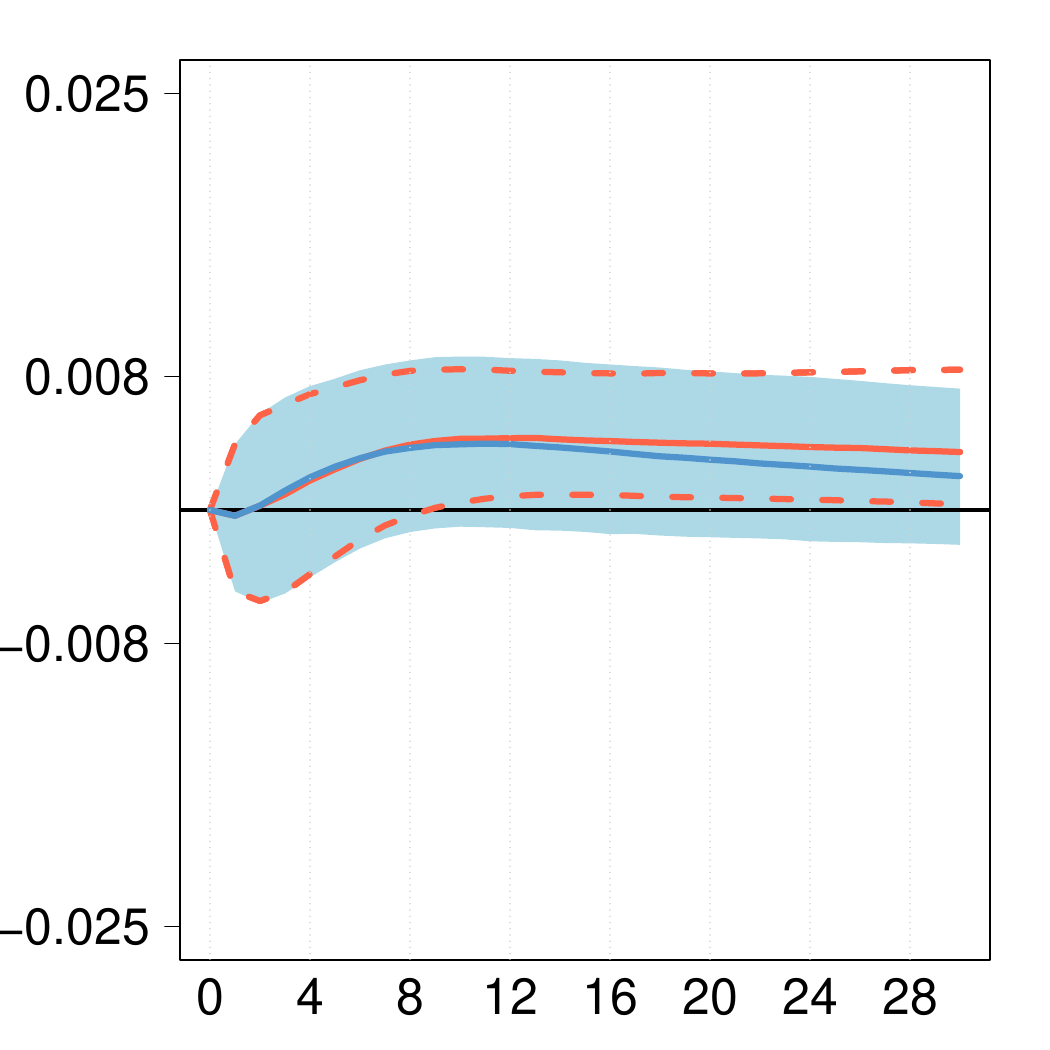}
\end{minipage}
\begin{minipage}[b]{00.162\linewidth}
No unemployment
\centering \includegraphics[scale=0.2]{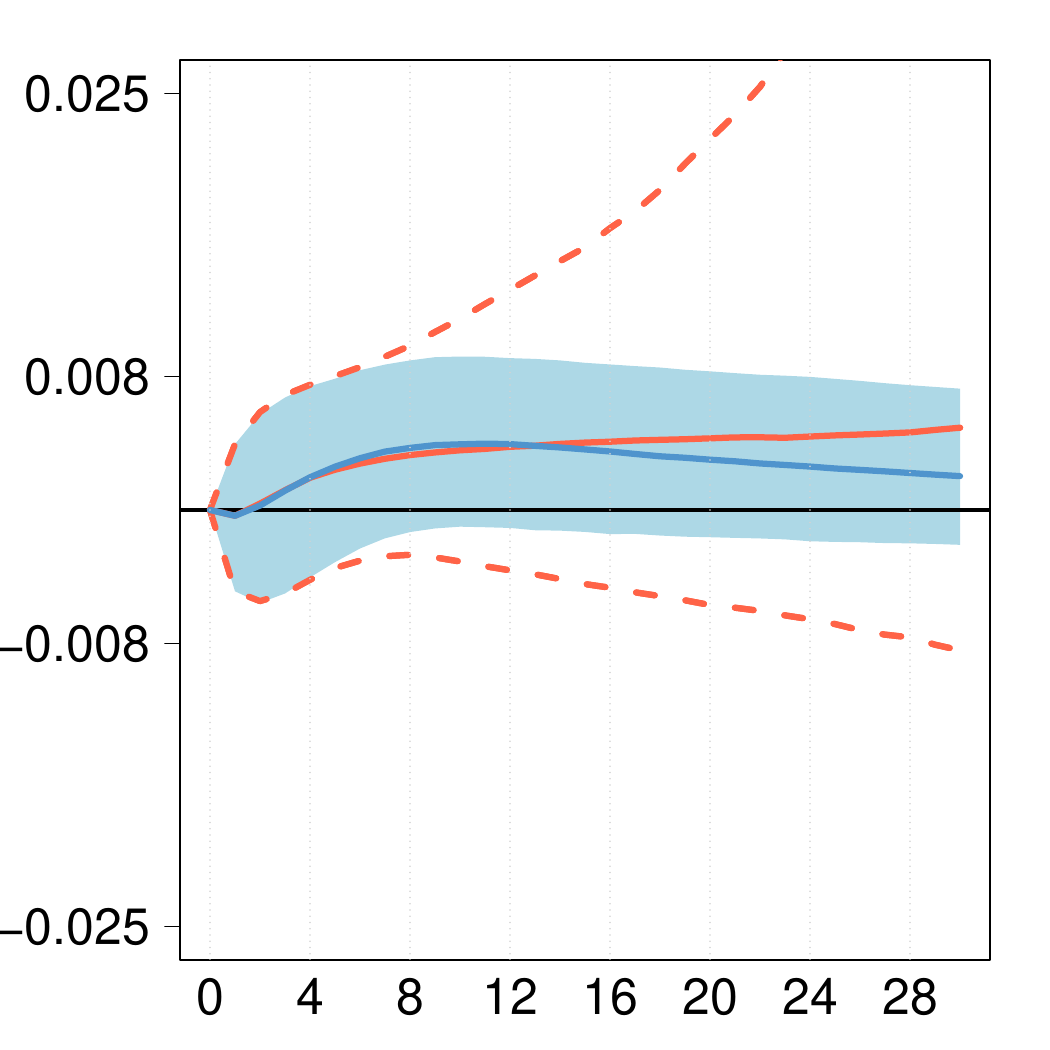}
\end{minipage}
\begin{minipage}[b]{00.162\linewidth}
No long rates
\centering \includegraphics[scale=0.2]{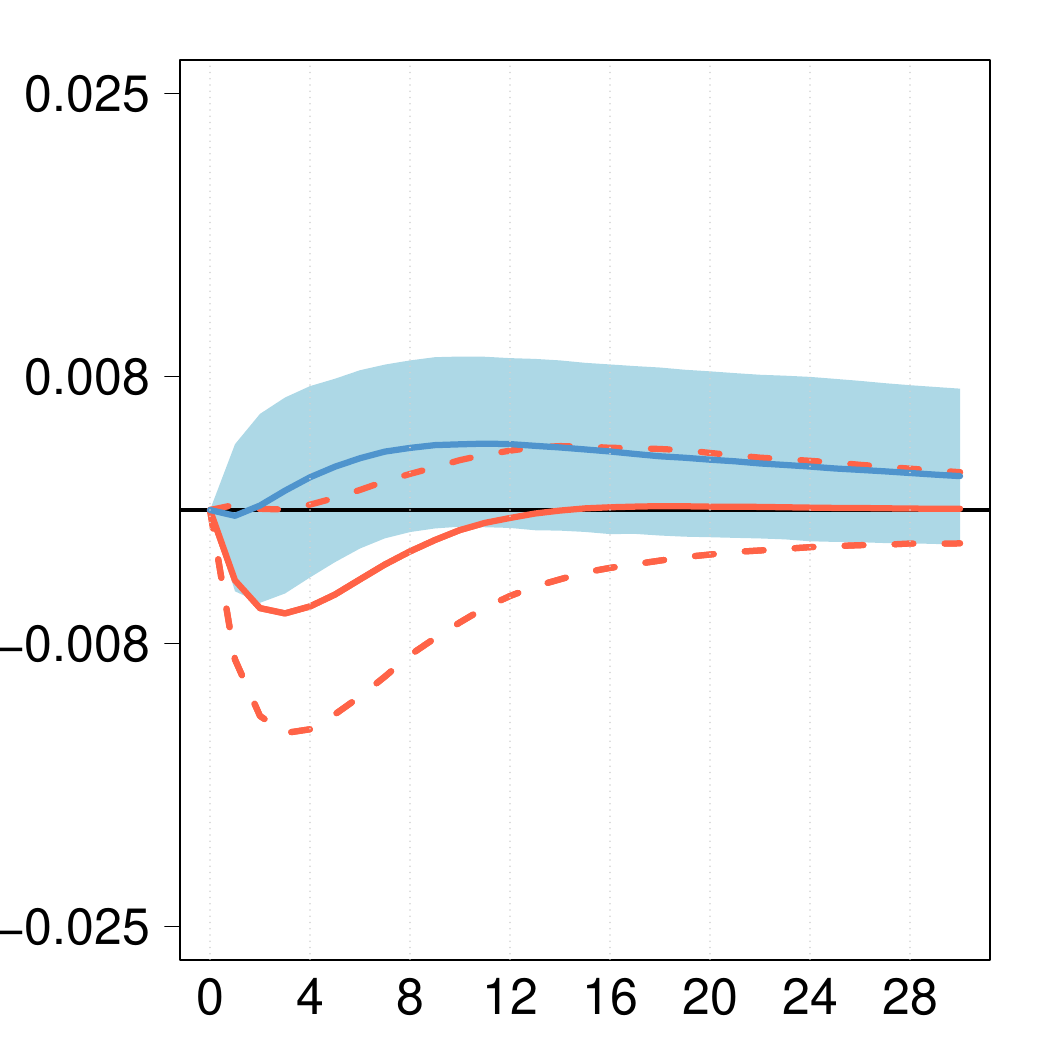}
\end{minipage}%
\begin{minipage}[b]{00.162\linewidth}
No real eff. exchange rate
\centering \includegraphics[scale=0.2]{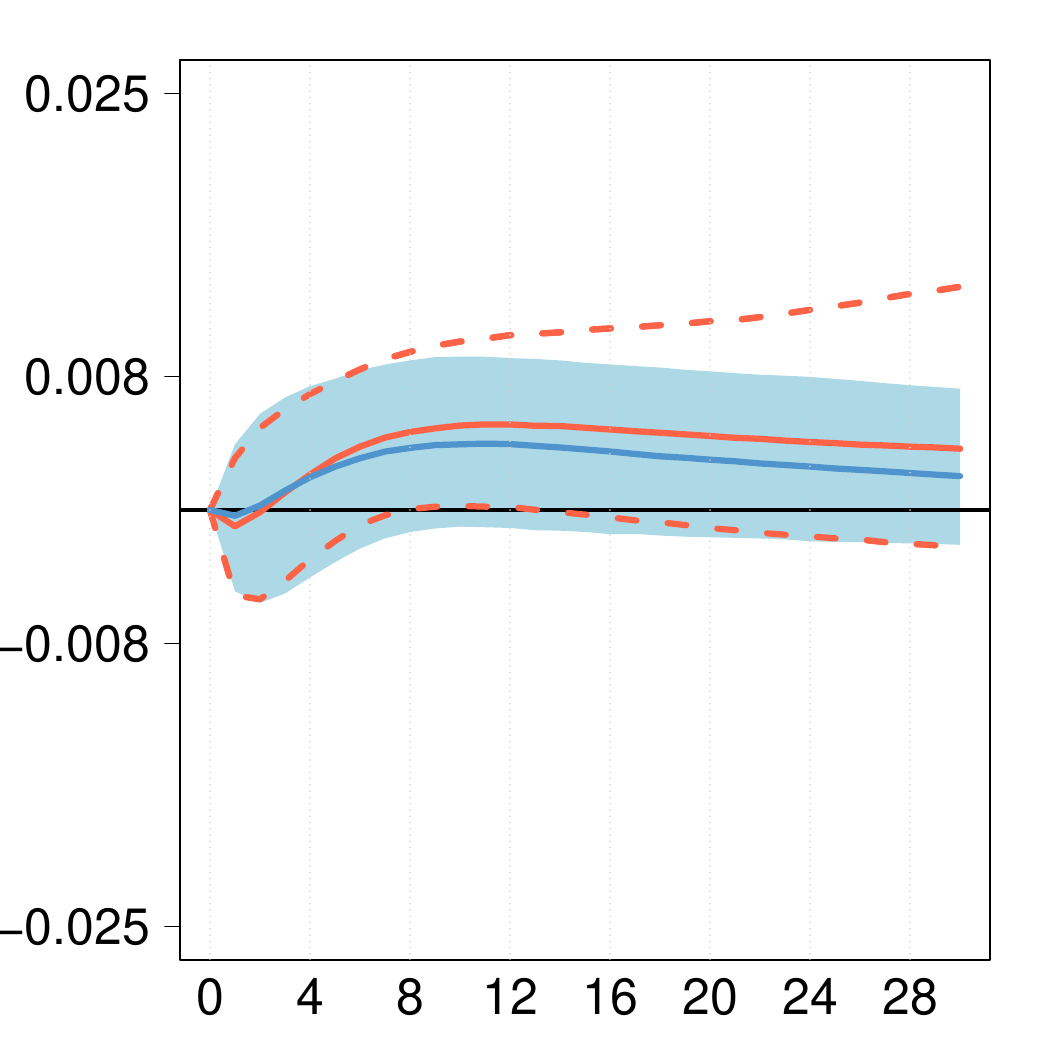}
\end{minipage}
\begin{minipage}[b]{00.162\linewidth}
No equity prices
\centering \includegraphics[scale=0.2]{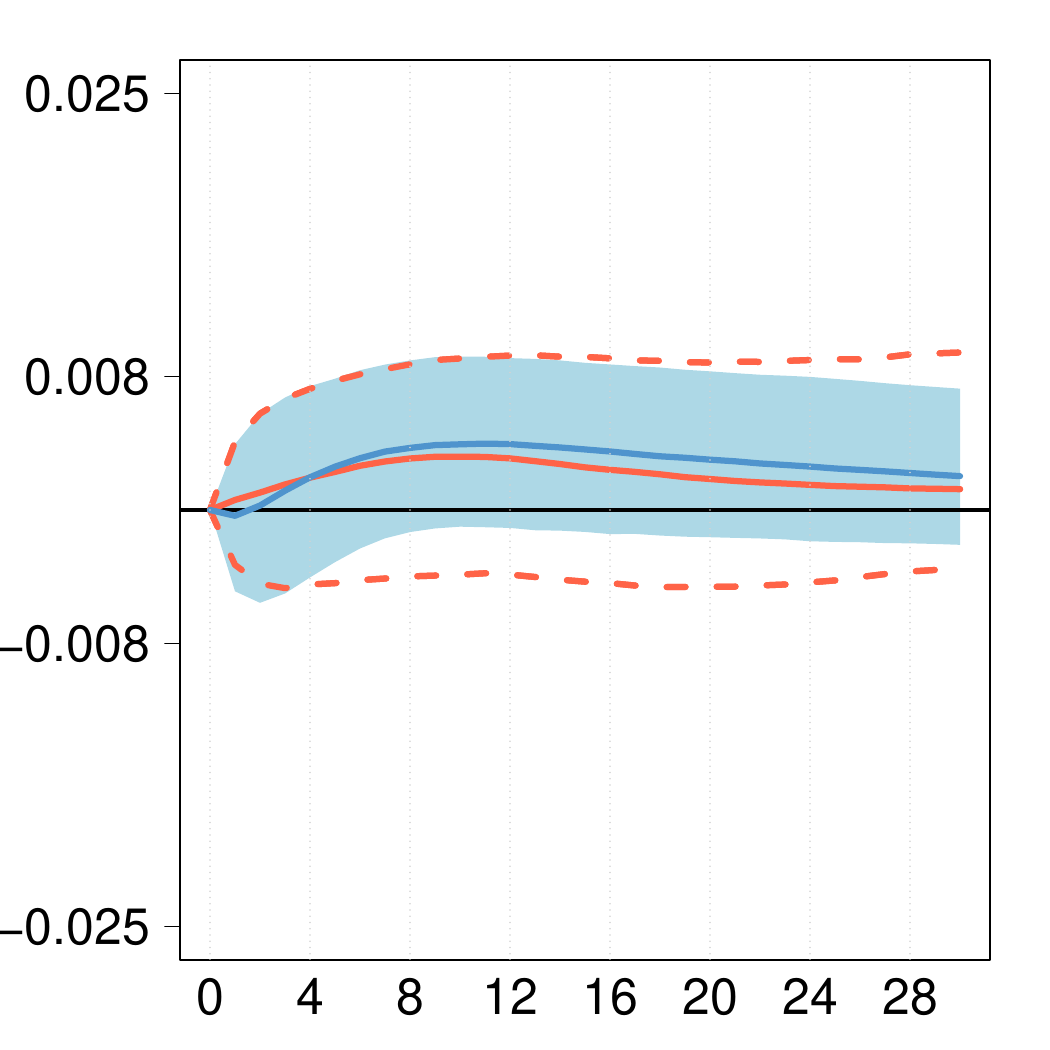}\\
\end{minipage}\\
\begin{minipage}{22cm}~\\
\footnotesize The plot shows the response of the Gini coefficient to a +100bp increase in the shadow rate based on a Cholesky decomposition. In blue, we show the unconditional response, in orange the one with a particular transmission channel switched-off. Posterior median along 68\% credible sets. An increase in the real effective exchange rate corresponds to an appreciation. 
\end{minipage}%	
\end{sidewaysfigure}

The results are depicted in \autoref{fig:shut}. The figure shows the counterfactual responses on the Gini coefficient with a particular variable / channel shut-down in orange and the unconditional response for comparison in blue.

\begin{center}[INCLUDE \autoref{fig:shut} HERE]\end{center}

Examining the relative importance of the channels for the broader household sample shows that the negative response persists for all counterfactual scenarios. In particular we see that counterfactual responses nullifying effects through either inflation or the real effective exchange rate are very similar to the unconditional responses, which leave these channels open. This indicates that these two variable do not play an important role in determining the effects on income inequality. We next consider the unemployment rate. Intuitively and from the perspective of an already unemployed household, other people losing their jobs equalizes the income distribution. Nullifying this effect would result into a slightly less negative effect on the Gini coefficient over the medium term as the orange solid line lies above the blue solid line. This implies that part of the effect of monetary policy on income inequality runs through the unemployment rate. Last and considering financial variables, we do find different adjustment patterns looking at the responses of long-term rates and equity prices. In both cases shutting off responses through financial variables would lead to a sharper but less persistent response of the Gini. That the financial channel is important in transmitting monetary policy is consistent with the fact that monetary policy in Japan during the sample period worked mostly through the longer end of the yield curve and financial markets.

The counterfactual exercise using the sub-sample of workers' households is contained in the bottom panel of \autoref{fig:shut}. Here, we see that the Gini tends to increase throughout all counterfactual scenarios. Moreover, counterfactual responses of inflation expectations, the unemployment rate, the exchange rate and equity prices all lead to very similar results compared to the unconditional effect on the Gini. Hence, for the income distribution of workers, these variables seem of less importance in the context of a monetary policy shock. %A modest difference arises considering effects through the unemployment rate over the longer term. This could be interpreted as evidence for  job-destruction channel for workers' households: as interest rates increase, unemployment rises which affects poor households (that can be  laid off more easily) more strongly triggering a rise in income inequality.
The largest difference between counterfactual and unconditional responses can be observed when looking at effects through long-term interest rates. Here, we see that -- in line with unconditional responses of all households -- the effect on the Gini would be negative once transmission through long-term rates would be shut-down. This implies that an important channel of monetary policy transmission works through long-term rates: if these increase, overall financing conditions deteriorate and for part of the population it is harder to get a loan (which is included as "spurious" income in the income data). This leads to more inequality. %Once this effect is nullified, a monetary tightening would lead to less inequality.

Summing up, we find that unemployment and a financial channel both play an important role in determining the effect of monetary policy an inequality for the broad set of households. Both channels render the more equalizing effect of monetary policy more persistent. Most importantly, we find that the difference of distributional effects of monetary policy between workers' and all households are driven by effects through long-term rates. %Once this channel is shut-down, an increase in the shadow rate would also trigger a decrease in inequality for the sample of households whose head is employed. 
This implies that tighter financing conditions in the unconditional model are a main driver of inequality. 

%Long-term interest rates might affect the spurious income part of total pre-tax income: As long-term rates increase, borrowing money gets more expensive resulting into less spurious cash. Since households who can apply for a loan are not evenly distributed among the population, inequality increases. That this effect is present for  workers' households only points at a more important role of spurious income for these compared to all households. 

\subsection{Robustness exercises}
\begin{figure}[p]
\caption{Alternative estimations}\label{fig:rob}
\begin{minipage}{1\linewidth}~\\
\centering \textbf{Sign restrictions}
\end{minipage}\\
\begin{minipage}[b]{0.23\linewidth}
Gini
\centering \includegraphics[scale=0.20]{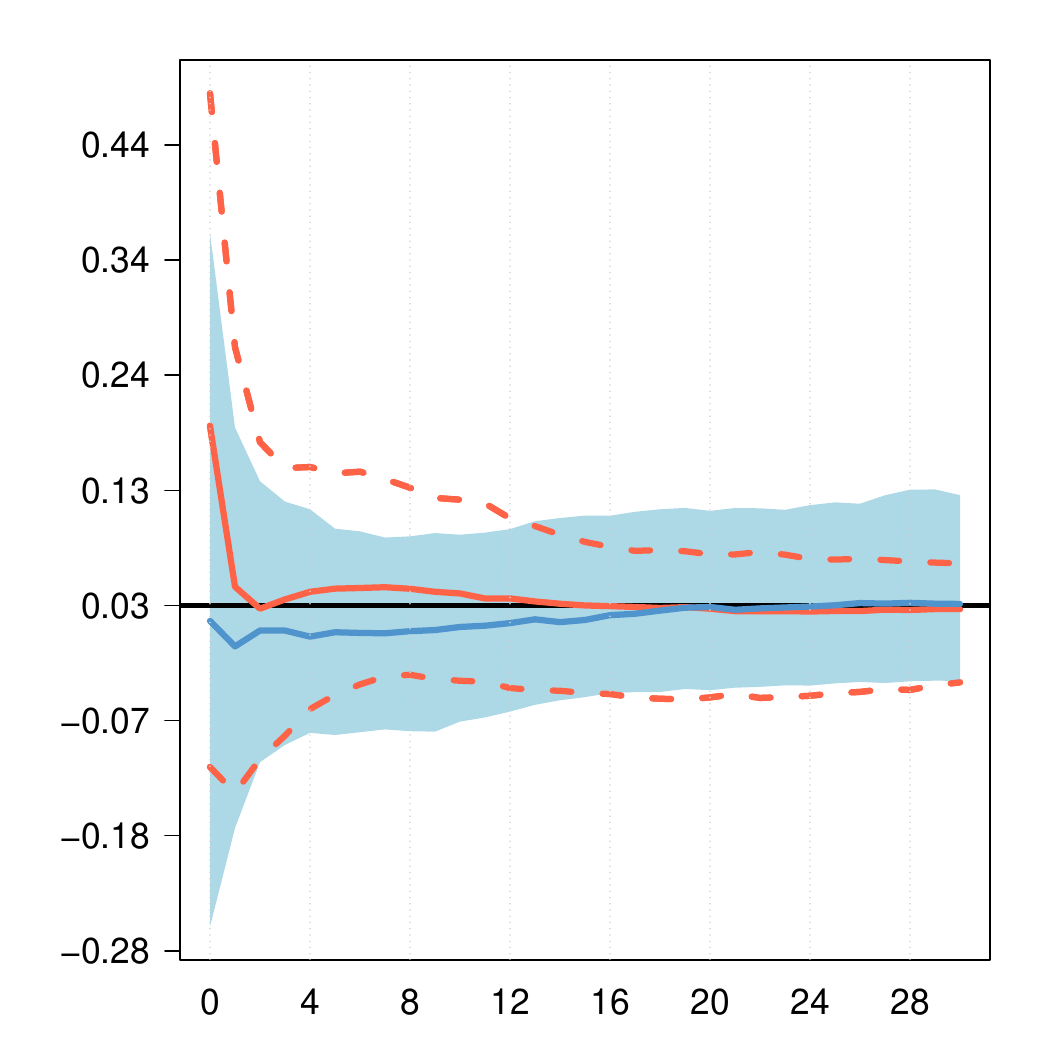}
\end{minipage}%
\begin{minipage}[b]{0.23\linewidth}
Real GDP
\centering\includegraphics[scale=0.20]{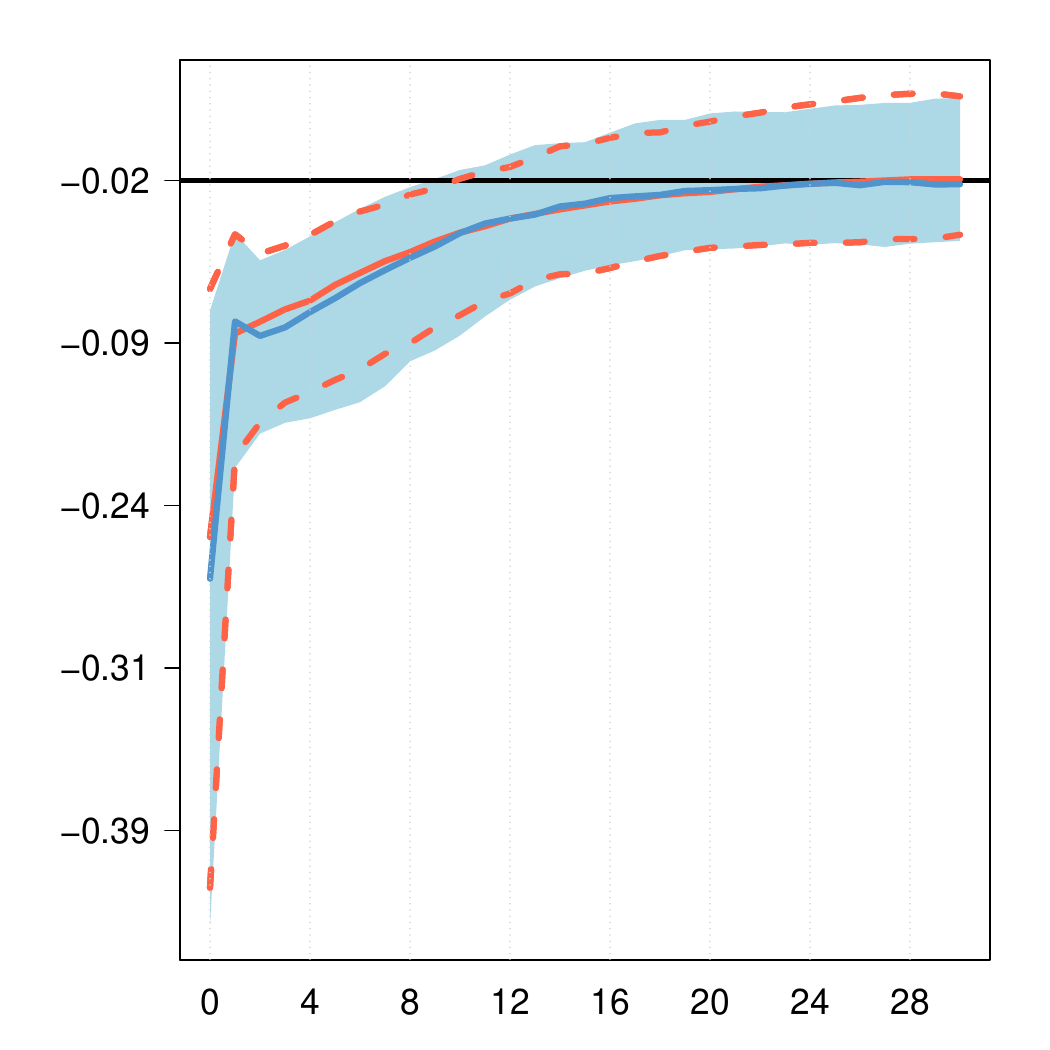}
\end{minipage}%
\begin{minipage}[b]{.23\textwidth}
Infl. expect.
\centering\includegraphics[scale=0.20]{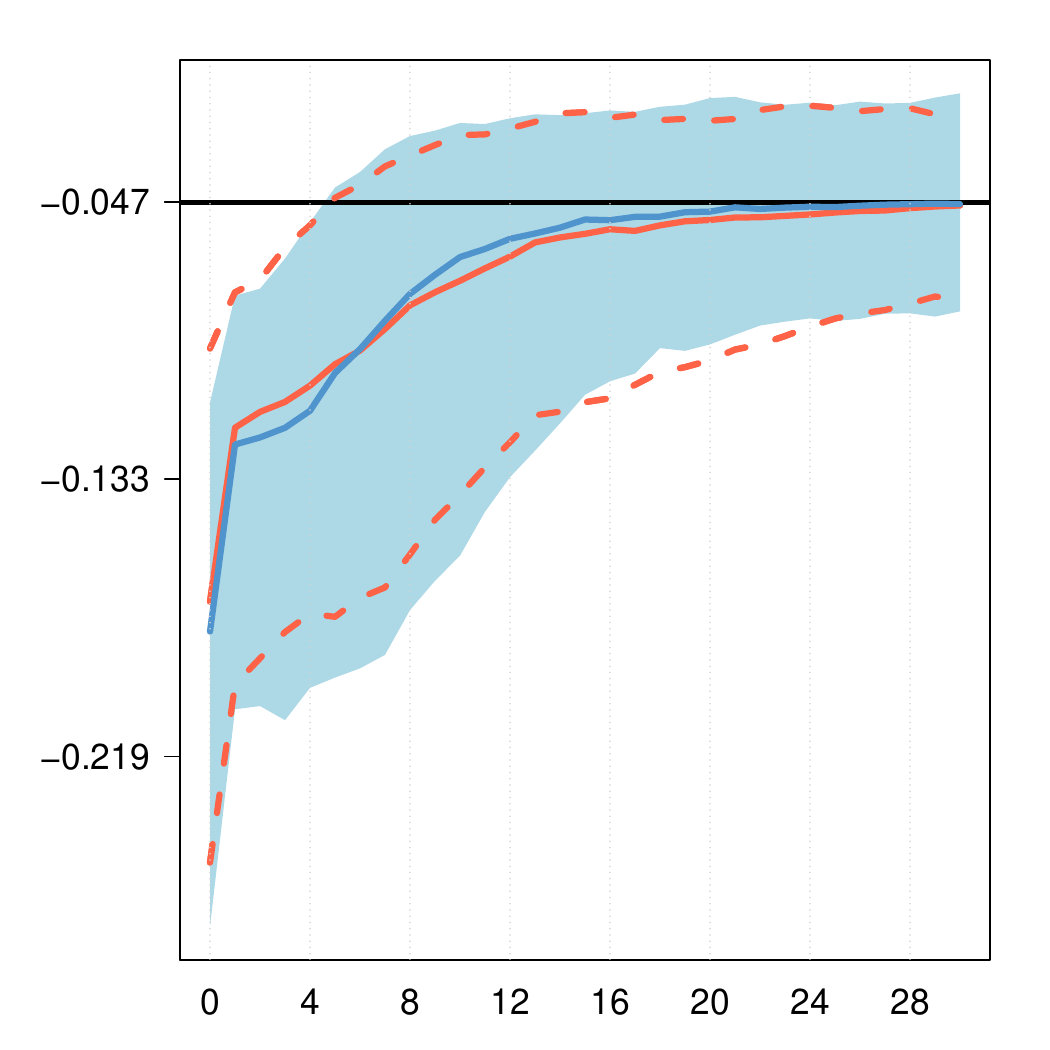}
\end{minipage}
\begin{minipage}[b]{.23\textwidth}
Unempl. rate
\centering\includegraphics[scale=0.20]{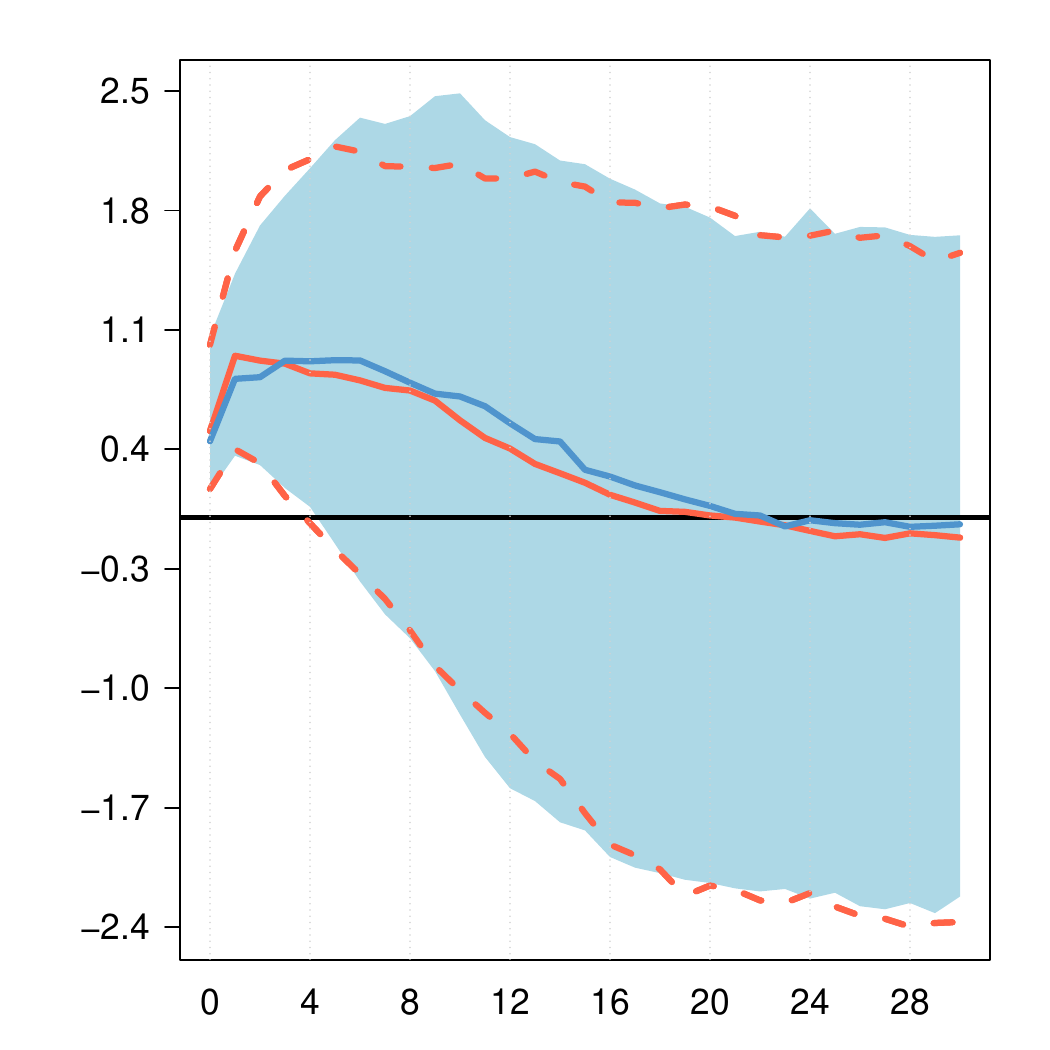}
\end{minipage}\\[.7em]
\begin{minipage}[b]{.23\textwidth}
Shadow rates
\centering\includegraphics[scale=0.20]{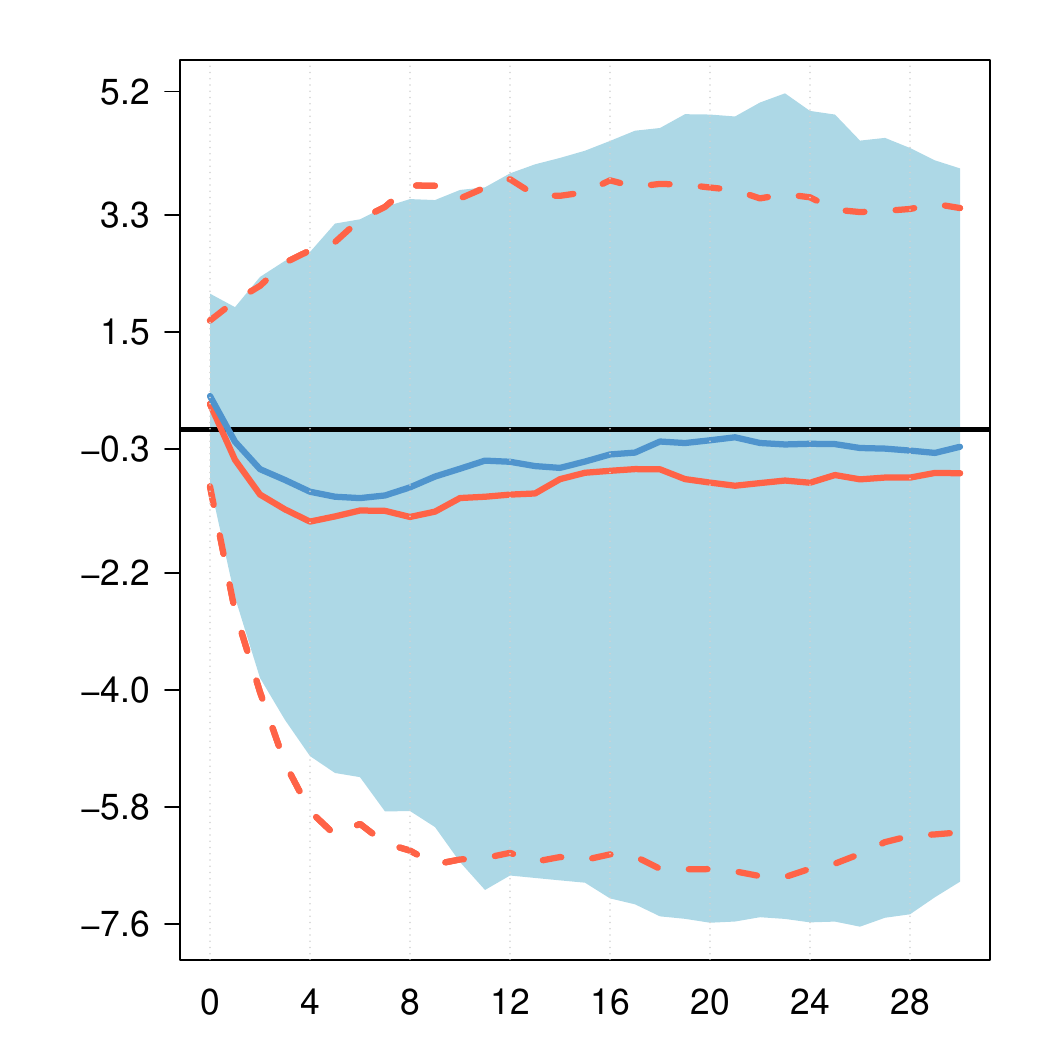}
\end{minipage}
\begin{minipage}[b]{.23\textwidth}
Long rates
\centering\includegraphics[scale=0.20]{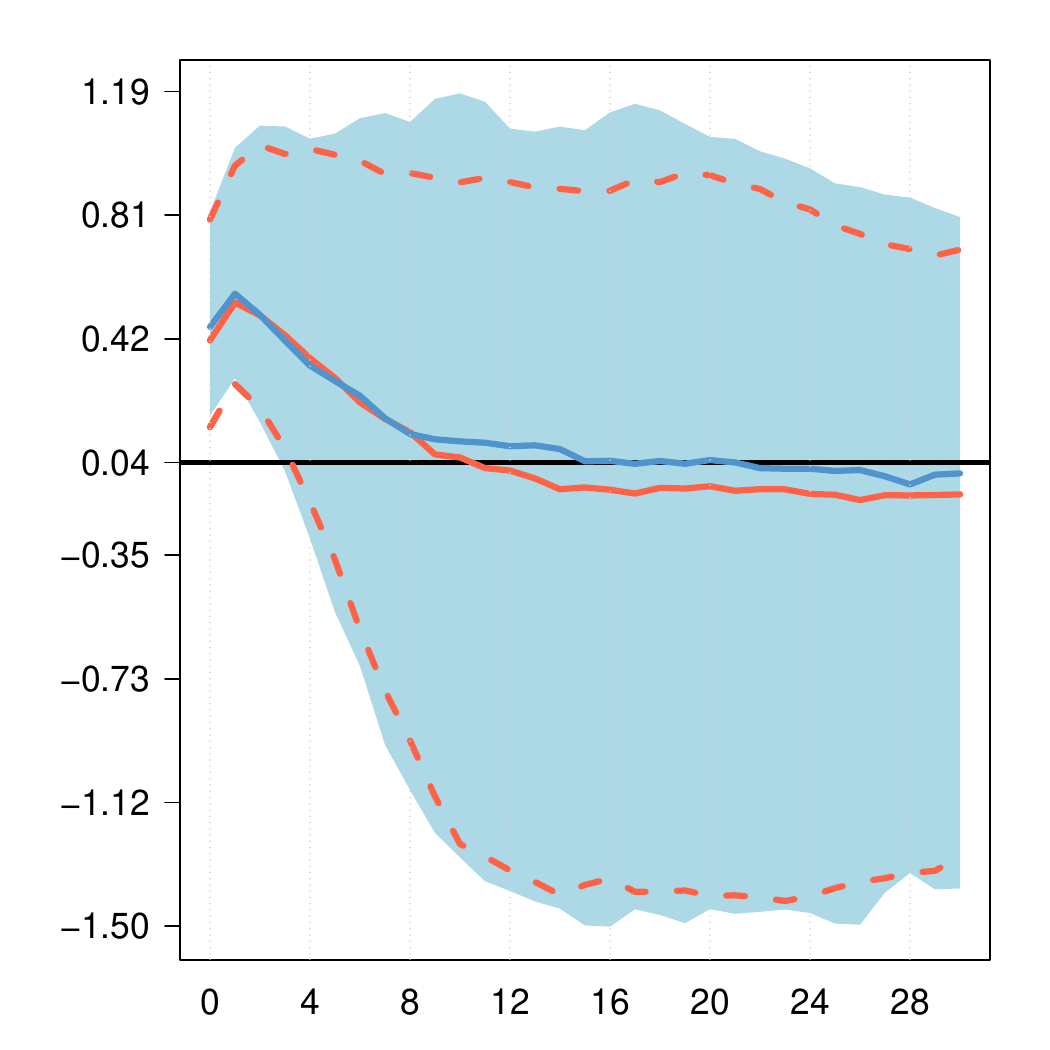}
\end{minipage}
\begin{minipage}[b]{.23\textwidth}
Real eff. exchange rate
\centering\includegraphics[scale=0.20]{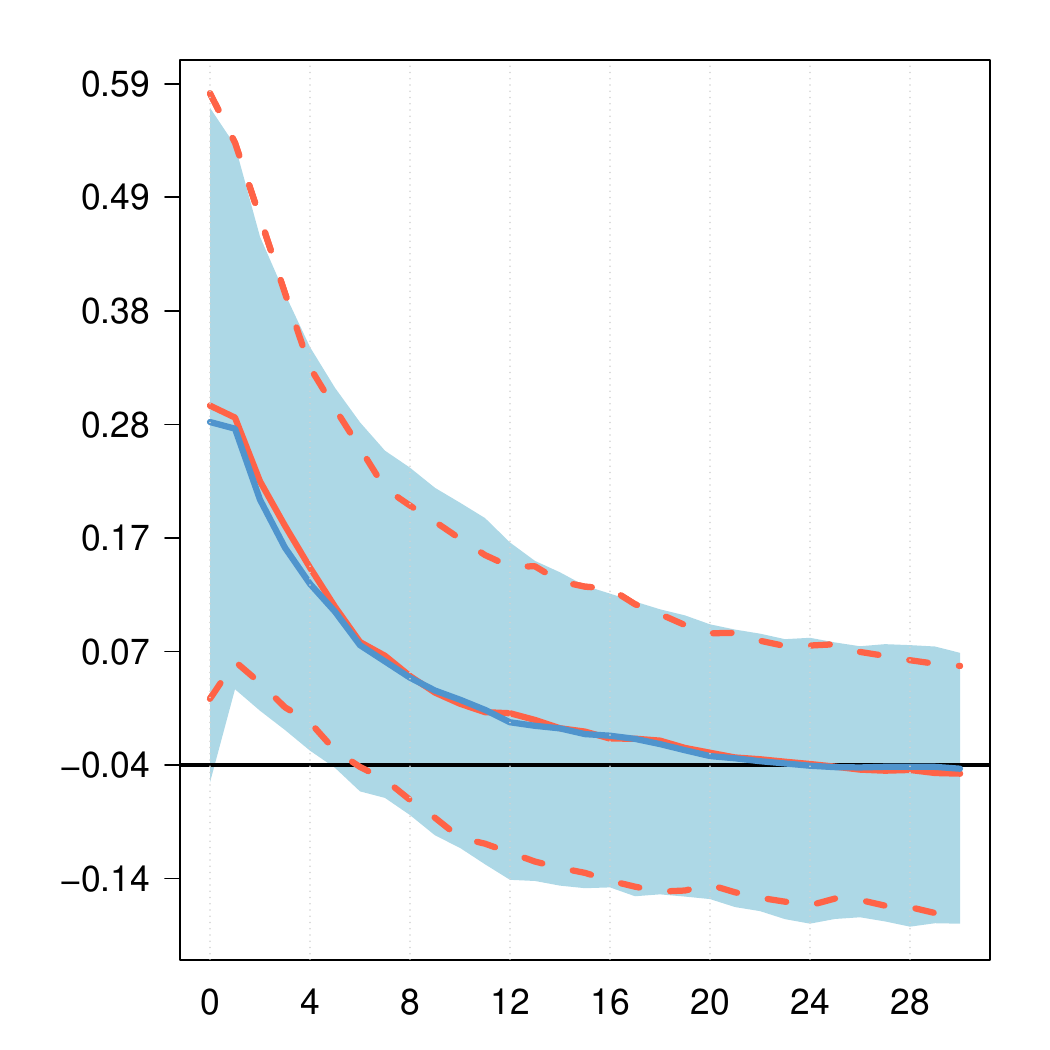}
\end{minipage}
\begin{minipage}[b]{.23\textwidth}
Equity prices
\centering\includegraphics[scale=0.20]{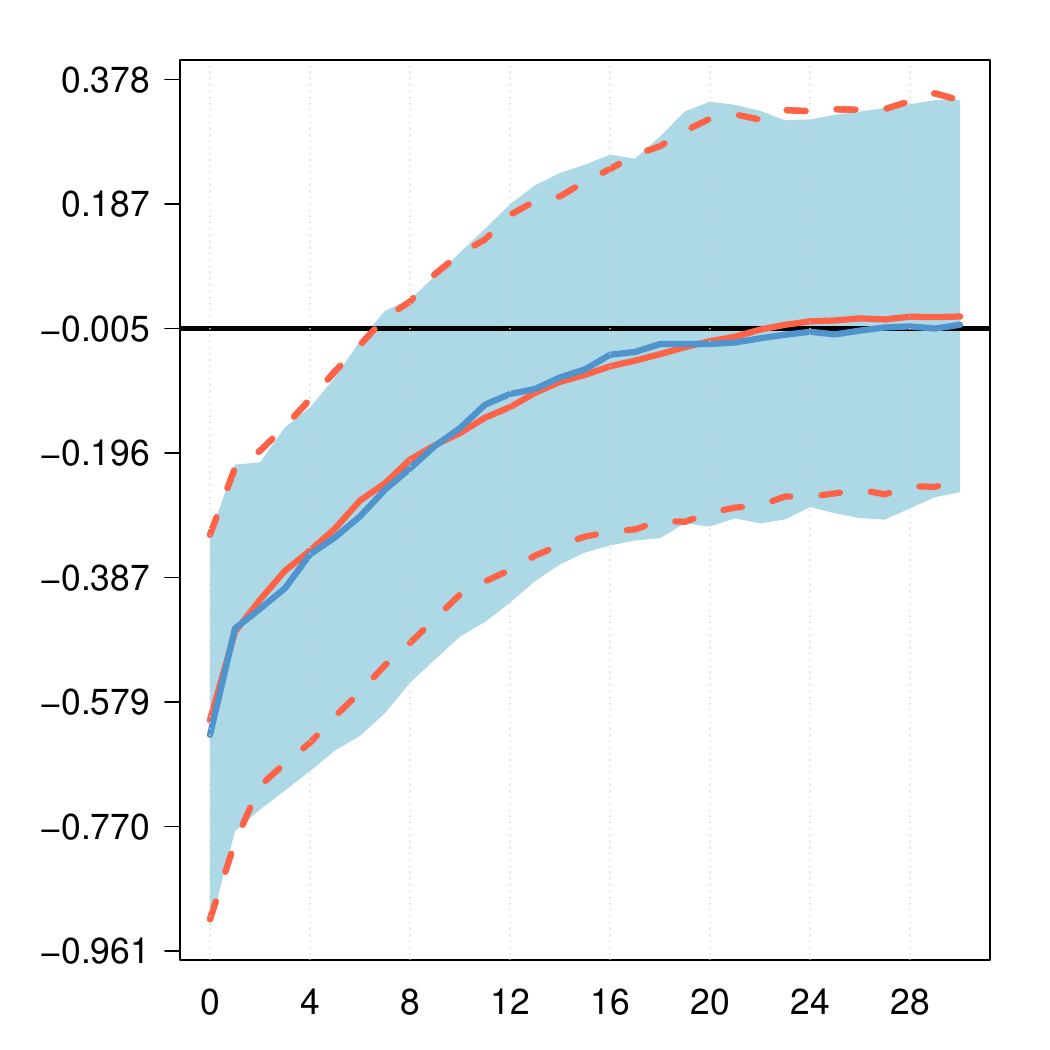}
\end{minipage}\\[.7em]
\begin{minipage}{1\linewidth}~\\
\centering \textbf{2-year government bond yields}
\end{minipage}\\
\begin{minipage}[b]{.23\textwidth}
Gini
\centering\includegraphics[scale=0.20]{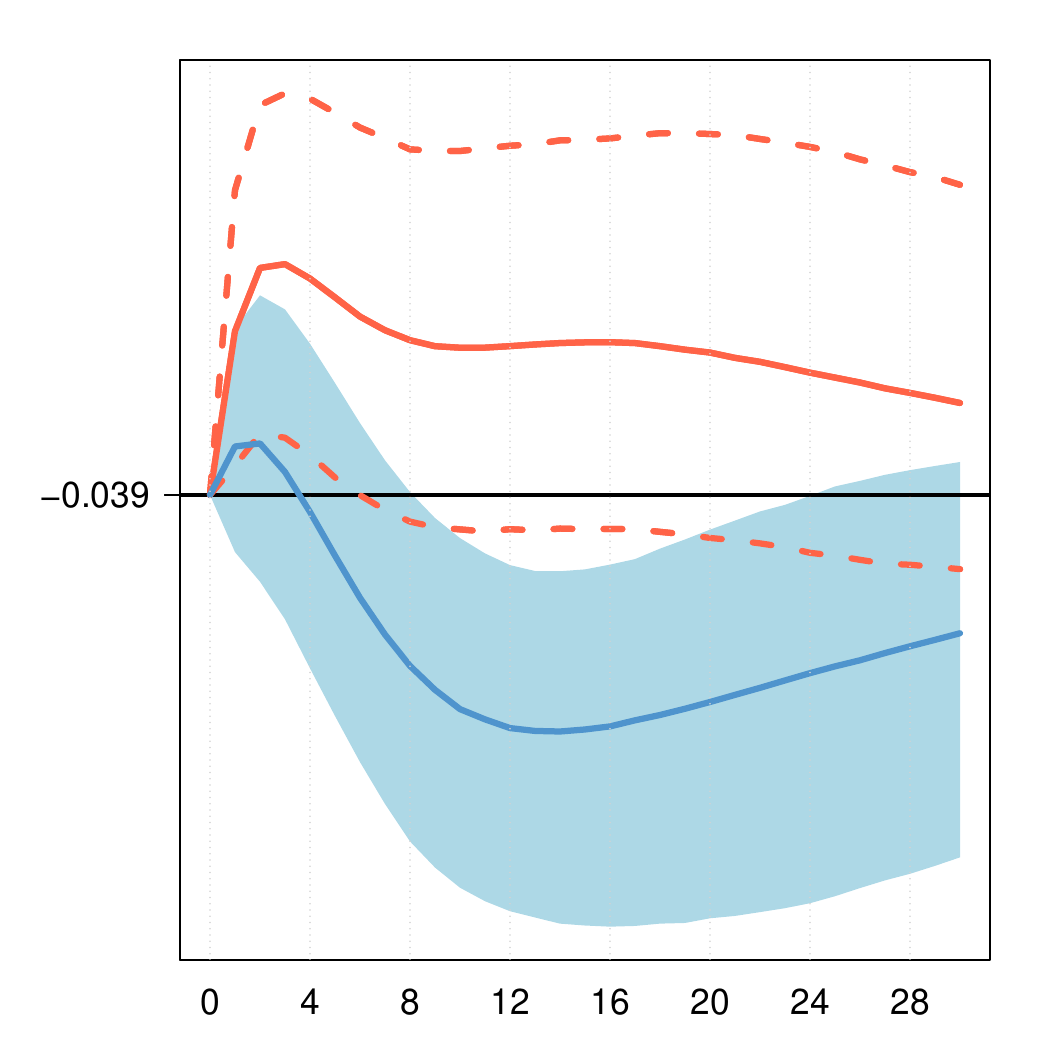}
\end{minipage}
\begin{minipage}[b]{.23\textwidth}
Real GDP
\centering\includegraphics[scale=0.20]{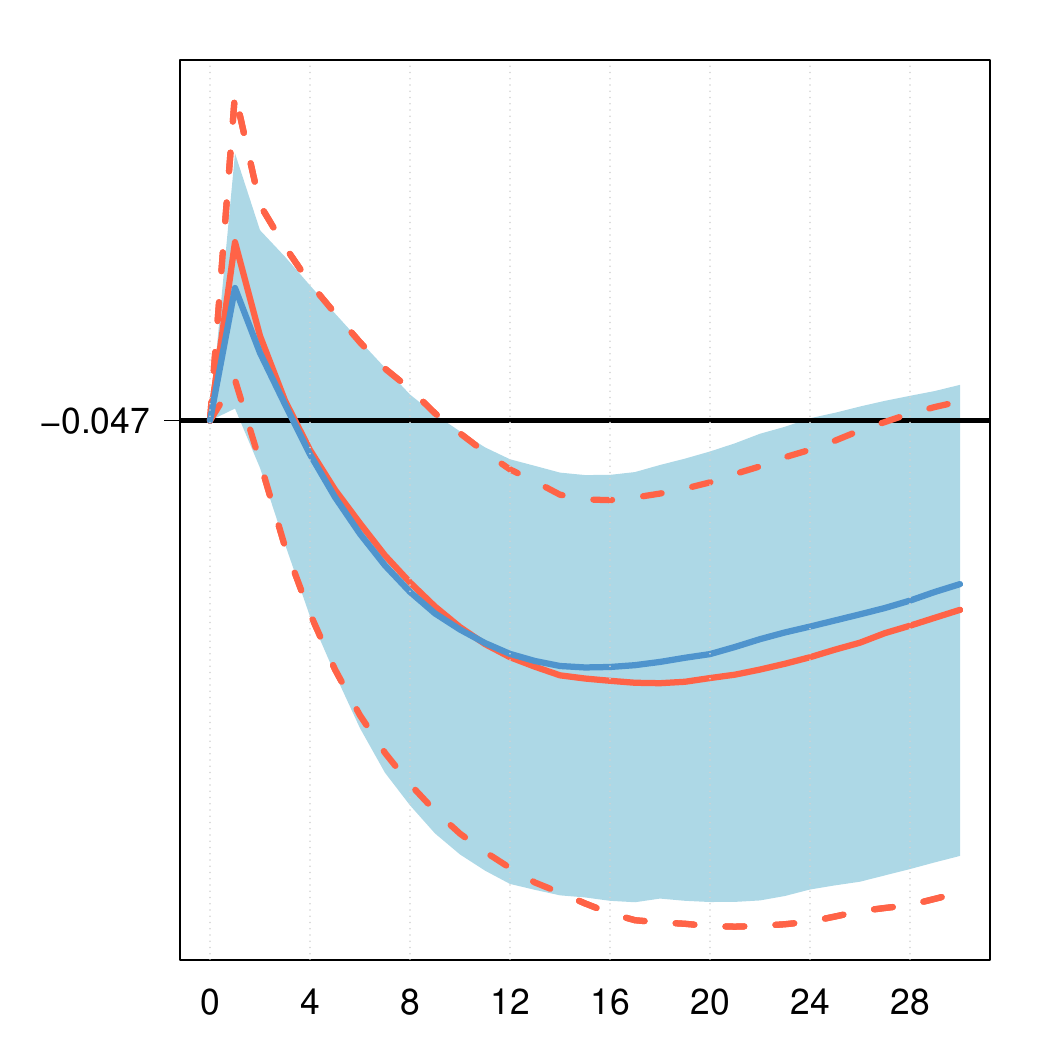}
\end{minipage}
\begin{minipage}[b]{.23\textwidth}
Infl. expect.
\centering\includegraphics[scale=0.20]{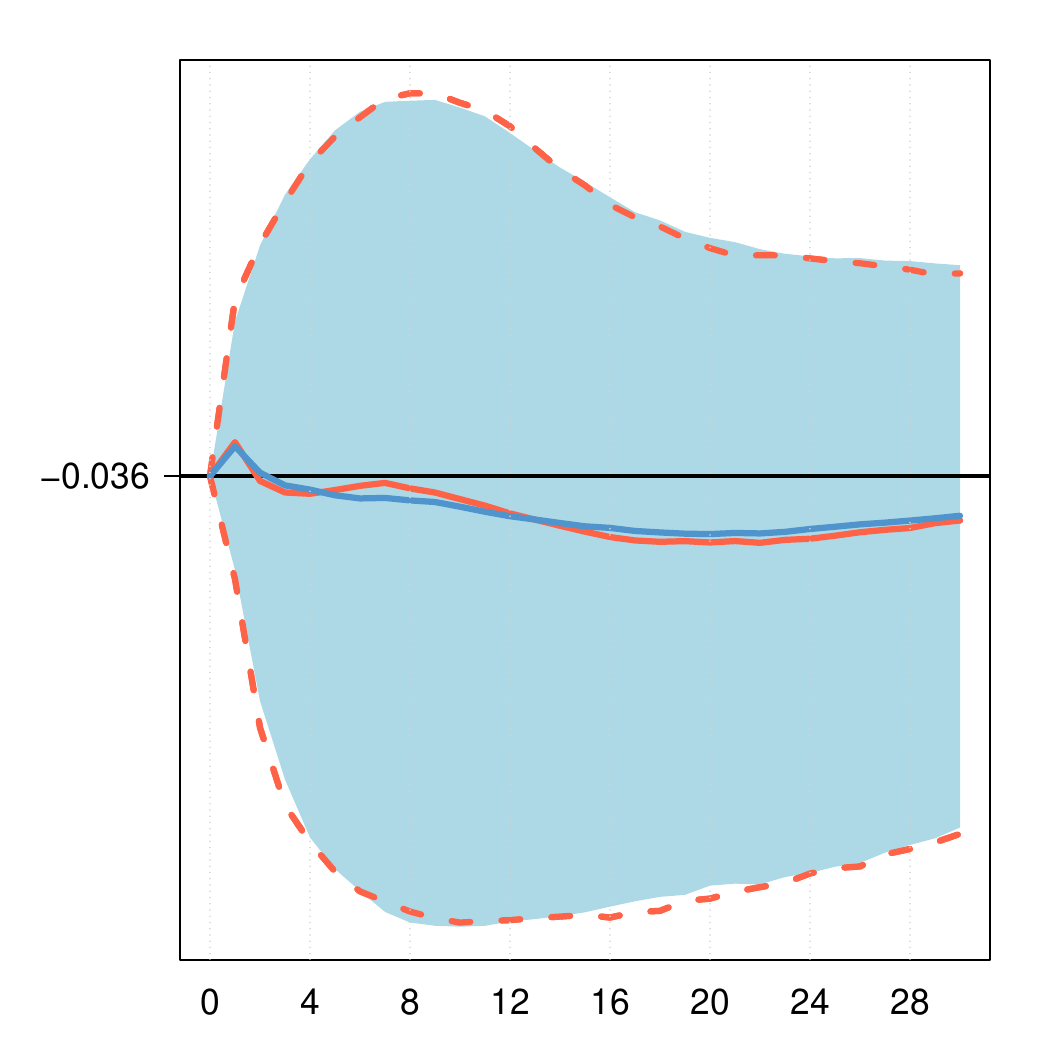}
\end{minipage}
\begin{minipage}[b]{.23\textwidth}
Unempl. rate
\centering\includegraphics[scale=0.20]{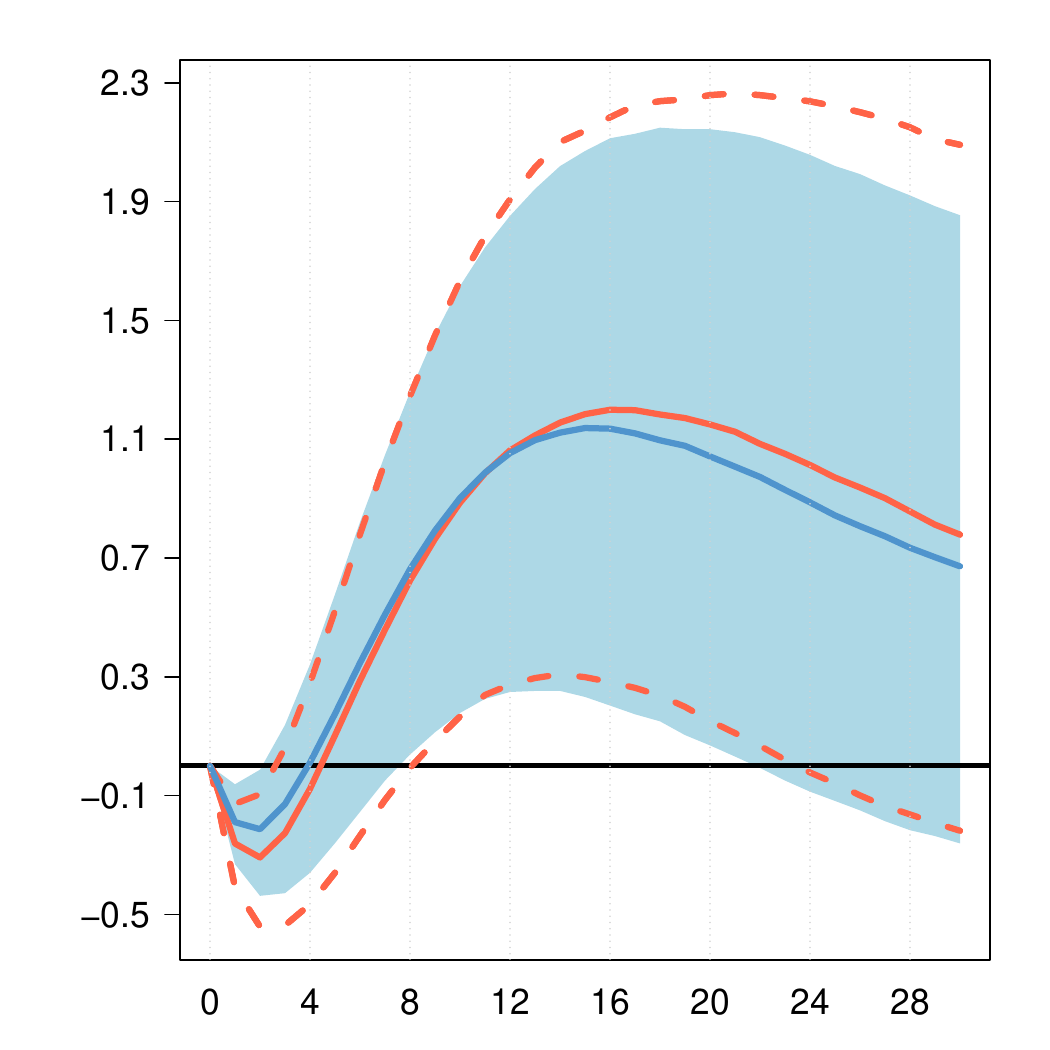}
\end{minipage}\\[.7em]
\begin{minipage}[b]{.23\textwidth}
2-year gov. bond yields
\centering\includegraphics[scale=0.20]{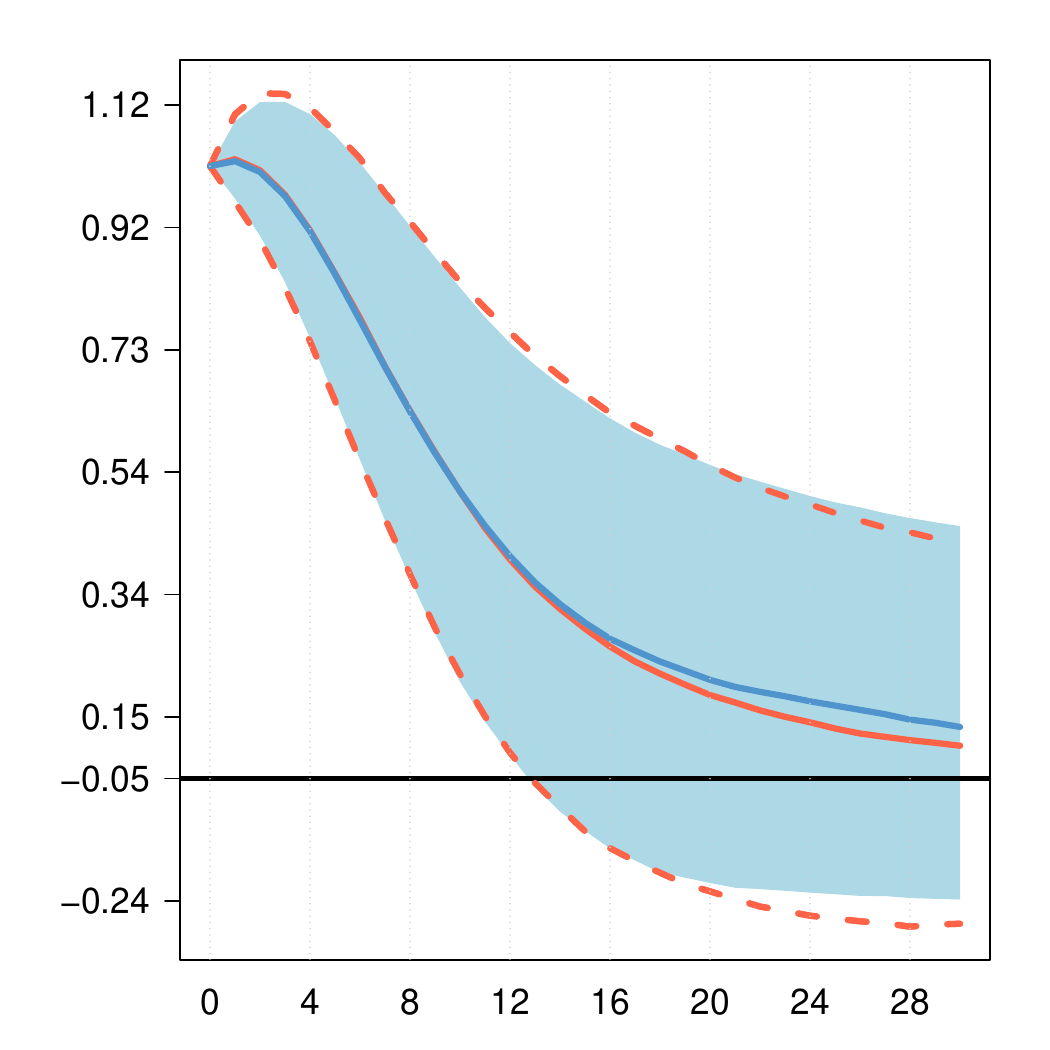}
\end{minipage}
\begin{minipage}[b]{.23\textwidth}
Long rates
\centering\includegraphics[scale=0.20]{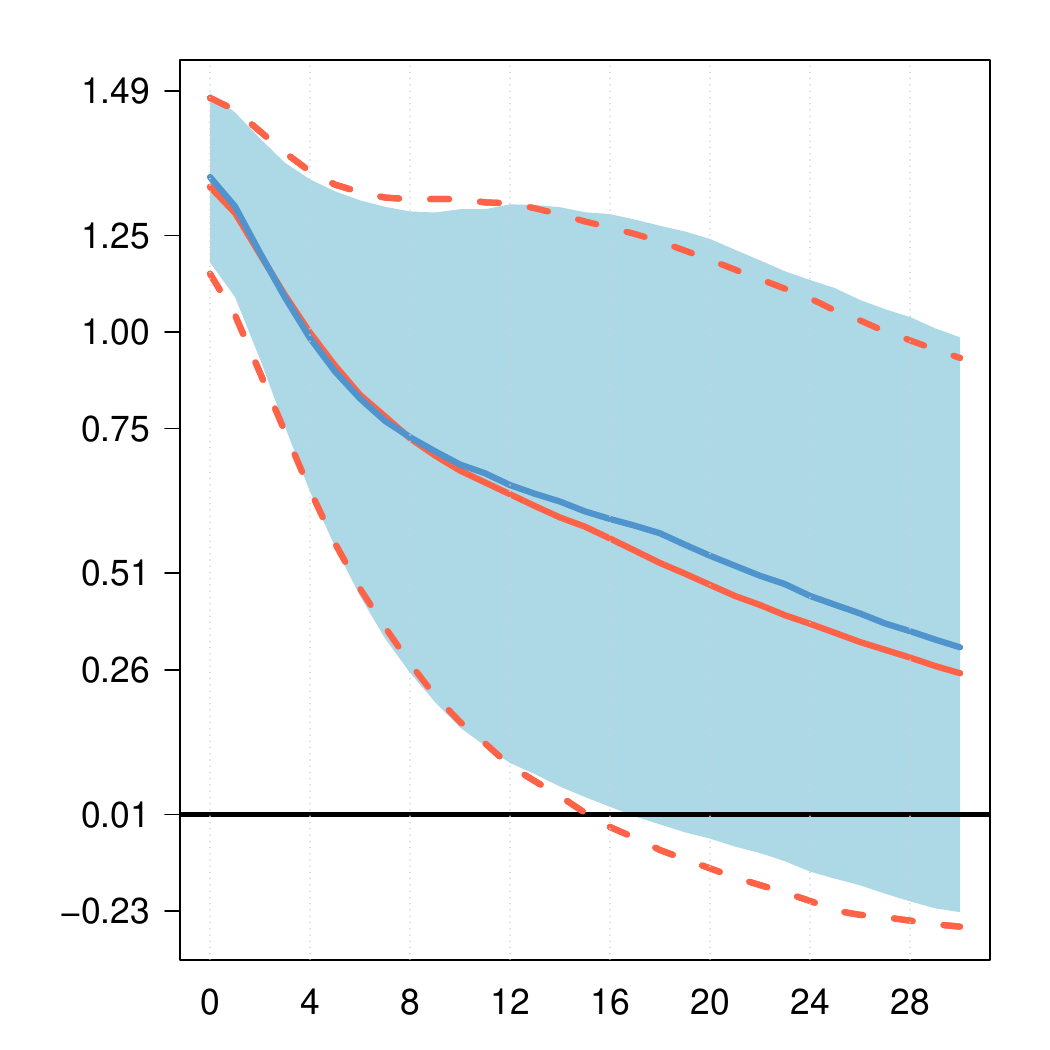}
\end{minipage}
\begin{minipage}[b]{.23\textwidth}
Real eff. exchange rate
\centering\includegraphics[scale=0.20]{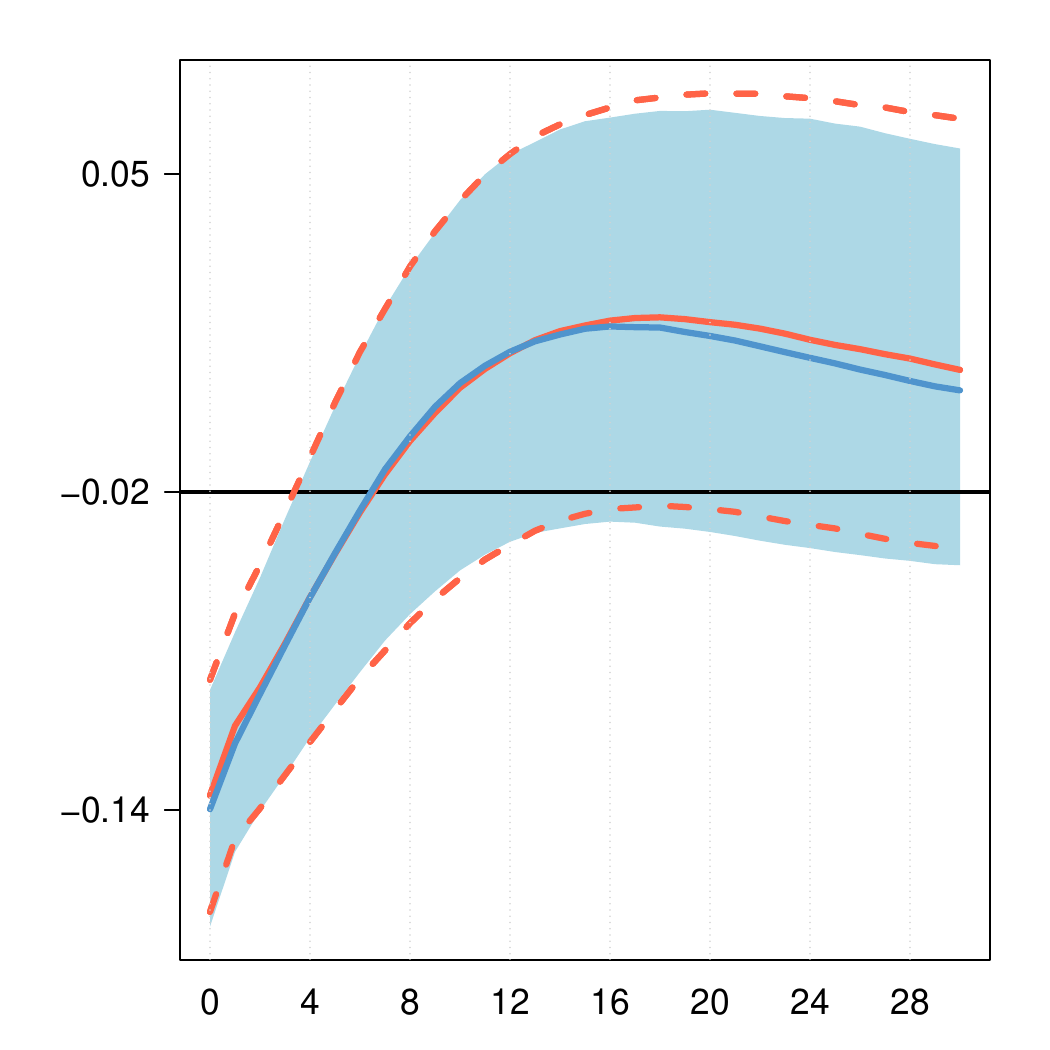}
\end{minipage}
\begin{minipage}[b]{.23\textwidth}
Equity prices
\centering\includegraphics[scale=0.20]{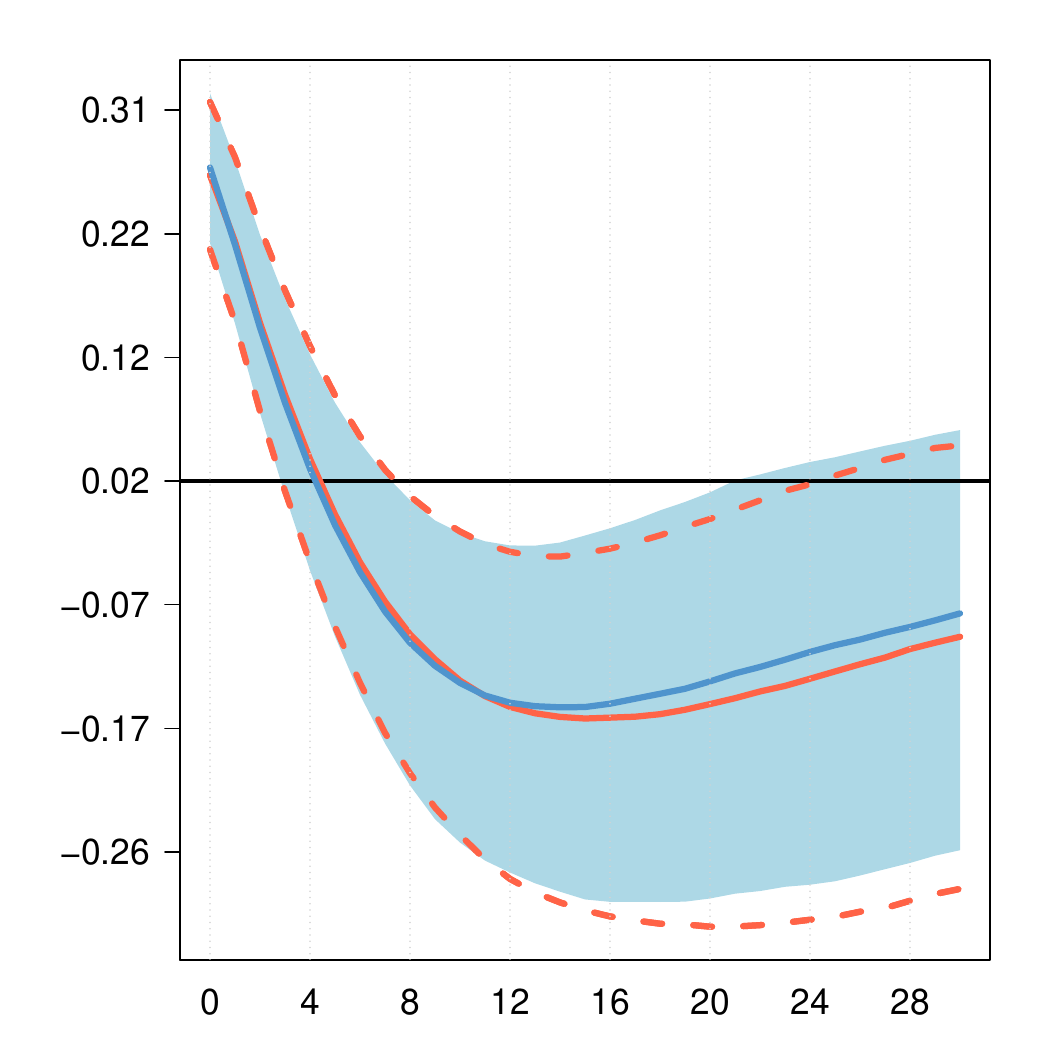}
\end{minipage}\\
\begin{minipage}{17cm}
\footnotesize \textit{Notes}:  The plot shows results of a robustness exercise. In the top panel, we show structural responses to a +100bp increase in the shadow rate identified through sign restrictions. In the bottom panel, we show responses to a +100bp increase in 2-year government bond yields using a Cholesky decomposition. 
For all plots, in blue, results are base on all households, in orange only for workers' households. Posterior median along  68\% credible sets. An increase in the real effective exchange rate corresponds to an appreciation. 
\end{minipage}%
\end{figure}

In this section, we carry out several robustness checks. We start with assessing the role of identification followed by the choice of the policy instrument. 
Our main results from section \ref{sec:emp} are based on a simple Cholesky decomposition using a policy instrument that covers monetary policy in a broad sense. As an alternative to recursive identification, several authors propose using sign restrictions to identify monetary policy shocks \citep[see e.g.,][]{Uhlig2005}. We assume that an unexpected increase in the shadow rate leads to a decrease in output, expected consumer price inflation and equity prices. The assumption on equity prices is based on empirical evidence for the reaction of stock markets to monetary policy-induced interest rate changes \citep{Thorbecke1997, Rigobon2004, Bernanke2005a, Bohl2007, Li2010}. Last, the unemployment rate is supposed to increase. These restrictions  imply that we rule out counterintuitive reactions of prices and output by construction. Naturally, we leave the variable of interest, the Gini, unrestricted. All restrictions are only binding on impact. The results are shown in \autoref{fig:rob}. We first note that credible intervals tend to be wider compared to the Cholesky-identified responses, which is a typical feature of responses identified through sign restrictions. More precisely, additional uncertainty arises through the need of sampling the rotation matrix necessary to implement sign restrictions. Qualitatively, we find the same effect on the Gini as in our baseline model: The Gini coefficient decreases in response to the monetary policy tightening if we consider the full sample of all households and increases for workers' households. In both cases, however, estimation uncertainty is considerable as the credible intervals contain zero throughout the response horizon.

\begin{figure}[p]
\caption{Impulse responses based on high frequency identification}\label{fig:ext}
\begin{minipage}{1\linewidth}~\\
\centering \textbf{Target shock}
\end{minipage}\\
\begin{minipage}[b]{0.23\linewidth}
Target
\centering \includegraphics[scale=0.20]{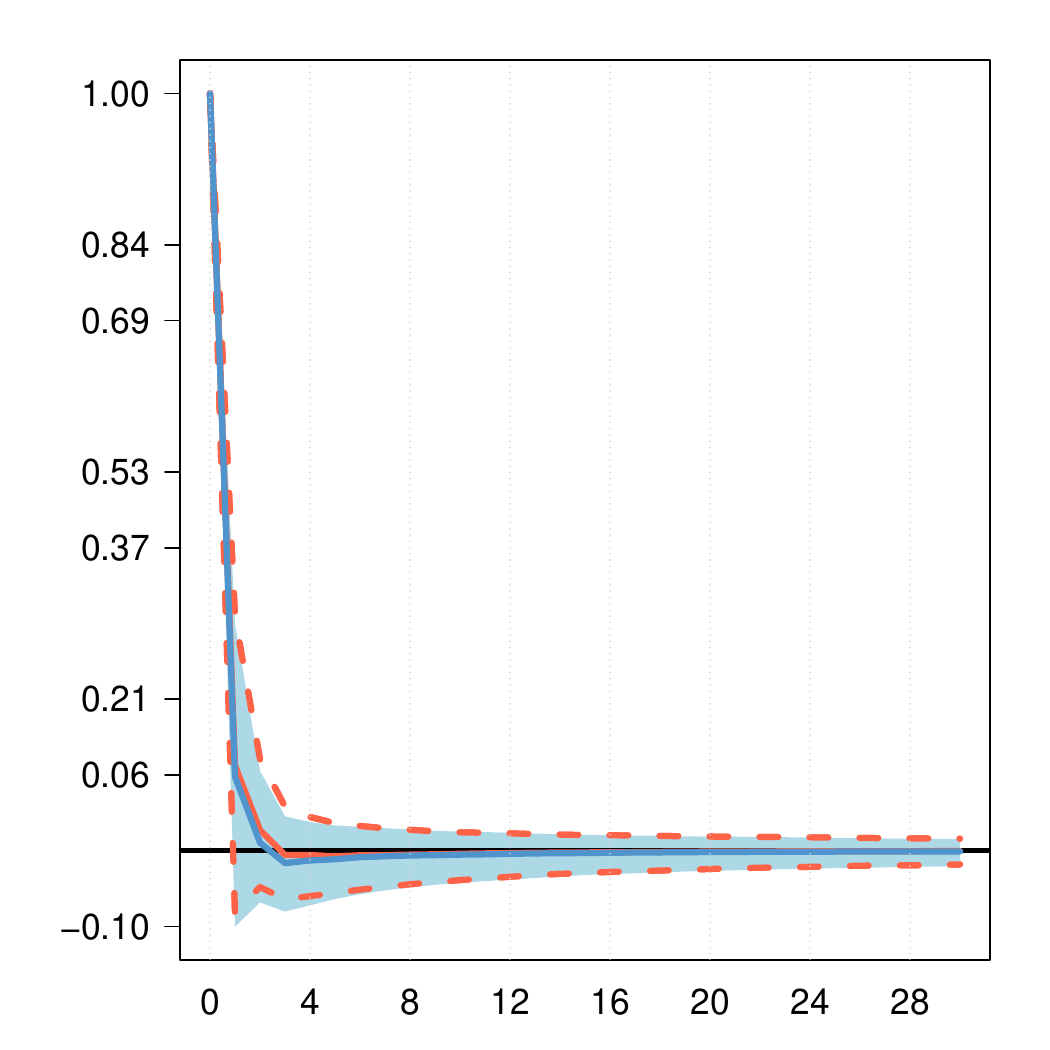}
\end{minipage}%
\begin{minipage}[b]{0.23\linewidth}
Gini
\centering\includegraphics[scale=0.20]{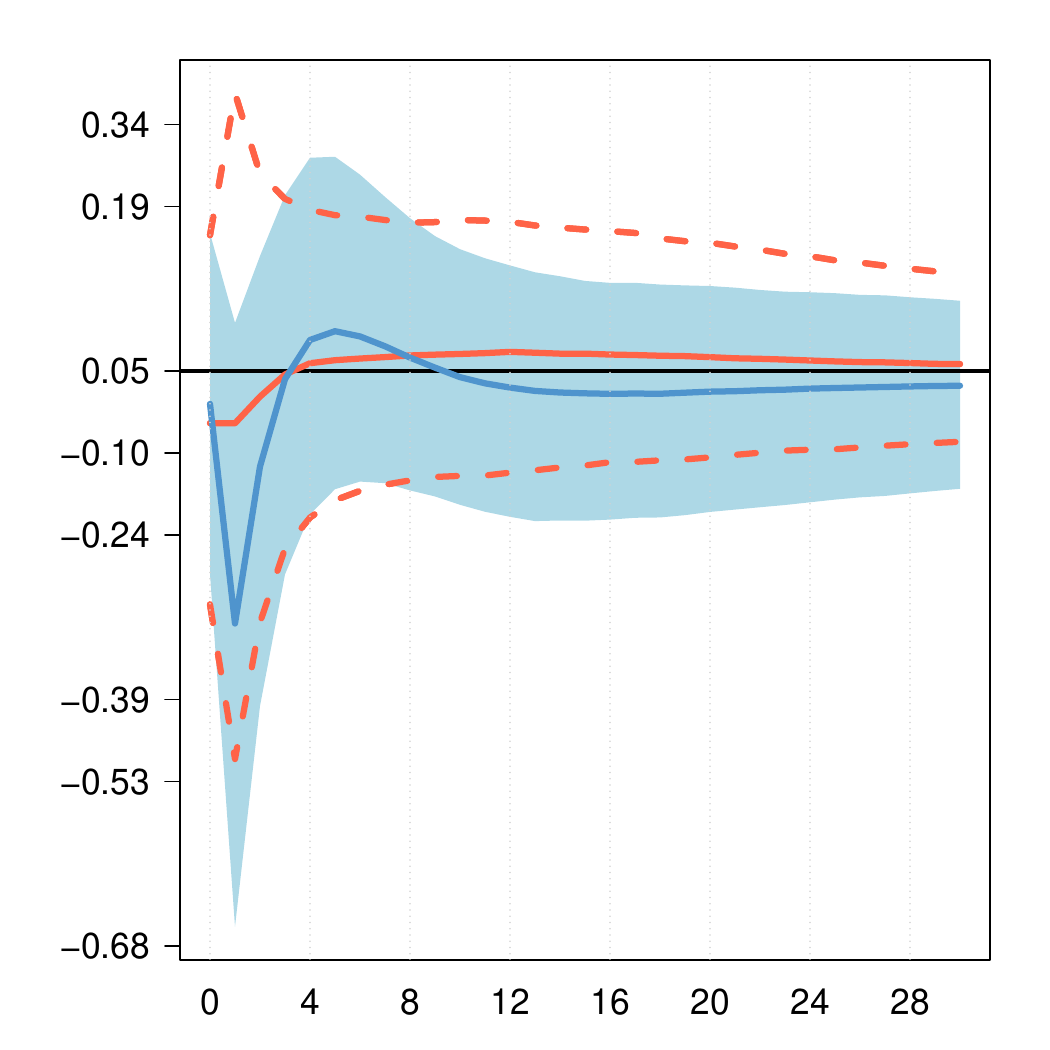}
\end{minipage}%
\begin{minipage}[b]{.23\textwidth}
Real GDP
\centering\includegraphics[scale=0.20]{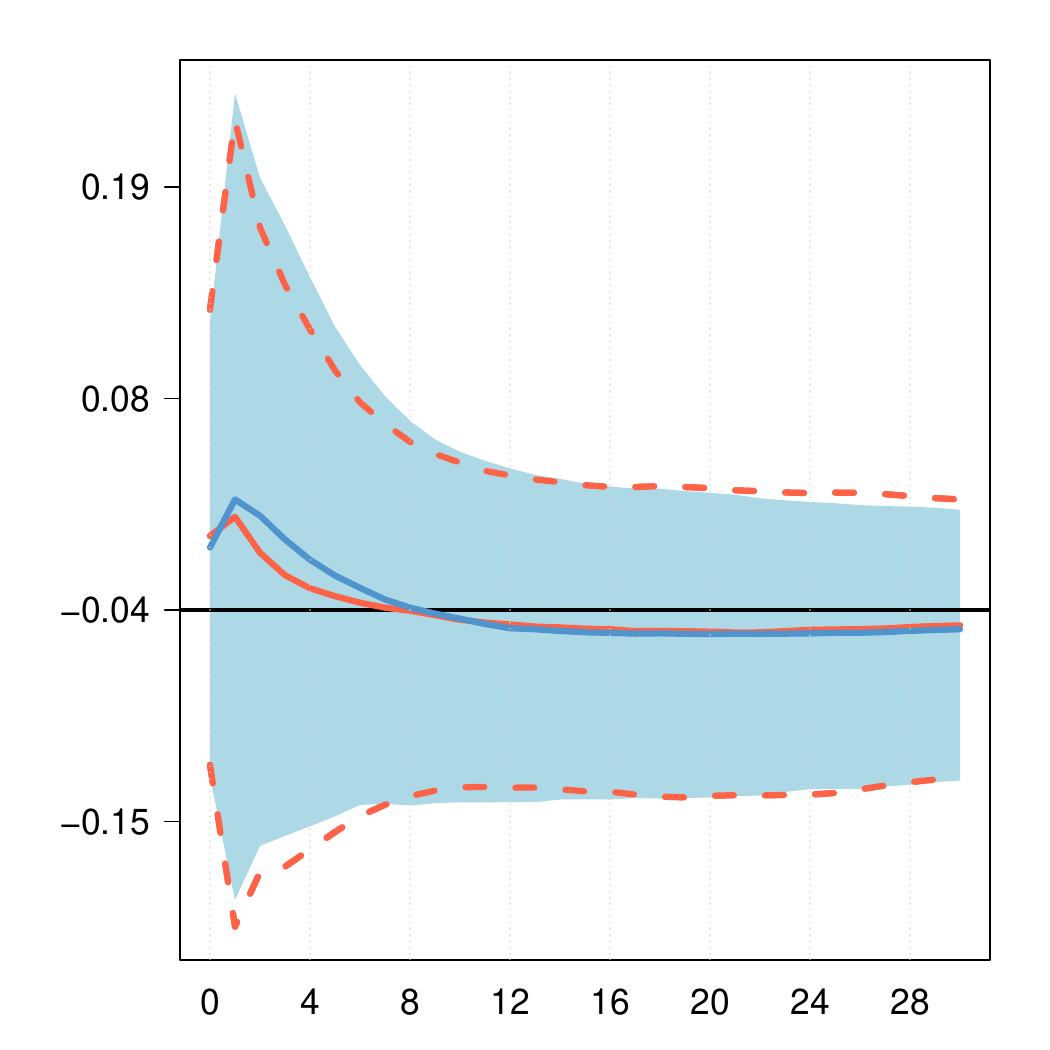}
\end{minipage}
\begin{minipage}[b]{.23\textwidth}
Infl. expect.
\centering\includegraphics[scale=0.20]{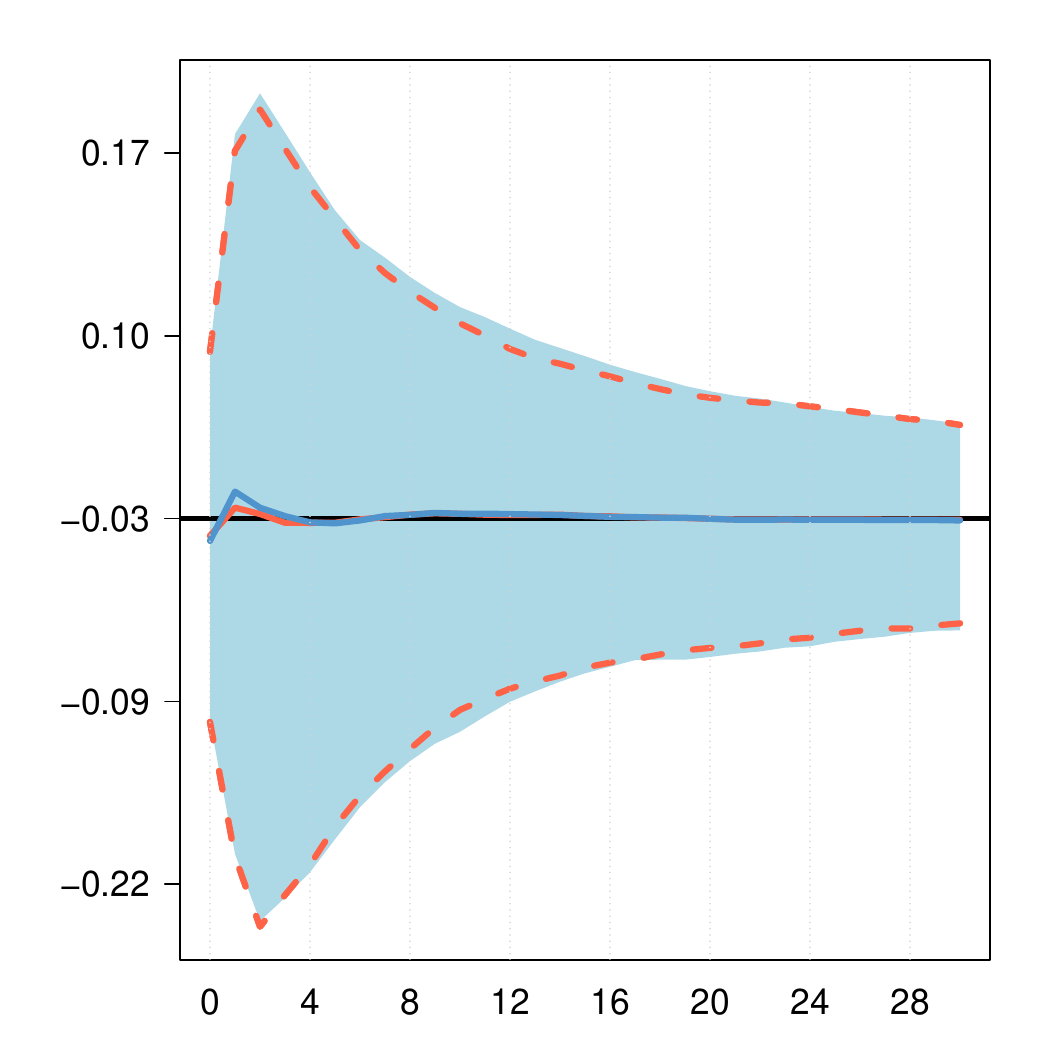}
\end{minipage}\\[.7em]
\begin{minipage}[b]{.23\textwidth}
Unempl. rate
\centering\includegraphics[scale=0.20]{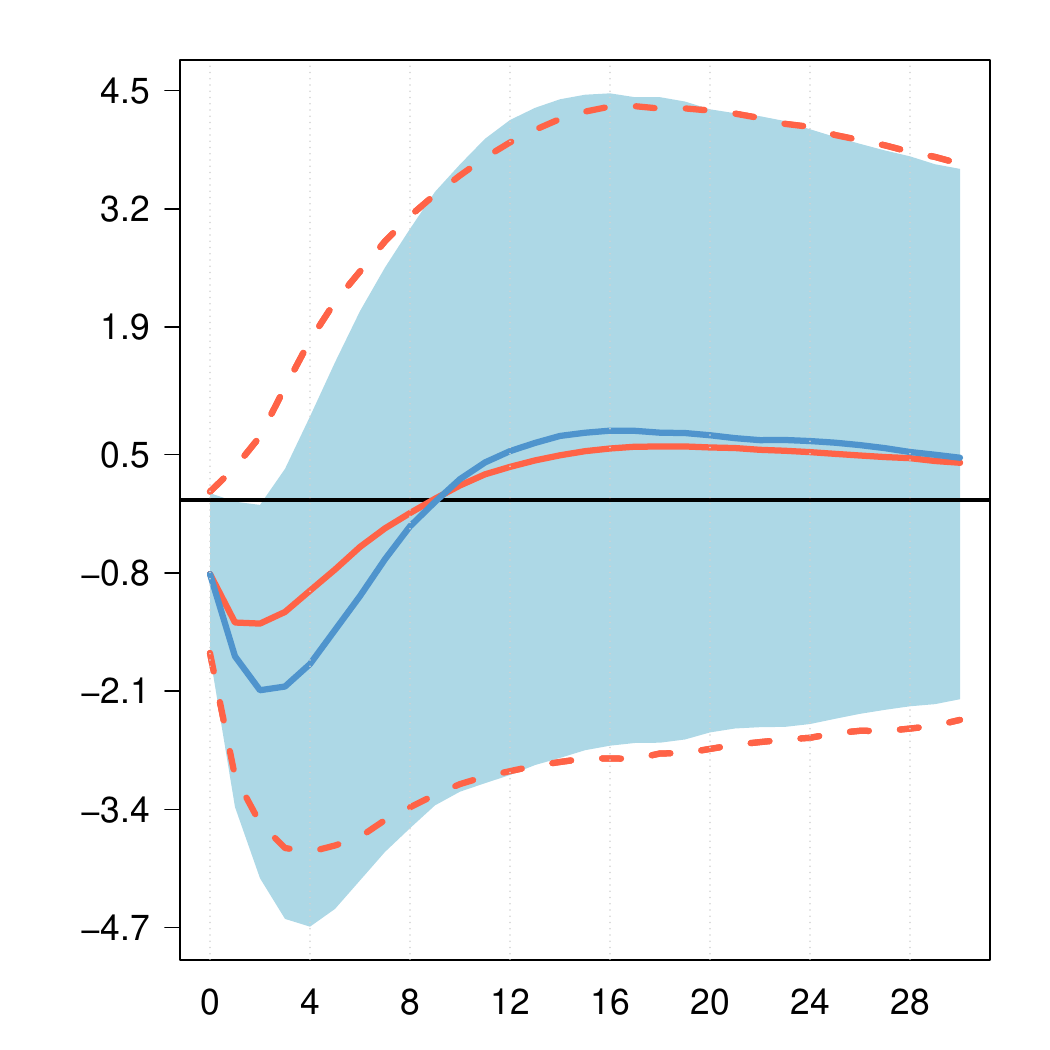}
\end{minipage}
\begin{minipage}[b]{.23\textwidth}
Long rates
\centering\includegraphics[scale=0.20]{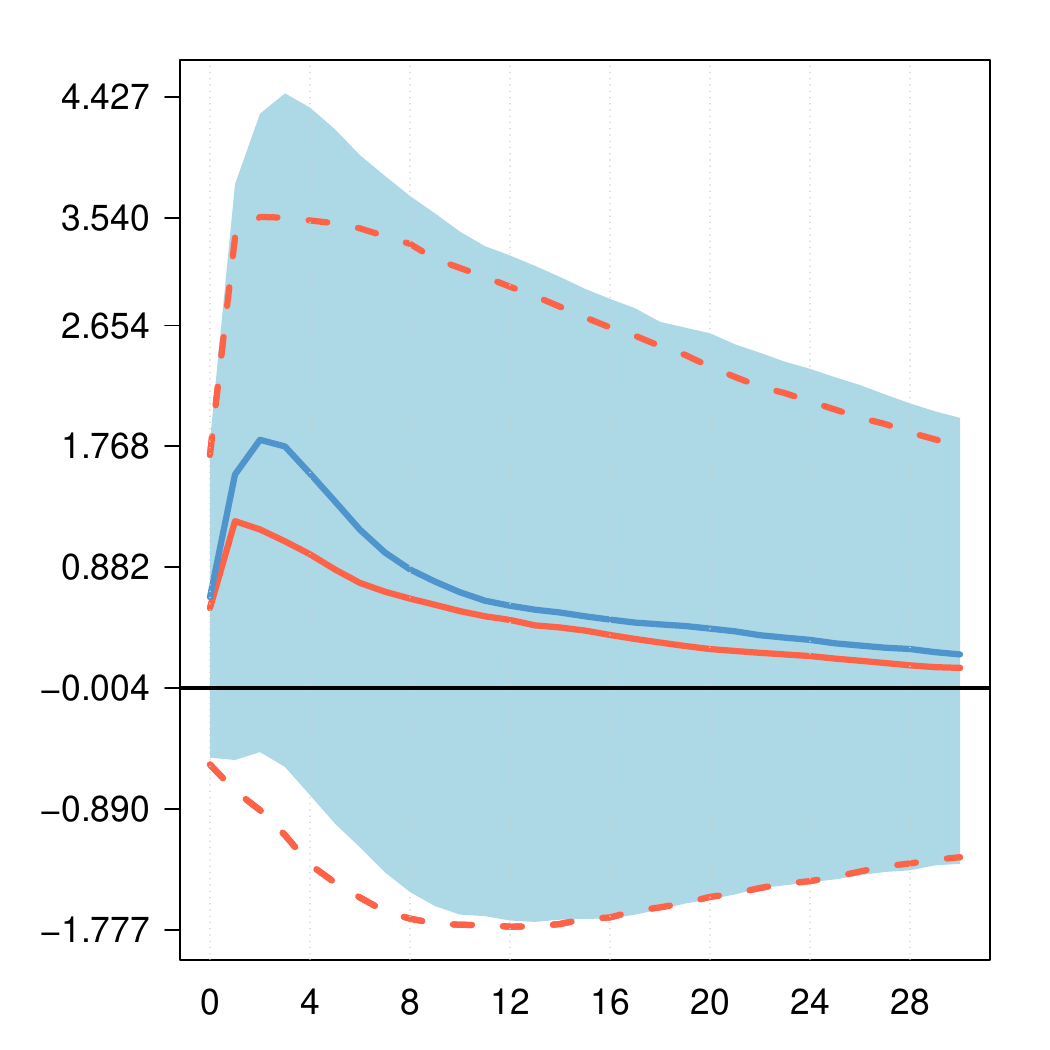}
\end{minipage}
\begin{minipage}[b]{.23\textwidth}
Real eff. exchange rate
\centering\includegraphics[scale=0.20]{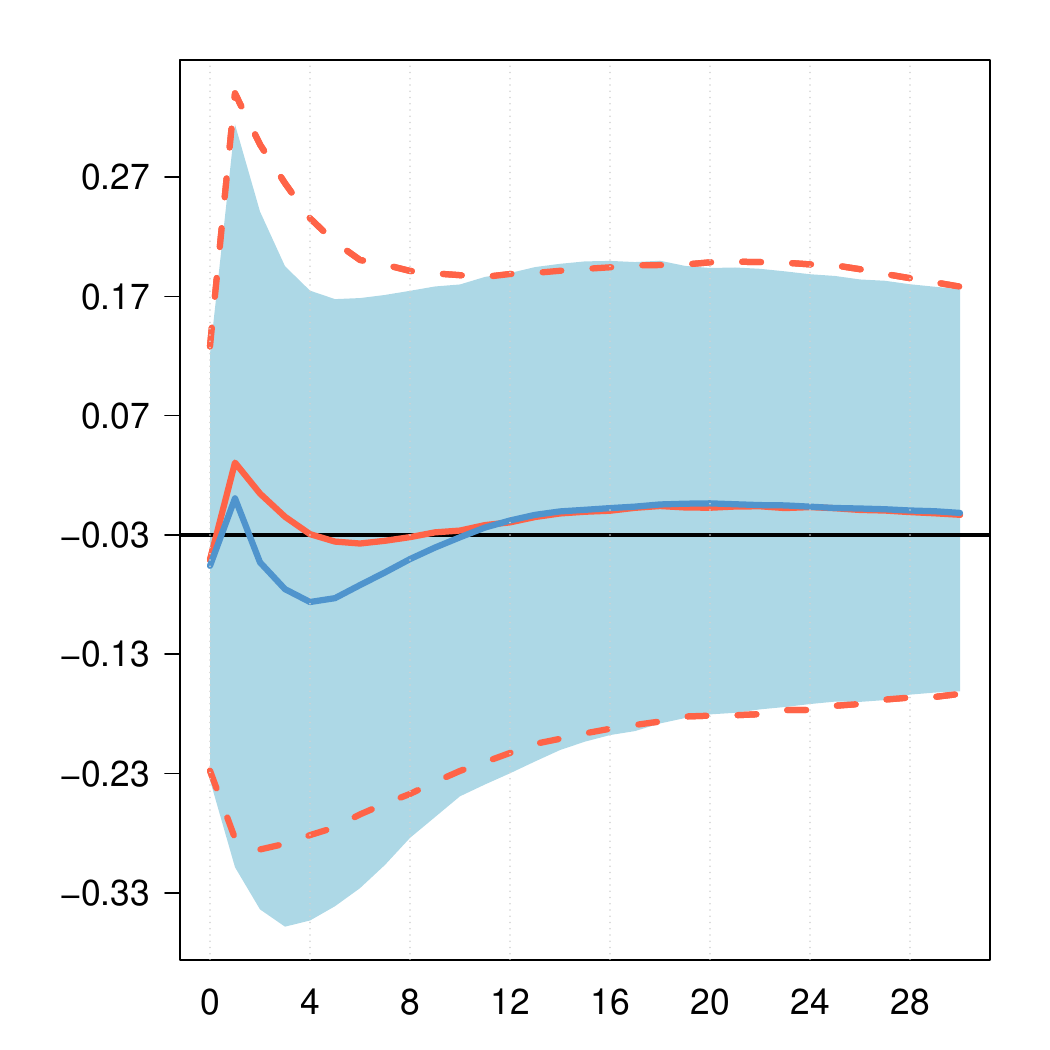}
\end{minipage}
\begin{minipage}[b]{.23\textwidth}
Equity prices
\centering\includegraphics[scale=0.20]{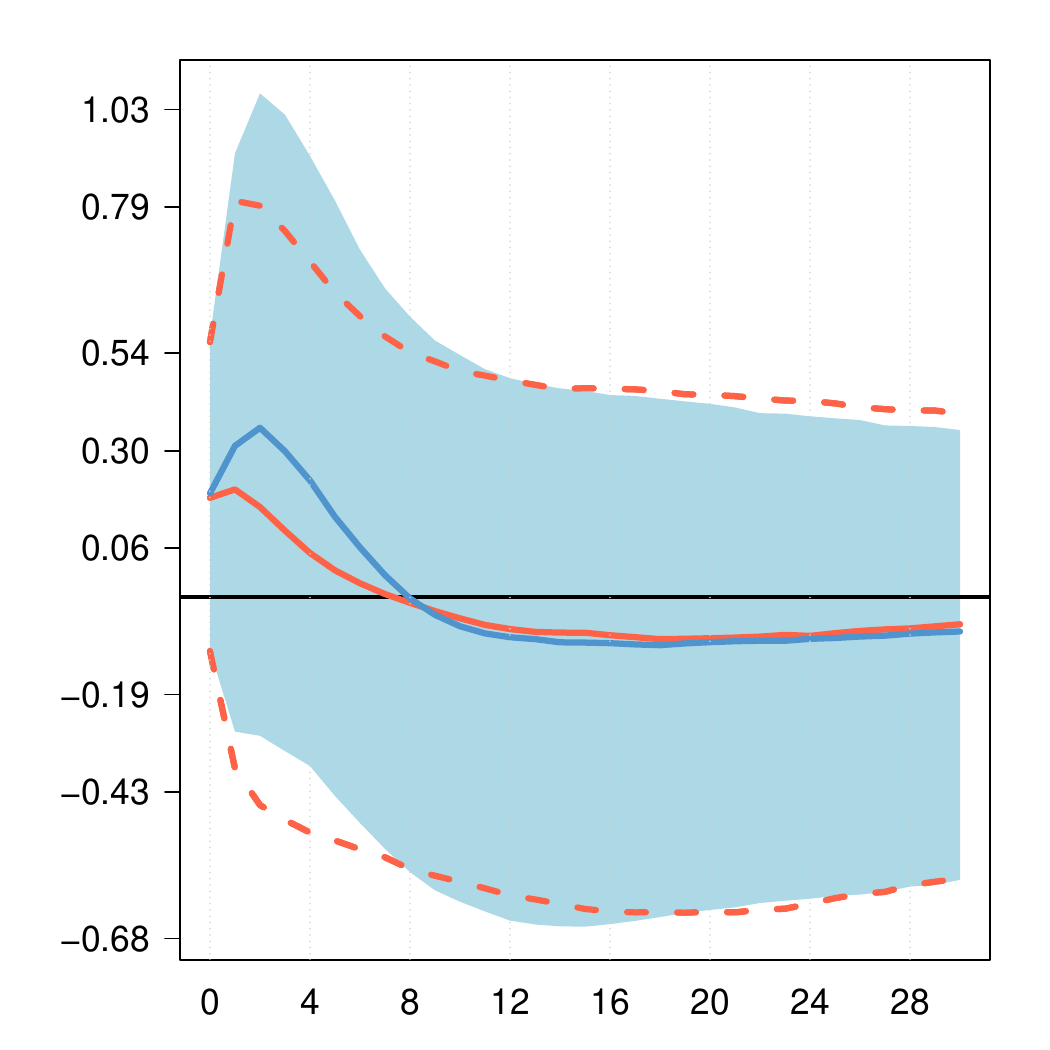}
\end{minipage}\\[.7em]
\begin{minipage}{1\linewidth}~\\
\centering \textbf{Path shock}
\end{minipage}\\
\begin{minipage}[b]{0.23\linewidth}
Path shock
\centering \includegraphics[scale=0.20]{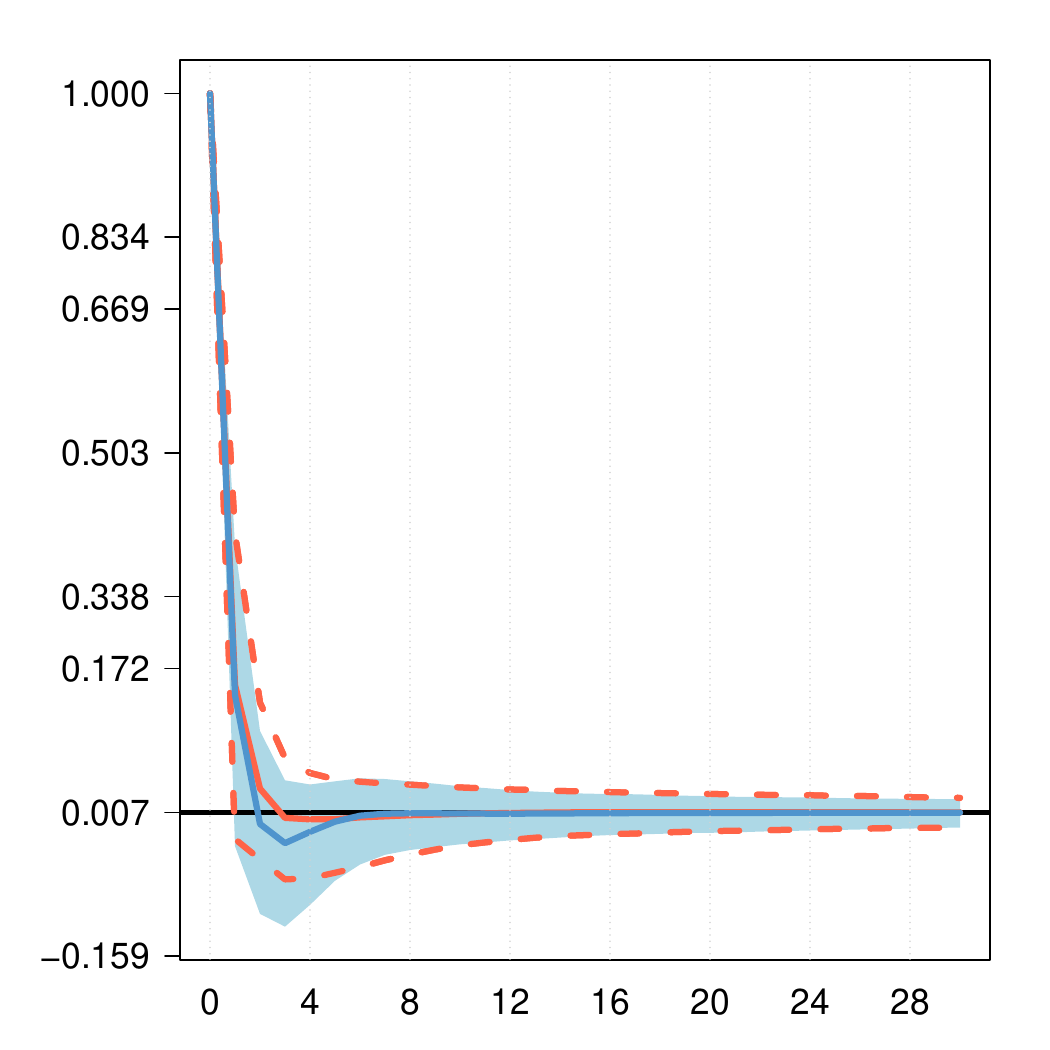}
\end{minipage}%
\begin{minipage}[b]{0.23\linewidth}
Gini
\centering\includegraphics[scale=0.20]{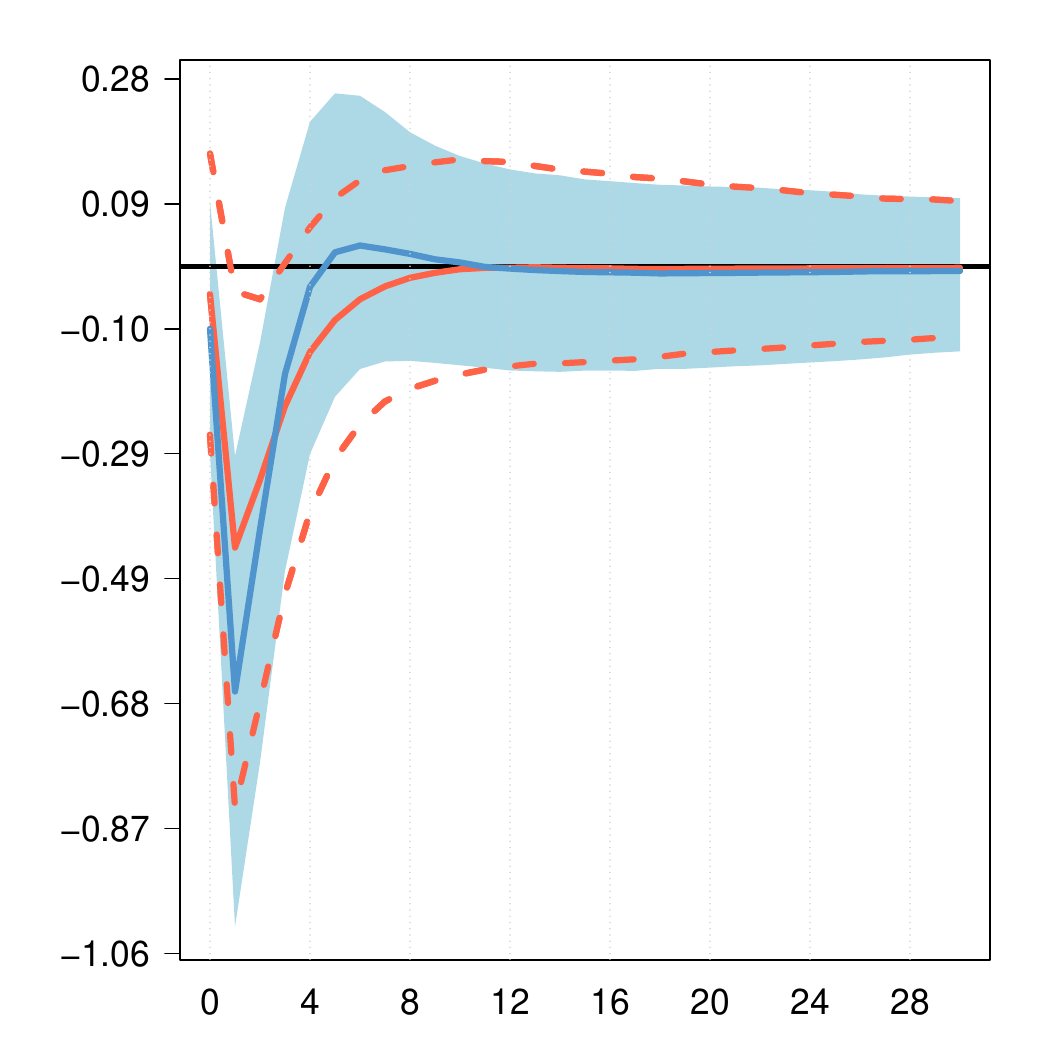}
\end{minipage}%
\begin{minipage}[b]{.23\textwidth}
Real GDP
\centering\includegraphics[scale=0.20]{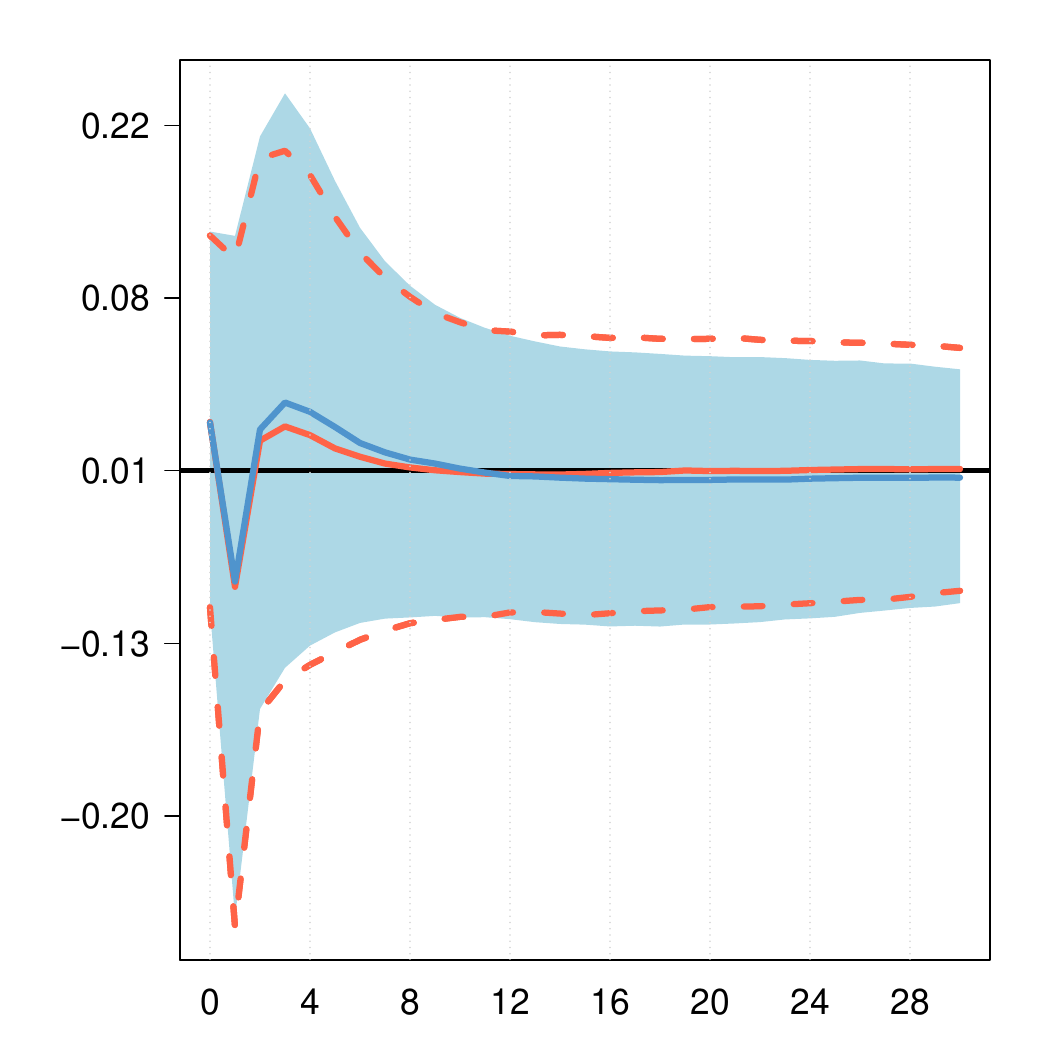}
\end{minipage}
\begin{minipage}[b]{.23\textwidth}
Infl. expect.
\centering\includegraphics[scale=0.20]{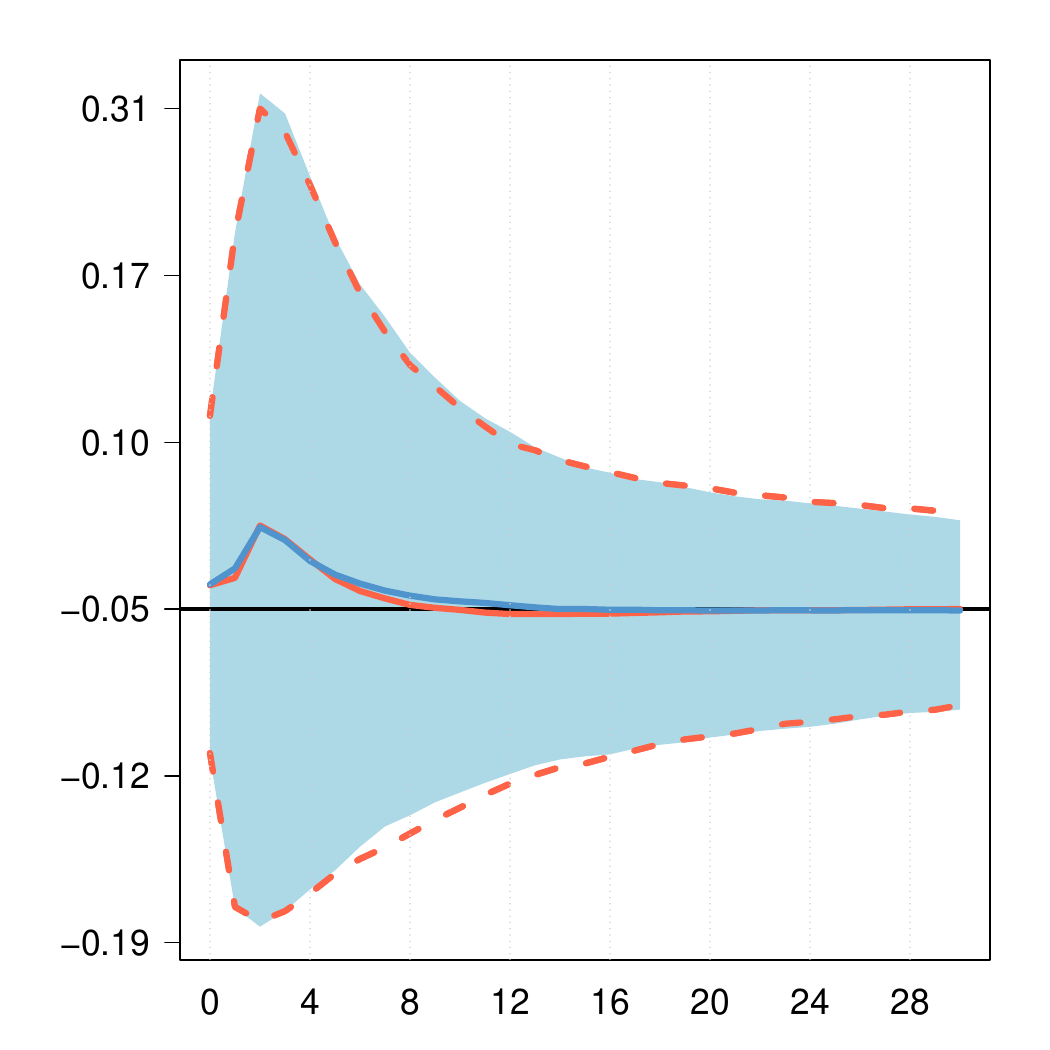}
\end{minipage}\\[.7em]
\begin{minipage}[b]{.23\textwidth}
Unempl. rate
\centering\includegraphics[scale=0.20]{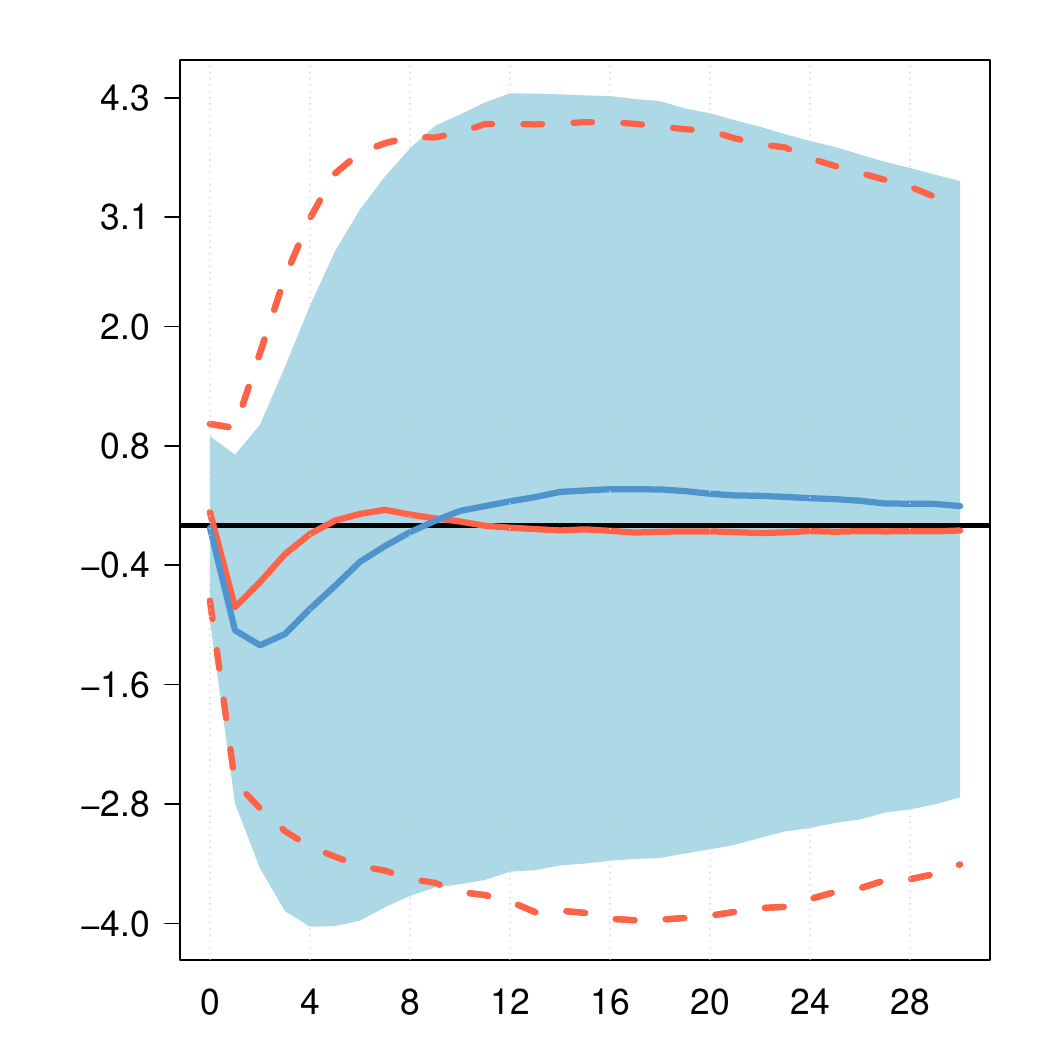}
\end{minipage}
\begin{minipage}[b]{.23\textwidth}
Long rates
\centering\includegraphics[scale=0.20]{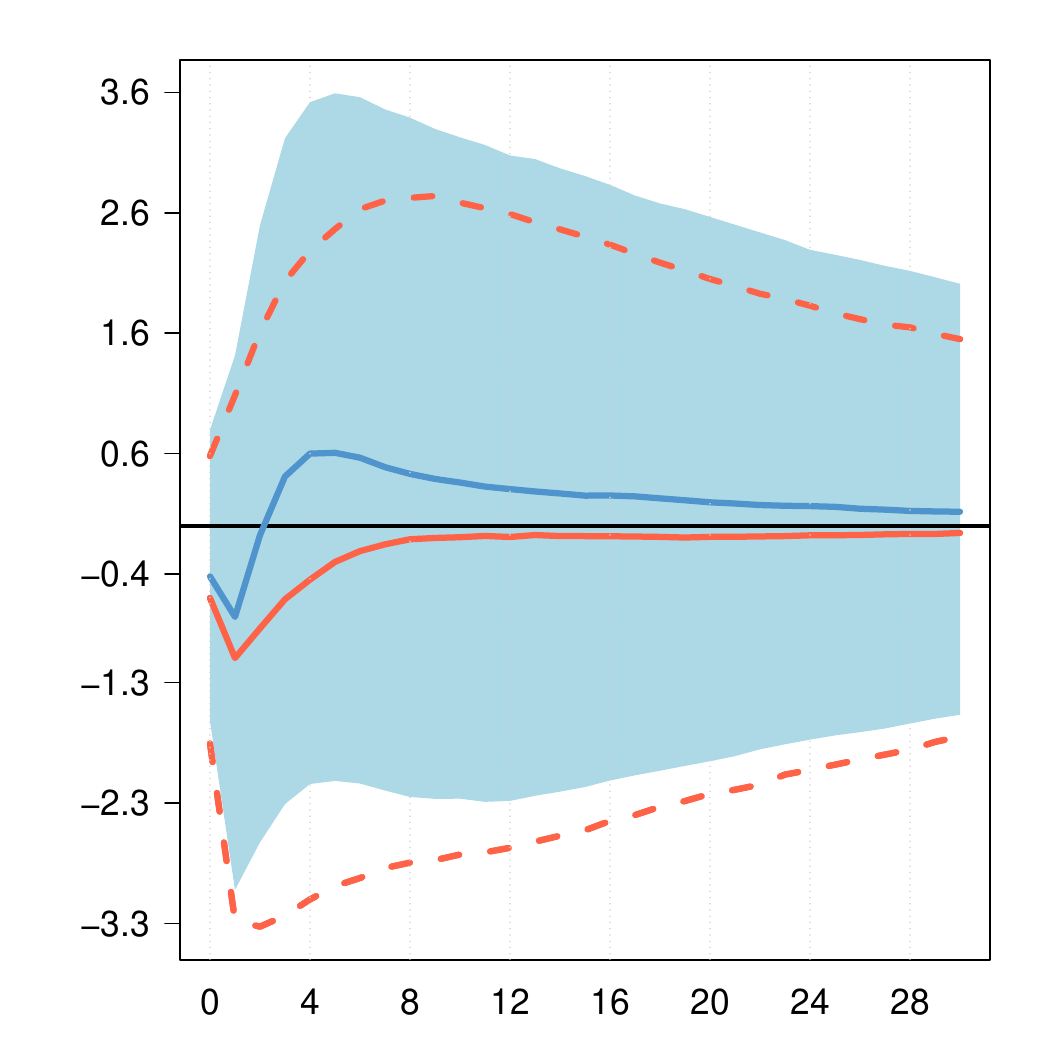}
\end{minipage}
\begin{minipage}[b]{.23\textwidth}
Real eff. exchange rate
\centering\includegraphics[scale=0.20]{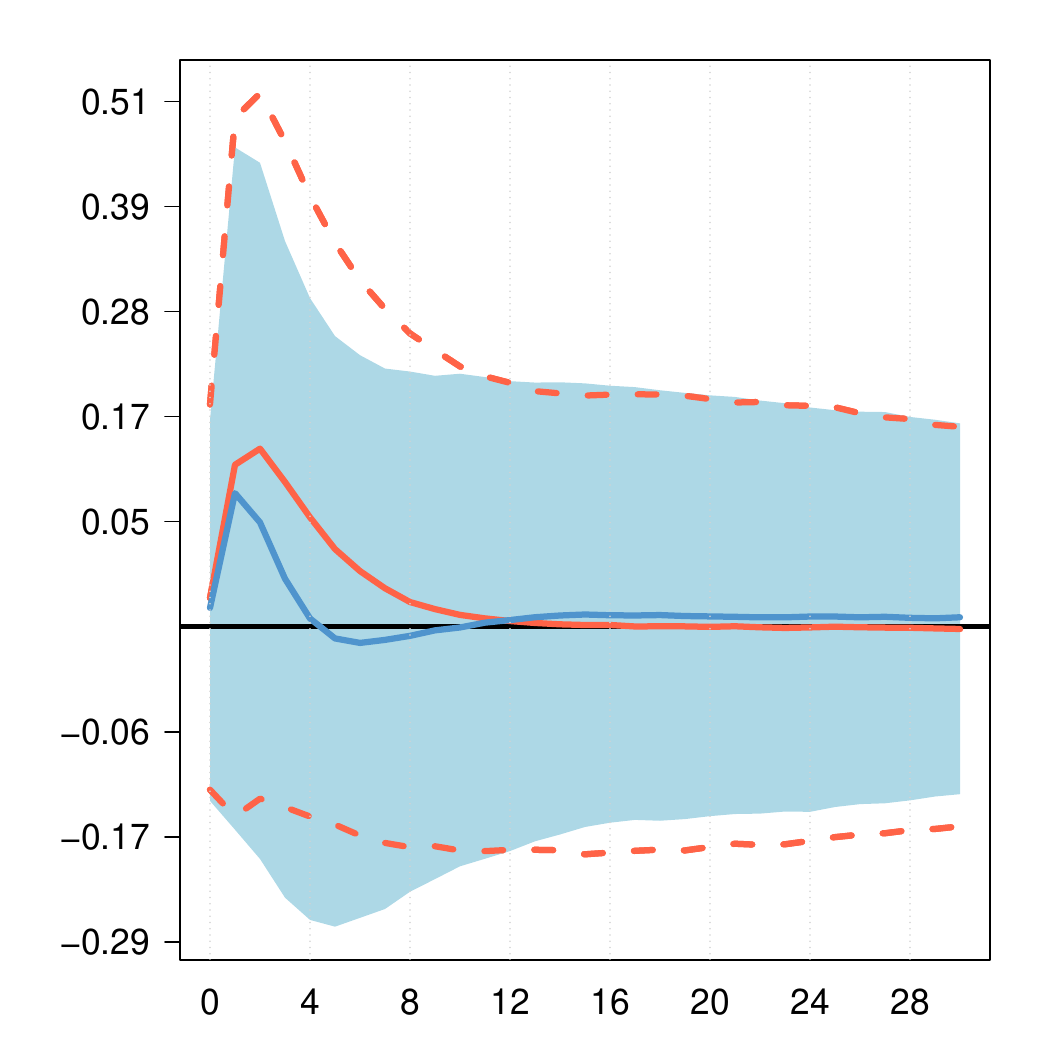}
\end{minipage}
\begin{minipage}[b]{.23\textwidth}
Equity prices
\centering\includegraphics[scale=0.20]{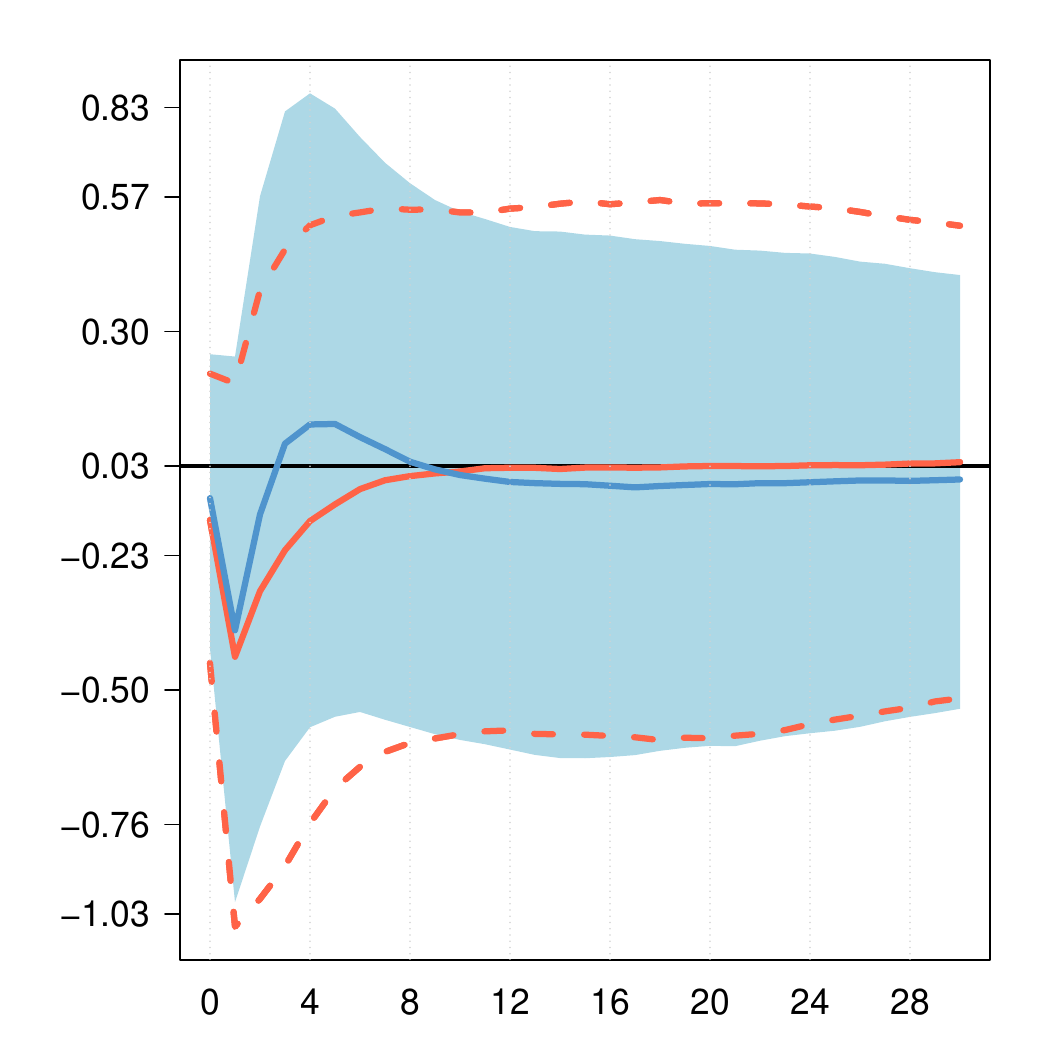}
\end{minipage}\\
\begin{minipage}{17cm}
\footnotesize \textit{Notes}: The plot shows results of a robustness exercise. In the top panel, we show structural responses to a one standard deviation increase in the target shock, in the bottom panel corresponding responses to a path shock. The derivation of these shocks is described in \citet{Kubota2020}. For all plots, in blue, results are base on all households, in orange only for workers' households. Posterior median along  68\% credible sets. An increase in the real effective exchange rate corresponds to an appreciation. 
\end{minipage}%
\end{figure}

Next, we assess how our results are affected when considering identification via external instruments \citep{Gertler2015, Altavilla2019}. The idea is to exploit high-frequency variation in interest rates during a tight window around policy announcements by the central bank. For Japan, \citet{Kubota2020} use changes in future contracts for the Tokyo interbank offered rate (TIBOR) at different maturities to construct two external monetary policy measures: a target factor which loads most strongly on the short-end of the yield curve and a path factor which mostly affects longer term yields. The target factor is thus reminiscent of conventional interest rate policy while the path factor covers quantitative easing programs. We estimate our model for both policy measures, where we have ordered the respective external instrument first and then used a Cholesky decomposition \citep{Jarocinski2020}. The results are depicted in \autoref{fig:ext}. If we examine results of the target shock first (top panel), we see that the Gini reacts negatively to a monetary policy tightening. Effects on the Gini, and in general also the other effects tend not to be precisely estimated, though. This could mirror the fact that the BoJ has rarely used conventional monetary policy over the considered sample period. Next, we consider effects to a path shock, depicted in the bottom panel of \autoref{fig:ext}. In line with the target shock, responses of macroeconomic variables are estimated with wide credible intervals. However, the results show a negative and precisely estimated effect on the Gini coefficient. This holds true for both samples, all households and workers' households.

Summing up, our main result namely that income inequality decreases when considering all households is robust to different identification schemes. The effect is less clear when considering workers' households. This implies that the occupation of the household plays an important role when assessing the effects of monetary policy on income inequality. High-frequency identification suggests that most of the effect on the Gini comes from monetary policy that targets the long-end of the yield curve, i.e., quantitative easing policies. This suggests that the shadow rate seems to pick up these policies rather than conventional changes in the policy rate.

\section{Conclusions}\label{sec:concl}

In this paper, we examine the relationship between monetary policy and income inequality in Japan. For that purpose, we propose a new framework that models jointly the volatility of the income distribution and the dynamics of key macroeconomic variables. Our model draws on relative frequencies of grouped income data which is continuously available over a longer period than comparable micro- / survey based individual income data, which makes it suitable to analyze with time series methods. 

The empirical literature assessing monetary policy effects on income inequality has revealed a large degree of cross-country heterogeneity. This is the case since the theoretical effect is ambiguous and depends on the relative strength of transmission channels as well as endowments and demographics of households -- all which might differ significantly across countries. Our analysis reveals that a monetary tightening in Japan leads to a decrease in output an increase in unemployment and an appreciation of the real effective exchange rate. We do not find any significant effects on (expected) inflation. As regards financial variables, we find that long-term rates respond positively to the tightening, whereas equity prices decrease significantly over the medium-term. 

We then examine the effect o tightened monetary policy on income inequality as measured by changes in an estimated Gini coefficient. Our main finding is that a monetary tightening leads to a significant decrease in inequality if we consider a broad representation of income data that also includes the unemployed and retired population. More unemployment,  potential lay-offs of exporting firms triggered by a real appreciation and tighter financing conditions tend to mostly affect the employed. In the broad household sample, this leads to a more equal income distribution. This finding is reversed, if we consider the sub-sample of households whose head is employed. Here tighter monetary policy leads to more inequality since the same effects now apply to the poorer population of the sample. This result is in line with recent findings of \citet{Coibion2017} for the USA, \citet{Mumtaz2017} for the UK and \citet{Furceri2017} for a broad set of countries. Using a counterfactual analysis reveals changes in long-term interest rates and -- to a lesser degree -- the unemployment rate as main drivers of inequality caused by monetary policy. Especially so for workers' households, long-term rates which resemble overall financing conditions are crucial in transmitting monetary policy. 

%Having established these findings we then show that not only the type of underlying data matters for the assessment of monetary policy effects on income inequality but also the estimation framework. Our proposed framework treats estimation of the Gini coefficient as an inequality measure and the dynamics of the macroeconomic system jointly. In contrast most studies on inequality derive the Gini coefficient in a first step and treat it as if observed in the subsequent econometric analysis. That this can lead to flawed inference has been pointed out by \citet{Carriero2018} in the context of macroeconomic uncertainty. Comparing our results with those based on the predominant two-step approach indeed leads to striking differences. While credible sets of the jointly estimated responses are wider since they also reflect estimation uncertainty related to the Gini coefficient, they point to significant effects of monetary policy on the distribution of income. Two-step inference, by contrast yields inconclusive results. 

\section*{Compliance with Ethical Standards}
\begin{description}
\item[Funding:] Kakamu acknowledges the financial support by KAKENHI (\#16KK0081, \#16K03592, \#25245035, \#20H00080 and \#20K01590).
\item[Conflict of Interest:] Author Feldkircher declares that he has no conflict of interest. Author Kakamu declares that he has no conflict of interest.
\item[Ethical approval:] This article does not contain any studies with human participants or animals performed by any of the authors.
\end{description}

%**********************************************************************************************
\normalsize
\singlespacing
\bibliographystyle{./bibtex/jae}
\bibliography{./bibtex/inequality}
\addcontentsline{toc}{section}{References}
%**********************************************************************************************
\newpage

\begin{appendices}
\section{Posterior Analysis}\label{sec:PA}
\subsection{Sampling $\mu_{t}$ for $t = 1, 2, \ldots, T$}
The full conditional distribution (FCD) of $\mu_{t}$ for $t = 1, 2, \ldots, T$ is given by
\begin{eqnarray*}
	\mu_{t}|h_{t}, \vx_{t} \sim \N\left( \hat{\mu}_{t}, \hat{\tau}_{t}^{2} \right),
\end{eqnarray*}
where $\displaystyle \hat{\tau}_{t}^{2} = (\exp(-h_{t}) \vone_{k}^{\prime}{\mW_{t}^{*}}^{-1} \vone_{k} + \tau_{0}^{-2})^{-1}$, $\displaystyle \hat{\mu}_{t} = \hat{\tau}_{t}^{2} \left\{ \exp(-h_{t}) \vone_{k}^{\prime}{\mW_{t}^{*}}^{-1} \left( \ln \vx_{t} - \exp(h_{t} / 2) \vu_{t} \right) + \tau_{0}^{-2} \mu_{0} \right\}$, and $\vone_{k}$ is a $k \times 1$ unit vector.

\subsection{Sampling $h_{t}$ for $t = 1, 2, \ldots, T$}
\begin{eqnarray*}
	\pi(h_{t} | \mu_{t}, \valpha, \mB, \mSigma, \vh_{-t}, \mY^{*},\vx_{t} ) \propto \exp(-h_{t} / 2) \exp\left( -\frac{\vv_{t}^{\prime} \mW_{t}^{*-1} \vv_{t}}{2 \exp(h_{t})} \right) f(h_{t}),
\end{eqnarray*}
where $\vh_{-t} = (h_{1}, h_{2}, \ldots, h_{t-1}, h_{t+1}, \ldots, h_{T})$ and
\begin{eqnarray*}
	f(h_{t}) = 
	\begin{cases}
		\displaystyle \exp\left( -\frac{\ve_{t}^{\prime} \mSigma^{-1} \ve_{t} + \ve_{t + 1}^{\prime} \mSigma^{-1} \ve_{t + 1}}{2} \right), & \text{if $t < T$}\\
		\displaystyle \exp\left( -\frac{\ve_{t}^{\prime} \mSigma^{-1} \ve_{t}}{2} \right), & \text{if $t =T$}
	\end{cases}.
\end{eqnarray*}
In the sampling of $h_{t}$ for $t = 1, 2, \ldots, T$, we use normal distributions as the proposal distributions along with tuned random-walk procedures suggested by \citet{HOLLOWAY2002383}.

\subsection{Sampling $\valpha$, $\mB$ and $\mSigma$}
Let $\vy = \text{vec}(\mY)$ and 
$\mZ = \left\{ ( \mI_{m} \otimes (1, \vy_{0}^{\prime} ) )^{\prime}, ( \mI_{m} \otimes (1, \vy_{1}^{\prime} ) )^{\prime}, \ldots, ( \mI_{m} \otimes (1, \vy_{T - 1}^{\prime} ) )^{\prime} \right\}^{\prime}$, respectively, where
$\mI_{m}$ is an $m \times m$ unit matrix.
Then,
\begin{eqnarray*}
	\vbeta | \mSigma, \mY &\sim& \N_{S}\left( \hat{\vbeta}, \hat{\mOmega} \right),\\
	\mSigma | \valpha, \mB, \mY &\sim& \IW\left( \hat{\nu}, \hat{\mSigma} \right)
\end{eqnarray*}
where
$\vbeta = \text{vec}\left\{ ( \valpha, \mB )^{\prime} \right\}$,
$S$ is the space, where $\mB$ satisfies the stationary condition,
$\hat{\mOmega} =\left\{ \mZ^{\prime} \left( \mI_{T} \otimes \mSigma^{-1} \right) \mZ + \mOmega_{0}^{-1} \right\}^{-1}$,
$\hat{\vbeta} = \hat{\mOmega} \left\{ \mZ^{\prime} \left( \mI_{T} \otimes \mSigma^{-1} \right) \vy + \mOmega_{0}^{-1} \vbeta_{0} \right\}$,
$\hat{\nu} = T + \nu_{0}$,
$\hat{\mSigma} = \mE \mE^{\prime} + \mSigma_{0}$,
and
$\mE = (\ve_{1}, \ve_{2}, \ldots, \ve_{T})$.

\end{appendices}

\end{document}